\documentclass[a4paper,onecolumn,11pt,accepted=2026-06-10]{quantumarticle}
\pdfoutput=1
\usepackage[utf8]{inputenc}
\usepackage[english]{babel}
\usepackage[T1]{fontenc}
\usepackage{amsmath}
\usepackage{booktabs}
\usepackage[hyperfootnotes=false]{hyperref}
\usepackage{mathtools, nccmath}
\usepackage{setspace}
\usepackage{outlines}
\usepackage{fancyvrb}
\usepackage{mathtools}
\usepackage{physics}
\usepackage{comment}
\usepackage{bbold}
\usepackage{amssymb}
\usepackage{cancel}
\usepackage{graphicx}
\usepackage[caption=false]{subfig}
\usepackage[usenames, dvipsnames]{color}
\usepackage{soul}
\usepackage{tensor}
\usepackage{amsthm}
\usepackage[caption=false]{subfig}
\usepackage{algorithm}

\usepackage{tikz}
\usepackage{lipsum}
\usepackage{tkz-graph}
\usetikzlibrary{hobby,backgrounds,calc,trees}
\usepackage{xcolor}
\usepackage{soul}
\usetikzlibrary{positioning}
\usepackage{ulem}
\usepackage{yquant}
\usepackage{caption}
\usepackage{subfig}
\useyquantlanguage{groups}
\usepackage{makecell}
\usepackage{tikz,esvect,tikz-3dplot}
\usetikzlibrary{3d,calc,intersections}
\usepackage{tikz}
\usepackage{tikz-3dplot}
\usepackage{adjustbox}
\usepackage{zx-calculus}
\usetikzlibrary{zx-calculus}
\usepackage{tikz-cd}
\usepackage[official]{eurosym}
\usepackage{lscape}

\makeatletter

\usepackage{tensor}
\usepackage{accents}
\usetikzlibrary{angles,quotes}

\definecolor{zx_red}{RGB}{232, 165, 165}
\definecolor{zx_green}{RGB}{216, 248, 216}

\usepackage[numbers,sort&compress]{natbib}
\usepackage{graphicx}
\usepackage{iftex}
\usepackage{braket}
\usepackage{tikzit}
\usepackage{amsfonts,amssymb}
\usepackage{tablefootnote}

\tikzstyle{gate}=[shape=rectangle, text height=1.5ex, text depth=0.25ex, yshift=0.5mm, fill=white, draw=black, minimum height=5mm, yshift=-0.5mm, minimum width=5mm, font={\small}, tikzit category=circuit]
\tikzstyle{big gate}=[shape=rectangle, text height=1.5ex, text depth=0.25ex, yshift=0.5mm, fill=white, draw=black, minimum height=10mm, yshift=-0.5mm, minimum width=5mm, font={\small}, tikzit category=circuit]
\tikzstyle{Z dot}=[inner sep=0mm, minimum size=2mm, shape=circle, draw=black, fill={rgb,255: red,221; green,255; blue,221}, tikzit category=zx]
\tikzstyle{Z phase dot}=[minimum size=5mm, font={\footnotesize\boldmath}, shape=rectangle, rounded corners=2mm, inner sep=0.2mm, outer sep=-2mm, scale=0.8, tikzit shape=rectangle, draw=black, fill={rgb,255: red,221; green,255; blue,221}, tikzit draw=blue, tikzit category=zx]
\tikzstyle{X dot}=[Z dot, shape=circle, draw=black, fill={rgb,255: red,255; green,136; blue,136}, tikzit category=zx]
\tikzstyle{X phase dot}=[Z phase dot, tikzit shape=rectangle, tikzit draw=blue, fill={rgb,255: red,255; green,136; blue,136}, font={\footnotesize\boldmath}, tikzit category=zx]
\tikzstyle{hadamard}=[fill=yellow, draw=black, shape=rectangle, inner sep=0.6mm, minimum height=1.5mm, minimum width=1.5mm, tikzit category=zx]
\tikzstyle{paulibox}=[fill={rgb,255: red,221; green,221; blue,255}, draw=black, shape=rectangle, inner sep=0.6mm, minimum height=5mm, minimum width=5mm, font={\footnotesize}, text height=1.5ex, text depth=0.25ex, tikzit category=zx]
\tikzstyle{vertex}=[inner sep=0mm, minimum size=1mm, shape=circle, draw=black, fill=black, tikzit category=misc]
\tikzstyle{vertex set}=[inner sep=0mm, minimum size=1mm, shape=circle, draw=black, fill=white, font={\footnotesize\boldmath}, tikzit category=misc]
\tikzstyle{small black dot}=[fill=black, draw=black, shape=circle, inner sep=0pt, minimum width=1.2mm, tikzit category=circuit]
\tikzstyle{cnot ctrl}=[fill=black, draw=black, shape=circle, inner sep=0pt, minimum width=1.2mm, tikzit category=circuit]
\tikzstyle{cnot targ}=[fill=white, draw=white, shape=circle, tikzit category=circuit, label={center:$\oplus$}, inner sep=0pt, minimum width=2.1mm, tikzit fill={rgb,255: red,102; green,204; blue,255}, tikzit draw=black]
\tikzstyle{ket}=[fill=white, draw=black, shape=regular polygon, regular polygon sides=3, regular polygon rotate=-30, scale=0.7, inner sep=1pt, tikzit category=circuit, tikzit shape=rectangle, tikzit fill=green]
\tikzstyle{bra}=[fill=white, draw=black, shape=regular polygon, regular polygon sides=3, regular polygon rotate=30, scale=0.7, inner sep=1pt, tikzit category=circuit, tikzit shape=rectangle, tikzit fill=red]
\tikzstyle{scalar}=[shape=rectangle, text height=1.5ex, text depth=0.25ex, yshift=0.5mm, fill=white, draw=black, minimum height=5mm, yshift=-0.5mm, minimum width=5mm, font={\small}]
\tikzstyle{clabel}=[fill=white, draw=none, shape=rectangle, tikzit fill={rgb,255: red,56; green,255; blue,242}, font={\footnotesize}, inner sep=1pt, tikzit category=labels]
\tikzstyle{empty diagram}=[draw={gray!40!white}, dashed, shape=rectangle, minimum width=1cm, minimum height=1cm, tikzit category=misc]
\tikzstyle{amap}=[fill=white, draw=black, shape=NEbox, tikzit category=asymmetric, tikzit fill=yellow, tikzit shape=rectangle]
\tikzstyle{amap conj}=[fill=white, draw=black, shape=NWbox, tikzit category=asymmetric, tikzit fill=green, tikzit shape=rectangle]
\tikzstyle{amap adj}=[fill=white, draw=black, shape=SEbox, tikzit category=asymmetric, tikzit fill=red, tikzit shape=rectangle]
\tikzstyle{amap trans}=[fill=white, draw=black, shape=SWbox, tikzit category=asymmetric, tikzit fill=orange, tikzit shape=rectangle]
\tikzstyle{astate}=[fill=white, draw=black, shape=NEtriangle, tikzit category=asymmetric, tikzit shape=circle, tikzit fill=yellow]
\tikzstyle{astate conj}=[fill=white, draw=black, shape=NWtriangle, tikzit category=asymmetric, tikzit shape=circle, tikzit fill=green]
\tikzstyle{astate adj}=[fill=white, draw=black, shape=SEtriangle, tikzit category=asymmetric, tikzit shape=circle, tikzit fill=red]
\tikzstyle{astate trans}=[fill=white, draw=black, shape=SWtriangle, tikzit category=asymmetric, tikzit shape=circle, tikzit fill=orange]

\tikzstyle{hadamard edge}=[-, dashed, dash pattern=on 2pt off 0.5pt, thick, draw={rgb,255: red,68; green,136; blue,255}]
\tikzstyle{star edge}=[-, dashed, dash pattern=on 2pt off 0.5pt, thick, draw={rgb,255: red,255; green,136; blue,68}]
\tikzstyle{box edge}=[-, dashed, dash pattern=on 2pt off 0.5pt, thick, draw={rgb,255: red,203; green,192; blue,225}]
\tikzstyle{brace edge}=[-, tikzit draw=blue, decorate, decoration={brace,amplitude=1mm,raise=-1mm}]
\tikzstyle{diredge}=[->]
\tikzstyle{double edge}=[-, double, shorten <=-1mm, shorten >=-1mm, double distance=2pt]
\tikzstyle{gray edge}=[-, {gray!60!white}]
\tikzstyle{pointer edge}=[->, very thick, gray]
\tikzstyle{boldedge}=[-, line width=1.6pt, shorten <=-0.17mm, shorten >=-0.17mm]
\tikzstyle{bidir edge}=[<->, very thick, draw={rgb,255: red,191; green,191; blue,191}]
\tikzstyle{separator edge}=[-, dashed, dash pattern=on 2pt off 0.5pt, thick, draw={rgb,255: red,153; green,153; blue,153}]

\tikzstyle{X Web}=[-, preaction={line width=1mm, draw={rgb,255: red,150; green,100; blue,100}}]
\tikzstyle{Z Web}=[-, preaction={line width=1.7mm, draw={rgb,255: red,158; green,200; blue,158}}]
\tikzstyle{XZ Web}=[-, preaction={line width=1.8mm, draw={rgb,255: red,158; green,200; blue,158}}, preaction={line width=1mm, draw={rgb,255: red,150; green,100; blue,100}}]

\usepackage{algorithm}
\usepackage{algpseudocode}
\algnewcommand{\To}{\textbf{To }}
\algnewcommand\Input{\item[\textbf{Input:}]}%
\algnewcommand\Output{\item[\textbf{Output:}]}%
\usepackage{pgffor}%

\begin{document}

\title{Simulating magic state cultivation with few Clifford terms}


\author{Kwok Ho Wan}
\orcid{0000-0002-1762-1001}
\affiliation{Blackett Laboratory, Imperial College London, South Kensington, London SW7 2AZ, UK}
\affiliation{Mathematical Institute, University of Oxford, Andrew Wiles Building, Woodstock Road, Oxford OX2 6GG, UK}
\email{((initials))1496((at))((9.81))mail.com}
\thanks{current affiliation: PsiQuantum, Palo Alto.}
\author{Zhenghao Zhong}
\orcid{0000-0001-5159-1013}
\affiliation{Mathematical Institute, University of Oxford, Andrew Wiles Building, Woodstock Road, Oxford OX2 6GG, UK}
\author{Ainhoa Zapirain}
\affiliation{Independent Researcher}

\begin{abstract}
Building upon [arXiv:2509.01224], we present a few methods on how to simulate the non-Clifford $d=5$ magic state cultivation circuits [arXiv:2409.17595] with a sum of $\approx 8$ Clifford ZX-diagrams on average, at $0.1\%$ noise. Compared to a magic cat state stabiliser decomposition of all $53$ non-Clifford spiders ($6{,}377{,}292$ terms required), this is more than $7 \times 10^{5}$ times reduction in the number of terms. Our stabiliser decomposition has the advantage of representing the final non-Clifford state (in light of circuit errors) as a sum of Clifford ZX-diagrams. This will be useful in simulating the escape stage of magic state cultivation, where one needs to port the resultant state of cultivation into a larger Clifford circuit with many more qubits. Still, it's necessary to only track $\approx 8$ Clifford terms. Our result sheds light on the simulability of operationally relevant, high $T$-count quantum circuits with some internal structure.

Finally, we provide numerical results for full non-Clifford stabiliser rank simulation based on $\mathtt{tsim}$ along with optimisations using our cutting decompositions. Nearly $4\times 10^{6}$ shots per second can be obtained on a laptop for the smaller $d = 3$ circuits at uniform circuit level noise $p=0.0005$, making it only $\sim$$1.1$ times slower than its (circuit-unspecific and un-optimised) fully Clifford proxy simulation via $\mathtt{stim}$ using $S$ gates.
\end{abstract}

\maketitle

\begin{figure*}[t]
\centering
    \scalebox{0.3}{\includegraphics{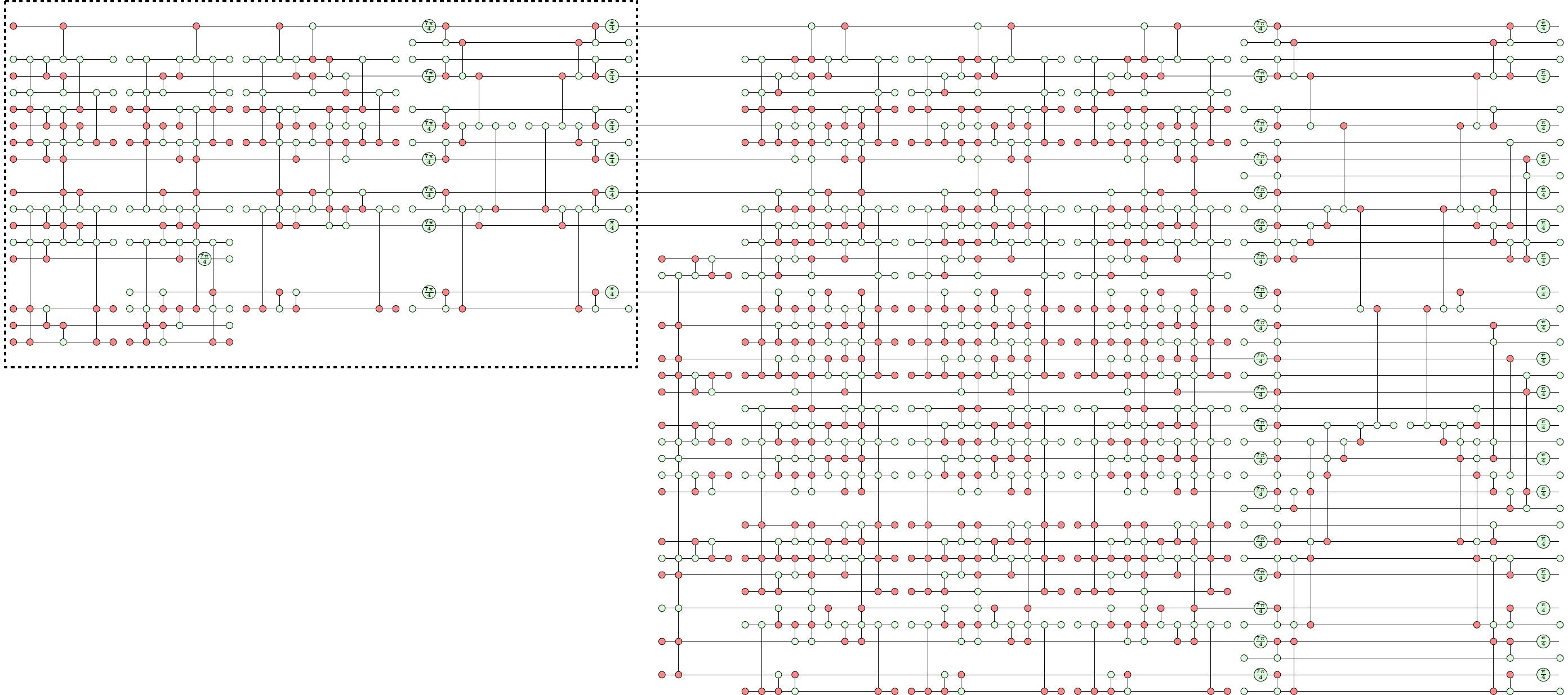}}
    \caption{The $d=5$ magic state cultivation circuit represented as a ZX-diagram. The sub-circuit inside the black dashed box corresponds to the $d=3$ variant. Green (Z-spider) and red (X-spider) nodes denote ZX-calculus spiders, with phases indicated inside each node. Nodes with phase $\pi/4$ or $7\pi/4$ are non-Clifford ($T/T^{\dagger}$) spiders.}
    \label{eq:d5_dashed}
\end{figure*}

\section{Introduction}

Fault-tolerant quantum computation with the Clifford $+ \; T \; \text{gate}$  set requires high-fidelity logical magic states $\ket{\bar{T}}$ to implement the logical non-Clifford $\bar{T}$ gate \cite{Bravyi_2005, Litinski_2019}. The standard approach, magic state distillation, carries a large spacetime overhead \cite{Litinski2019game}. Recently, Gidney, Shutty and Jones~\cite{gidney2024magicstatecultivationgrowing} proposed magic state cultivation, which grows a single $\ket{\bar{T}}$ state within a 2D colour code at a fraction of the cost (${\sim}3\times 10^{3}$ expected qubit-rounds for the un-grown distance $d=5$ colour code variant at ${\sim}6\times 10^{-10}$ logical error rate). Classical simulation of these cultivation circuits is essential for benchmarking their performance at low logical error rate.

The $d=5$ magic state cultivation circuit \cite{gidney2024magicstatecultivationgrowing}, shown in Figure~\ref{eq:d5_dashed}, is difficult to simulate due to its high $T$-count of $53$ $T(T^{\dagger})$ gates. The authors of \cite{gidney2024magicstatecultivationgrowing} simulated most of the larger $d=5$ circuits by replacing $T(T^{\dagger})$ gates entirely with $S(S^{\dagger})$ gates. Applying the same replacement to the $d=3$ sub-circuit (inside the black dashed box of Figure~\ref{eq:d5_dashed}) reveals an $\approx 2\times$ discrepancy in logical error rate (LER) when benchmarked against brute-force state vector simulation. This demonstrates the need for simulation techniques that handle the non-Clifford gates exactly while remaining compatible with Monte Carlo error sampling.

In this work, we use techniques from \cite{wan2025cuttingstabiliserdecompositionsmagic} to decompose the $d=5$ magic state cultivation circuit from Figure~\ref{eq:d5_dashed} into $\sim 8$ Clifford ZX-diagrams on average, while fully accounting for Pauli errors on every edge in the ZX-diagram at error rates in the operationally relevant regime of $10^{-4}$ to $10^{-3}$. A meaningful benchmark of logical performance\footnote{And any extension to simulate the escape stage end-to-end.} without the $T(T^{\dagger}) \to S(S^{\dagger})$ substitution may therefore be feasible, given sufficient integration of ZX-calculus-based stabiliser decomposition into tools such as $\mathtt{tsim}$ \cite{QuEraComputing_tsim_2026,haenel2026tsimfastuniversalsimulator}.

For the smaller $d=3$ sub-circuit, we complement the analytical decomposition with a full numerical simulation pipeline. By combining spider cutting~\cite{Sutcliffe2025thesis} with BSS stabiliser decomposition~\cite{BSS2016}, the circuit compiles into $120$ Clifford graphs with a total of $28$ node terms, $192$ sign terms, and $1{,}408$ phase-pair terms. We implement several optimisations on top of $\mathtt{tsim}$ \cite{QuEraComputing_tsim_2026}: BLAS-accelerated evaluation via matmul-based binary row sums, enumeration-based sampling over the $32$ measurement-outcome combinations, noiseless outcome caching, and full-program JIT compilation. 

We further accelerate sampling by replacing per-shot channel sampling with a sparse geometric-skip sampler that generates events only at non-identity fire positions, deduplicating repeated noise-channel patterns with a persistent cross-batch evaluation cache, and compiling the combo-evaluation loop via \texttt{jax.lax.scan}. Together, these yield a throughput exceeding $4$ million shots per second on a laptop (Apple M4 Macbook Pro) at circuit-level noise $p = 5\times10^{-4}$, only ${\sim}\,1.1\times$ slower than a fully Clifford proxy simulation via $\mathtt{stim}$, partially addressing the open problem of simulating magic state cultivation on common day-to-day hardware posed by \cite{gidney2024magicstatecultivationgrowing}. Note that our pipeline is optimised specifically for this circuit, while $\mathtt{stim}$ is a general-purpose Clifford simulator. This confirms a logical error rate of $\sim\!10^{-6}$ at $p = 10^{-3}$ for the $d=3$ circuit without resorting to the $T \to S$ substitution or statevector simulations.

This paper is organised as follows. We first provide preliminaries on ZX-calculus, Pauli webs, and magic state distillation. We then review the magic cat state and BSS stabiliser decompositions, before applying them to the $d=3$ and $d=5$ cultivation circuits. The cutting decomposition is revisited with numerical evidence for its efficiency under a Pauli error noise model, where by each edge in ZX-diagram's can acquire a Pauli error. Subsequent sections treat randomised measurement spiders, post-selected closed Pauli webs, and an end-to-end simulation pipeline. The numerical methods and accelerated sampling pipeline with full benchmarks follow, and we conclude with a summary and discussion.

\section{Preliminaries}
\label{sec:zx-notation}

\subsection{The Clifford$+T$ gate set}

Fault-tolerant quantum computation can be built on the \textit{Clifford$+T$} gate set. The Clifford group is generated by:
\begin{equation}
\label{eq:clifford_gp}
\Bigg\langle H = \frac{1}{\sqrt{2}}\begin{pmatrix} 1 & 1 \\ 1 & -1 \end{pmatrix}, \quad
S = \begin{pmatrix} 1 & 0 \\ 0 & i \end{pmatrix}, \quad
\mathrm{CNOT} = \ketbra{0}{0}\!\otimes\! I + \ketbra{1}{1}\!\otimes\! \begin{pmatrix} 0 & 1 \\ 1 & 0 \end{pmatrix} \nonumber \Bigg\rangle \ .
\end{equation}
Circuits composed solely of Clifford gates are efficiently classically simulable \cite{gottesman1998heisenbergrepresentationquantumcomputers, Aaronson_2004}. Adding the non-Clifford $T = \mathrm{diag}(1,e^{i\pi/4})$ gate yields a universal gate set. By the Eastin--Knill theorem \cite{eastin2009restrictions}, 
no quantum error-correcting code admits a continuous symmetry acting transversally on its physical qubits; in particular, the logical $\bar{T}$ gate is not transversal in the surface and colour codes. The logical $\bar{T}$ gate can instead be performed via consuming high quality pre-prepared magic states $\ket{\bar{T}} = \bar{T}\ket{\bar{+}}$ \cite{Bravyi_2005,Litinski2019game,Litinski_2019}.

\subsection{ZX-calculus notation}

We briefly summarise the ZX-calculus notation used in this paper; for comprehensive treatments see \cite{coecke2017picturing, vandewetering2020zxcalculus, KissingerWetering2024Book}. ZX-calculus is a graphical language for reasoning about quantum processes, in which quantum circuits are represented as diagrams (ZX-diagrams) built from fundamental objects called \textit{spiders}. The basic spiders of ZX-calculus are summarised in Table~\ref{tab:zx_generators}.

\begin{table}[h]
\centering
\caption{The spiders of ZX-calculus: Z-spiders, X-spiders (defined via Hadamard conjugation), and Hadamard gates (with yellow-box and blue-dashed-edge notations).}
\label{tab:zx_generators}
\begin{tabular}{lll}
\toprule
\textbf{Generator} & \textbf{Diagram} & \textbf{Tensor} \\
\midrule
Z-spider (green) &
\raisebox{-0.3\height}{\scalebox{1.0}{\input{zx_intro_Z_spider.tikz}}} &
$\ket{0\cdots 0} + e^{i\alpha}\ket{1\cdots 1}$ \\[1em]
X-spider (red) &
\raisebox{-0.3\height}{\scalebox{1.0}{\input{zx_intro_X_spider.tikz}}} $\equiv$ \raisebox{-0.3\height}{\scalebox{1.0}{\input{zx_intro_X_spider_Hdef.tikz}}} &
$\ket{\overbrace{+}^{i_1}\cdots \overbrace{+}^{i_m}} + e^{i\alpha}\ket{-\cdots -}$ \\[1em]
Hadamard gate &
\raisebox{-0.3\height}{\scalebox{0.7}{\input{zx_intro_hadamard_box.tikz}}} $\equiv$ \raisebox{-0.3\height}{\scalebox{0.7}{\input{zx_intro_hadamard_line.tikz}}} &
$\frac{1}{\sqrt{2}}\begin{psmallmatrix}1 & 1\\ 1 & -1\end{psmallmatrix}$ \\
\bottomrule
\end{tabular}
\end{table}

A key rule is \textit{spider fusion}: two connected spiders of the same colour merge into one whose phase is the sum of the two phases (see figure~\ref{fig:zx_rules}). Standard quantum gates translate directly into ZX-diagrams: CNOT becomes a Z-spider connected to an X-spider. The $T$ gate is a Z-spider with phase $\pi/4$ (i.e.\ $\scalebox{0.7}{
\begin{tikzpicture}
    \begin{pgfonlayer}{nodelayer}
        \node [style=Z phase dot] (2) at (8.50, -0.00) {$\frac{\pi}{4}$};
        \node [style=none] (3) at (7.50, -0.00) {};
        \node [style=none] (4) at (9.50, -0.00) {};
    \end{pgfonlayer}
    \begin{pgfonlayer}{edgelayer}
        \draw (2) to (3);
        \draw (2) to (4);
    \end{pgfonlayer}
\end{tikzpicture}
}$), and a $S$ gates becomes a Z-spider with phase $\pi/2$ (i.e.\ $\scalebox{0.7}{
\begin{tikzpicture}
    \begin{pgfonlayer}{nodelayer}
        \node [style=Z phase dot] (2) at (8.50, -0.00) {$\frac{\pi}{2}$};
        \node [style=none] (3) at (7.50, -0.00) {};
        \node [style=none] (4) at (9.50, -0.00) {};
    \end{pgfonlayer}
    \begin{pgfonlayer}{edgelayer}
        \draw (2) to (3);
        \draw (2) to (4);
    \end{pgfonlayer}
\end{tikzpicture}
}$). The full Clifford group generators~\eqref{eq:clifford_gp} can be re-written as ZX-diagrams:
\begin{equation}
\label{eq:clifford_gp}
\Bigg\langle H = \scalebox{0.7}{
\begin{tikzpicture}
    \begin{pgfonlayer}{nodelayer}
        \node [style=hadamard] (0) at (0.00, 0.00) {};
        \node [style=none] (1) at (-1.00, 0.00) {};
        \node [style=none] (2) at (1.00, 0.00) {};
    \end{pgfonlayer}
    \begin{pgfonlayer}{edgelayer}
        \draw (0) to (1);
        \draw (0) to (2);
    \end{pgfonlayer}
\end{tikzpicture}
}, \quad
S = \scalebox{0.7}{
\begin{tikzpicture}
    \begin{pgfonlayer}{nodelayer}
        \node [style=Z phase dot] (2) at (8.50, -0.00) {$\frac{\pi}{2}$};
        \node [style=none] (3) at (7.50, -0.00) {};
        \node [style=none] (4) at (9.50, -0.00) {};
    \end{pgfonlayer}
    \begin{pgfonlayer}{edgelayer}
        \draw (2) to (3);
        \draw (2) to (4);
    \end{pgfonlayer}
\end{tikzpicture}
}, \quad
\mathrm{CNOT} = \scalebox{0.7}{\input{zx_intro_cnot.tikz}} \nonumber \Bigg\rangle \ .
\end{equation}
Next, the remaining gate required for universal quantum computation, the $T$ gate, can be consumed via a magic state $\ket{T} = \scalebox{0.7}{
\begin{tikzpicture}
    \begin{pgfonlayer}{nodelayer}
        \node [style=Z phase dot] (0) at (-2.00, -0.00) {$\frac{\pi}{4}$};
        \node [style=none] (1) at (-1.00, -0.00) {};
    \end{pgfonlayer}
    \begin{pgfonlayer}{edgelayer}
        \draw (0) to (1);
    \end{pgfonlayer}
\end{tikzpicture}
}$ through the following gadget \cite{Bravyi_2005,Litinski_2019}:
\begin{equation}
    \scalebox{1}{\input{zx_intro_tgate_consume.tikz}} = \scalebox{1}{
\begin{tikzpicture}
    \begin{pgfonlayer}{nodelayer}
        \node [style=Z phase dot] (2) at (8.50, -0.00) {$\frac{\pi}{4}$};
        \node [style=none] (3) at (7.50, -0.00) {};
        \node [style=none] (4) at (9.50, -0.00) {};
    \end{pgfonlayer}
    \begin{pgfonlayer}{edgelayer}
        \draw (2) to (3);
        \draw (2) to (4);
    \end{pgfonlayer}
\end{tikzpicture}
} \nonumber
\end{equation}
\noindent where $a \in \{0,1\}$ is the measurement outcome and the $S^a$ correction is applied classically. A ZX-diagram with no open legs (a \textit{closed diagram}) evaluates to a complex scalar. Non-Clifford spiders are those with phases $\in \{\pi/4, 3\pi/4, 5\pi/4, 7\pi/4\}$. The stabiliser decomposition techniques in this work aim to express a non-Clifford ZX-diagram as a sum of Clifford diagrams, each of which can be efficiently evaluated.

The power of ZX-calculus lies in its rewrite rules, which allow diagrams to be simplified while preserving their meaning. Figure~\ref{fig:zx_rules} summarises the key identities used throughout this work, including spider fusion (top left), hadamard push (middle left), $\pi$-copy (top right), the bialgebra rule (middle middle), identity removal, and scalar evaluation rules~\cite{KissingerWetering2024Book}.

\begin{figure}[h]
\centering
    \scalebox{1}{\input{zx_intro_basic_rules.tikz}}
    \caption{Standard ZX-calculus rewrite rules used in this work. Top left: spider fusion. Middle left: colour change via Hadamard push. Top right and middle middle respectively: $\pi$-copy and bialgebra rules. Adapted from \cite{kissinger2022reduced, sutcliffe2024fast}.}
    \label{fig:zx_rules}
\end{figure}

These basic rules can be composed to derive more powerful simplification procedures. The \textit{local complementation} rule (Figure~\ref{fig:lcomp_pivot}, left) and the \textit{pivoting} rule (Figure~\ref{fig:lcomp_pivot}, right). These derived rules, together with \textit{phase gadget} fusion, form the core of the $\mathtt{full\_reduce}$ in $\mathtt{pyzx}$ simplification routine \cite{kissinger2020Pyzx, haenel2026tsimfastuniversalsimulator} are used extensively in our pipeline.

\begin{figure}[h]
\centering
    \raisebox{-1cm}{\scalebox{0.6}{\input{zx_intro_local_comp.tikz}}}
    \quad
    \scalebox{0.6}{\input{zx_intro_pivot.tikz}}
    \caption{Left: local complementation rule. An internal $(\pi/2 + b\pi)$-phase spider is removed, complementing the Hadamard-edge connectivity among its neighbours and shifting each neighbour's phase by $-\pi/2 - b\pi$. Right: pivoting rule. A connected pair of $a\pi$- and $b\pi$-phase spiders is removed, with phases and connectivity updated accordingly. The scalar exponent $E$ is related to the graph complementation of the neighbourhood; see \cite{kissinger2022reduced} for its precise definition. Adapted from \cite{kissinger2022reduced, haenel2026tsimfastuniversalsimulator}.}
    \label{fig:lcomp_pivot}
\end{figure}

\subsection{Pauli webs}
\label{sec:pauli-webs-prelim}

A \textit{Pauli web} \cite{Bombin_2024, rodatz2025faulttoleranceconstruction} is a consistent highlighting of edges in a ZX-diagram by Pauli operators (${\color{red}X}$, ${\color{green}Z}$, or ${\color{blue}Y} \propto {\color{red}X}{\color{green}Z}$), subject to compatibility rules at each spider determined by its phase. An \textit{open Pauli web} is one that highlights at least one input or output edge (i.e.\ it has free legs). A \textit{closed Pauli web} (or \textit{detecting region}) is a Pauli web that highlights no input or output edges --- it is entirely internal to the circuit (see Figure~\ref{fig:pauli_web_types}). When the ZX-diagram represents a quantum error-correcting circuit, closed Pauli webs correspond to parity checks. See \cite{Bombin_2024,rodatz2025faulttoleranceconstruction} for a more comprehensive notation on Pauli webs, whereby we use the colouring convention from \cite{rodatz2025faulttoleranceconstruction}.

\begin{figure}[h]
\centering
    \scalebox{1}{\input{zx_intro_pauli_web_types.tikz}}
    \caption{Left: a closed Pauli web (detecting region) --- the green highlighting is entirely internal, with no input or output edges highlighted. Right: an open Pauli web --- the highlighting extends to input/output edges (free legs). Adapted from \cite{Bombin_2024, rodatz2025faulttoleranceconstruction}.}
    \label{fig:pauli_web_types}
\end{figure}

\subsection{Magic state distillation}
\label{sec:msd}

Historically, a key way to prepare high quality logical $\ket{\bar{T}}$ states from faulty injected $\ket{\bar{T}}$ states~\cite{Li_2015} was through \textit{magic state distillation} \cite{Bravyi_2005}, where multiple noisy magic states are consumed through a post-selected circuit to produce fewer, higher-fidelity copies. The most common protocol is the 15-to-1 distillation circuit, which takes 15 noisy $\ket{T}$ states and produces one output state with suppressed error rate $O(p^3)$, where $p$ is the input error rate and succesful post-selection probabilty of $\mathcal{O}(p)$. Figure~\ref{fig:15to1} shows this circuit in ZX form, containing 15 non-Clifford ($\pi/4$) spiders.
\begin{figure}[h]
\centering
    \scalebox{1}{\input{zx_intro_15_to_1.tikz}}
    \caption{The 15-to-1 logical magic state distillation circuit in ZX form \cite{Litinski_2019}. The 15 $\pi/4$ phase spiders are the non-Clifford $T$-state inputs. 
    }
    \label{fig:15to1}
\end{figure}
Distillation works by post-selecting upon all $+1$ parity (or $0$ in binary) final measurement values ($a_j\pi = 0$ in figure~\ref{fig:15to1}) \cite{Bravyi_2005}. While effective, it carries a large spacetime overhead \cite{Litinski_2019, Litinski2019game}. An alternative is magic state cultivation \cite{gidney2024magicstatecultivationgrowing}, which grows a $\ket{\bar{T}}$ state within a 2D colour code at a fraction of the cost. In either case, to benchmark these protocols one needs to simulate circuits with many non-Clifford gates exactly, which motivates the stabiliser decomposition techniques in this work.

\section{Magic cat state stabiliser decomposition}
Magic cat states \cite{Qassim_2021,Codsi2022Masters,Kiss2022} are a family of states parametrised by integer $m$:
\begin{equation}
    \ket{\text{cat}_m} = \frac{1}{\sqrt{2}} (I^{\otimes m}+Z^{\otimes m})\ket{T}^{\otimes m} \ .
\end{equation}
This state can be written as the following ZX-diagram
\begin{equation}
    \label{eq:cat-n}
    \ket{\text{cat}_m} = \scalebox{1}{\input{cat-state.tikz}} \ ,
\end{equation}
\noindent with $m$ legs. The magic cat state stabiliser decomposition is the expansion of this non-Clifford ZX-diagram as a sum of Clifford ZX-diagrams.
\begin{equation}
    \label{eq:cat-m}
    \scalebox{1}{\input{cat-state.tikz}} = \sum_{j=1}^{\chi} a_j \begin{pmatrix}
        \scalebox{0.9}{\input{clifford_zx_diag.tikz}}
    \end{pmatrix}_j
\end{equation}
where $a_j \in \mathbb{C}$ are complex coefficients and each term in the sum is a Clifford ZX-diagram. This is a stabiliser rank decomposition \cite{BSS2016, bravyi2019simulationquantumcircuitslowrank} expressed in ZX-calculus notations. Explicit number of terms ($\chi$) involved in this sum for various $\ket{\text{cat}_m}$ is available at \cite{Codsi2022Masters}
Note that $\ket{T}^{\otimes (m-1)}$ can be obtained via a $\ket{\text{cat}_m}$ state by measuring one of its legs by fusing it with a $\scalebox{1}{
\begin{tikzpicture}
    \begin{pgfonlayer}{nodelayer}
        \node [style=Z phase dot] (0) at (-2.00, -0.00) {$\frac{7\pi}{4}$};
        \node [style=none] (1) at (-1.00, -0.00) {};
    \end{pgfonlayer}
    \begin{pgfonlayer}{edgelayer}
        \draw (0) to (1);
    \end{pgfonlayer}
\end{tikzpicture}
}$ \cite{Kiss2022}:
\begin{equation}
    \label{eq:magic_cat_decomp}
    \Big(\scalebox{1}{
\begin{tikzpicture}
    \begin{pgfonlayer}{nodelayer}
        \node [style=Z phase dot] (0) at (-2.00, -0.00) {$\frac{\pi}{4}$};
        \node [style=none] (1) at (-1.00, -0.00) {};
    \end{pgfonlayer}
    \begin{pgfonlayer}{edgelayer}
        \draw (0) to (1);
    \end{pgfonlayer}
\end{tikzpicture}
}\Big)^{\otimes (m-1)} = \scalebox{1}{\input{cat_state_to_Ts.tikz}} \ .
\end{equation} 
We will call this the magic cat state stabiliser decomposition of $\ket{T}^{\otimes (m-1)}$. See examples of $\ket{\text{cat}_{4}} $ and $ \ket{\text{cat}_{6}}$ \cite{Kiss2022}\footnote{Tikz figures \eqref{eq:cat4} and \eqref{eq:cat6} taken from \cite{Kiss2022}.}:
\begin{equation}
\label{eq:cat4}
    \scalebox{1}{\input{cat-4-decomp.tikz}} \ ,
\end{equation}
\begin{equation}
\label{eq:cat6}
    \scalebox{0.65}{\input{cat-6-decomp.tikz}} \ .
\end{equation}
These will be used to decompose the $d=3$ magic state cultivation circuit into Clifford terms.

\section{The BSS decomposition}
\label{sec:bss-intro}

An alternative stabiliser decomposition is the Bravyi-Smith-Smolin (BSS) decomposition \cite{BSS2016}, a foundational stabiliser rank technique that expresses six $T$-states as a sum of seven stabiliser (Clifford) terms. In ZX-calculus form \cite{kissinger2022reduced}, this is shown in Figure~\ref{fig:bss_decomp}. The BSS decomposition serves as a stand-alone decomposition method and as a secondary fallback in our cutting-based pipeline (Section~\ref{sec:zx-decomposition}).

\begin{figure}[h]
\centering
    \scalebox{0.85}{\input{zx_intro_bss_decomp.tikz}}
    \caption{The BSS stabiliser decomposition \cite{BSS2016} in ZX-calculus form \cite{kissinger2022reduced, wan2025cuttingstabiliserdecompositionsmagic}. Six $T$-states (top left, each with phase $\pi/4$) are decomposed into a sum of seven Clifford ZX-diagrams with scalar coefficients.}
    \label{fig:bss_decomp}
\end{figure}

\section{$d=3$ magic state cultivation circuit}
Now, let's see how one can use the magic cat state stabiliser decomposition \cite{Codsi2022Masters} to strongly simulate the $d=3$ magic state cultivation circuit (\eqref{eq:MSC_d_3_full} or ZX-diagram inside the black dashed box of Figure~\ref{eq:d5_dashed}) exactly.
We first un-fuse all the $\scalebox{1}{
\begin{tikzpicture}
    \begin{pgfonlayer}{nodelayer}
        \node [style=Z phase dot] (0) at (-2.00, -0.00) {$\frac{\pi}{4}$};
        \node [style=none] (1) at (-1.00, -0.00) {};
    \end{pgfonlayer}
    \begin{pgfonlayer}{edgelayer}
        \draw (0) to (1);
    \end{pgfonlayer}
\end{tikzpicture}
}/\scalebox{1}{
\begin{tikzpicture}
    \begin{pgfonlayer}{nodelayer}
        \node [style=Z phase dot] (0) at (-2.00, -0.00) {$\frac{7\pi}{4}$};
        \node [style=none] (1) at (-1.00, -0.00) {};
    \end{pgfonlayer}
    \begin{pgfonlayer}{edgelayer}
        \draw (0) to (1);
    \end{pgfonlayer}
\end{tikzpicture}
}$ of
\begin{equation}
\centering
    \label{eq:MSC_d_3_full}
    \scalebox{0.38}{\includegraphics{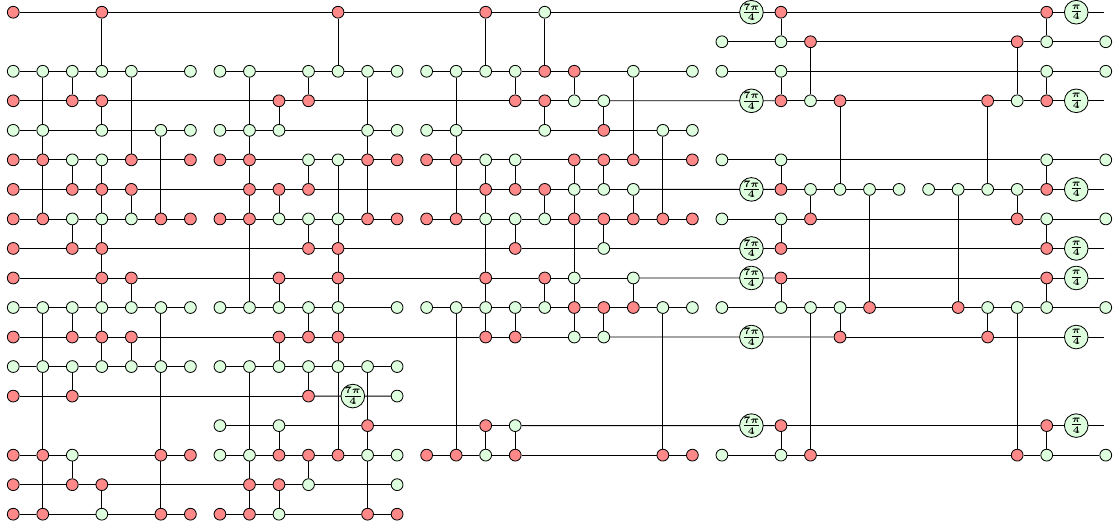}} \ ,
\end{equation}
resulting in:
\begin{equation}
\centering
    \label{eq:d3_unfused_non_clifford}
    \scalebox{0.38}{\includegraphics{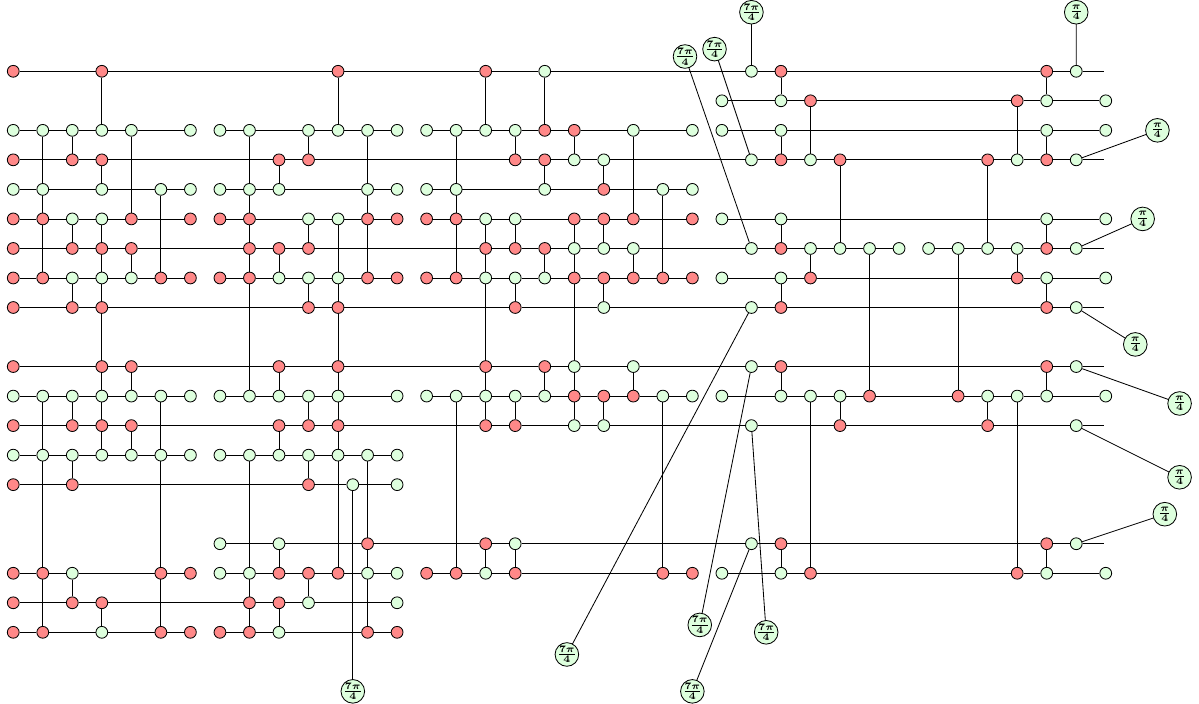}} \ .
\end{equation}
Here, we counted $15$ $\scalebox{1}{
\begin{tikzpicture}
    \begin{pgfonlayer}{nodelayer}
        \node [style=Z phase dot] (0) at (-2.00, -0.00) {$\frac{\pi}{4}$};
        \node [style=none] (1) at (-1.00, -0.00) {};
    \end{pgfonlayer}
    \begin{pgfonlayer}{edgelayer}
        \draw (0) to (1);
    \end{pgfonlayer}
\end{tikzpicture}
}/\scalebox{1}{
\begin{tikzpicture}
    \begin{pgfonlayer}{nodelayer}
        \node [style=Z phase dot] (0) at (-2.00, -0.00) {$\frac{7\pi}{4}$};
        \node [style=none] (1) at (-1.00, -0.00) {};
    \end{pgfonlayer}
    \begin{pgfonlayer}{edgelayer}
        \draw (0) to (1);
    \end{pgfonlayer}
\end{tikzpicture}
}$ spiders. We can chop these $\scalebox{1}{
\begin{tikzpicture}
    \begin{pgfonlayer}{nodelayer}
        \node [style=Z phase dot] (0) at (-2.00, -0.00) {$\frac{\pi}{4}$};
        \node [style=none] (1) at (-1.00, -0.00) {};
    \end{pgfonlayer}
    \begin{pgfonlayer}{edgelayer}
        \draw (0) to (1);
    \end{pgfonlayer}
\end{tikzpicture}
}/\scalebox{1}{
\begin{tikzpicture}
    \begin{pgfonlayer}{nodelayer}
        \node [style=Z phase dot] (0) at (-2.00, -0.00) {$\frac{7\pi}{4}$};
        \node [style=none] (1) at (-1.00, -0.00) {};
    \end{pgfonlayer}
    \begin{pgfonlayer}{edgelayer}
        \draw (0) to (1);
    \end{pgfonlayer}
\end{tikzpicture}
}$ spiders off and fuse them with the legs of the magic cat stabiliser decomposition of $\ket{T}^{\otimes 15}$ up to some local Clifford ($S=\text{diag}(1,i)$) transformations on the $8$ $\scalebox{1}{
\begin{tikzpicture}
    \begin{pgfonlayer}{nodelayer}
        \node [style=Z phase dot] (0) at (-2.00, -0.00) {$\frac{7\pi}{4}$};
        \node [style=none] (1) at (-1.00, -0.00) {};
    \end{pgfonlayer}
    \begin{pgfonlayer}{edgelayer}
        \draw (0) to (1);
    \end{pgfonlayer}
\end{tikzpicture}
}$ spiders. The underlying $\ket{\text{cat}_{16}}$ decomposition uses the $\ket{\text{cat}_{4}}\otimes\ket{\text{cat}_{6}}^{\otimes 3}$ decompositions fused together at some legs to obtain $\ket{\text{cat}_{16}}$.

Using the table from \cite[Table A.4]{Codsi2022Masters}, we  see that this sub-circuit can be represented with $2\times 54=108$ pure Clifford ZX-diagrams, whilst still maintaining its circuit structure without any ZX-rewrite/reduction/contractions. Hence, one can still insert errors in the same locations in each expanded term's circuit for Monte-Carlo simulations. This is useful in the context of computing logical error rates via sampling. Note that the number of terms here are double that of those in the table of \cite{Codsi2022Masters} since we are counting pure Clifford ZX-diagrams, whereas each term in \cite{Codsi2022Masters} has a single $\scalebox{1}{
\begin{tikzpicture}
    \begin{pgfonlayer}{nodelayer}
        \node [style=Z phase dot] (0) at (-2.00, -0.00) {$\frac{7\pi}{4}$};
        \node [style=none] (1) at (-1.00, -0.00) {};
    \end{pgfonlayer}
    \begin{pgfonlayer}{edgelayer}
        \draw (0) to (1);
    \end{pgfonlayer}
\end{tikzpicture}
}$ spider. Perhaps more theoretical insights  regarding the logical error rate deviation observed in \cite{gidney2024magicstatecultivationgrowing} can be gained from looking at how error propagates across terms in the stabiliser decomposition of this circuit. This is left for future work. 

This stabiliser decomposition of $\ket{T}^{\otimes 15}$ with an additional overhead of $108\times$, can be similarly applied to the 15-to-1 magic state distillation circuit \cite{Bravyi_2005} with encoded surface code patches \cite{Litinski_2019} for example.

\section{$d=5$ circuit via magic cat state stabiliser decomposition}
The $d=5$ magic state cultivation circuit is just the $d=3$ variant with additional Clifford gates followed by a larger double checking circuit at the end (shown in \eqref{eq:cultiv_DC_d5_pi_by_4}, see also appendix \ref{appendix:dc} for an extended discussion). This sub-routine has $38$ $\scalebox{1}{
\begin{tikzpicture}
    \begin{pgfonlayer}{nodelayer}
        \node [style=Z phase dot] (0) at (-2.00, -0.00) {$\frac{\pi}{4}$};
        \node [style=none] (1) at (-1.00, -0.00) {};
    \end{pgfonlayer}
    \begin{pgfonlayer}{edgelayer}
        \draw (0) to (1);
    \end{pgfonlayer}
\end{tikzpicture}
}/\scalebox{1}{
\begin{tikzpicture}
    \begin{pgfonlayer}{nodelayer}
        \node [style=Z phase dot] (0) at (-2.00, -0.00) {$\frac{7\pi}{4}$};
        \node [style=none] (1) at (-1.00, -0.00) {};
    \end{pgfonlayer}
    \begin{pgfonlayer}{edgelayer}
        \draw (0) to (1);
    \end{pgfonlayer}
\end{tikzpicture}
}$ spiders.
Naively, if we un-fuse all the spiders of the $d=5$ double checking sub-circuit (from \eqref{eq:cultiv_DC_d5_pi_by_4}) inside Figure~\ref{eq:d5_dashed} and replace them with magic cat state stabiliser decomposition of $\ket{T}^{\otimes 38}$, we will arrive at $2\times39{,}366 = 78{,}732$ pure Clifford terms \cite{Codsi2022Masters}. Taking into account the $d=3$ magic state cultivation circuit prior to the $d=5$ double checking circuit, the full $d=5$ circuit (in Figure~\ref{eq:d5_dashed}) will have $108\times 78{,}732=8{,}503{,}056$ terms. 

Looking at the $d=5$ circuit as a whole, with all the $53$ total non-Clifford spiders, we can actually break this down into $\ket{\text{cat}_{17}}$ and $\ket{\text{cat}_{38}}$, converted to $\ket{T}^{\otimes 16}\otimes\ket{T}^{\otimes 37}$ via fusing each magic cat state to $\scalebox{1}{
\begin{tikzpicture}
    \begin{pgfonlayer}{nodelayer}
        \node [style=Z phase dot] (0) at (-2.00, -0.00) {$\frac{7\pi}{4}$};
        \node [style=none] (1) at (-1.00, -0.00) {};
    \end{pgfonlayer}
    \begin{pgfonlayer}{edgelayer}
        \draw (0) to (1);
    \end{pgfonlayer}
\end{tikzpicture}
}$ spiders. This implies  a lower number: $6{,}377{,}292$ total pure Clifford terms. We emphasise that we need to simulate each ZX-reduced ZX-diagram with random Pauli-X/Y/Z errors on each edge, in the context of Monte Carlo sampling for computing logical error rates. 

\begin{equation}
    \label{eq:cultiv_DC_d5_pi_by_4}
    \scalebox{0.4}{\input{cultiv_DC_d5_pi_by_4.tikz}} 
\end{equation} 

Simulating the $d=5$ cultivation circuit in this manner would require $6$ to $8$ million additional terms per shot compared to an equivalent stabiliser circuit\footnote{Swapping $T \rightarrow S$.}. The only advantage is the inclusion of Pauli/Clifford errors comes at no additional cost, as they do not increase the number of terms in the stabiliser decomposition. In the next section, we will show how to reduce this massive overhead with the cutting decomposition even when Pauli errors are taken into account.

\section{The cutting decomposition re-visited}
In an earlier manuscript \cite{wan2025cuttingstabiliserdecompositionsmagic} we studied the cutting decomposition \cite{Codsi2022Masters,Sutcliffe_2024,Sutcliffe2025thesis} and showed that the error-free $d=3$ and $d=5$ magic cultivation circuits can be represented with a stabiliser decomposition of $4$ and $8$ terms respectively \cite{wan2025cuttingstabiliserdecompositionsmagic}. We summarised the results in Figure~\ref{fig:large_d3_d5} (see Appendix~\ref{appendix:cutting_diagrams}, presented in landscape format for readability) for the $d=3$ and $d=5$ cases respectively.
A natural question to ask is whether this number of terms persists when random errors are applied to each edge in the ZX-diagram. For a feasible simulation, we require that Pauli errors be included on each edge of the ZX-diagram while still keeping the number of terms in its cutting decomposition manageable. Furthermore, how does the number of terms scale with the error rate?

In order to study this, we subject every edge of the cultivation circuits to Pauli-${\color{red}X}/{\color{blue}Y}/{\color{green}Z}$ error, where each Pauli-${\color{red}X}/{\color{blue}Y}/{\color{green}Z}$ error occurs with probability $p/3$, governed by the channel:
\begin{equation}
    \mathcal{E}(\rho) = (1-p)\rho + \frac{p}{3}(X\rho X+Y\rho Y+Z\rho Z) \ .
\end{equation}
This is the standard single qubit depolarising error channel applied to every edge. For example, in the $d=3$ circuit (from \eqref{eq:MSC_d_3_full}), an error realisation is shown in \eqref{eq:d3_error},
\begin{equation}
\centering
    \label{eq:d3_error}
    \scalebox{0.38}{\includegraphics{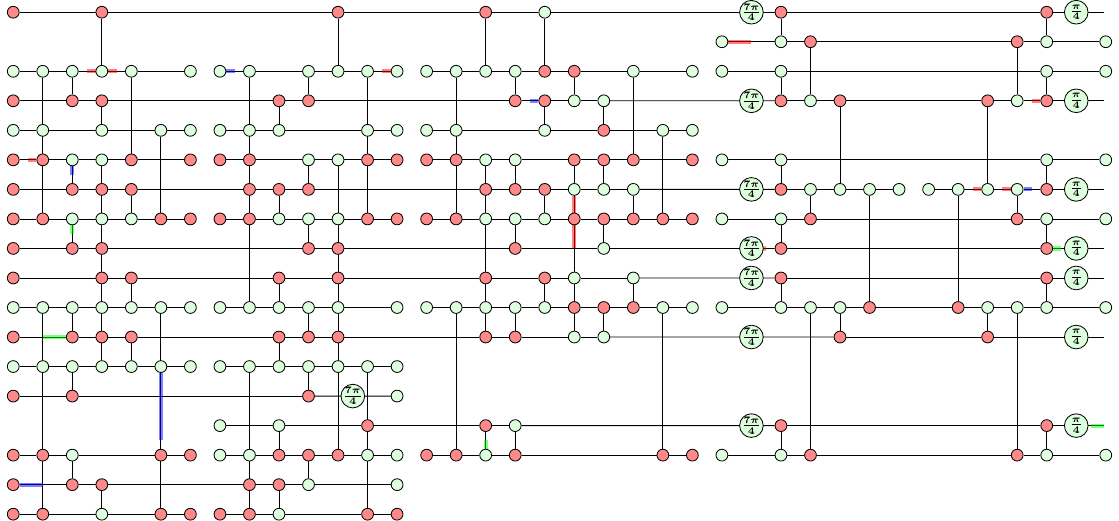}} \ ,
\end{equation}
where coloured half-edges are used to mark Pauli errors on its ZX-diagram edges. This edge decoration notation follows the opposite colouring Pauli web notation of \cite{Bombin_2024} and is consistent with notations from $\mathtt{pyzx}$ \cite{kissinger2020Pyzx}. Note that this error model has a higher error density in its ZX-diagram compared to the uniform circuit-level noise model used in \cite{gidney2024magicstatecultivationgrowing} (see Appendix~\ref{appendix:error_model} for a detailed comparison). Errors happening `in-between CNOTs' shown in \eqref{eq:cnot_err}: 
\begin{equation}
    \label{eq:cnot_err}
    \scalebox{1}{\input{cnot_err_l.tikz}} = \scalebox{1}{\input{cnot_err_r.tikz}}
\end{equation}
are permissible in our error model.

\begin{figure}[h]
    \centering
    \includegraphics[width=0.8\linewidth]{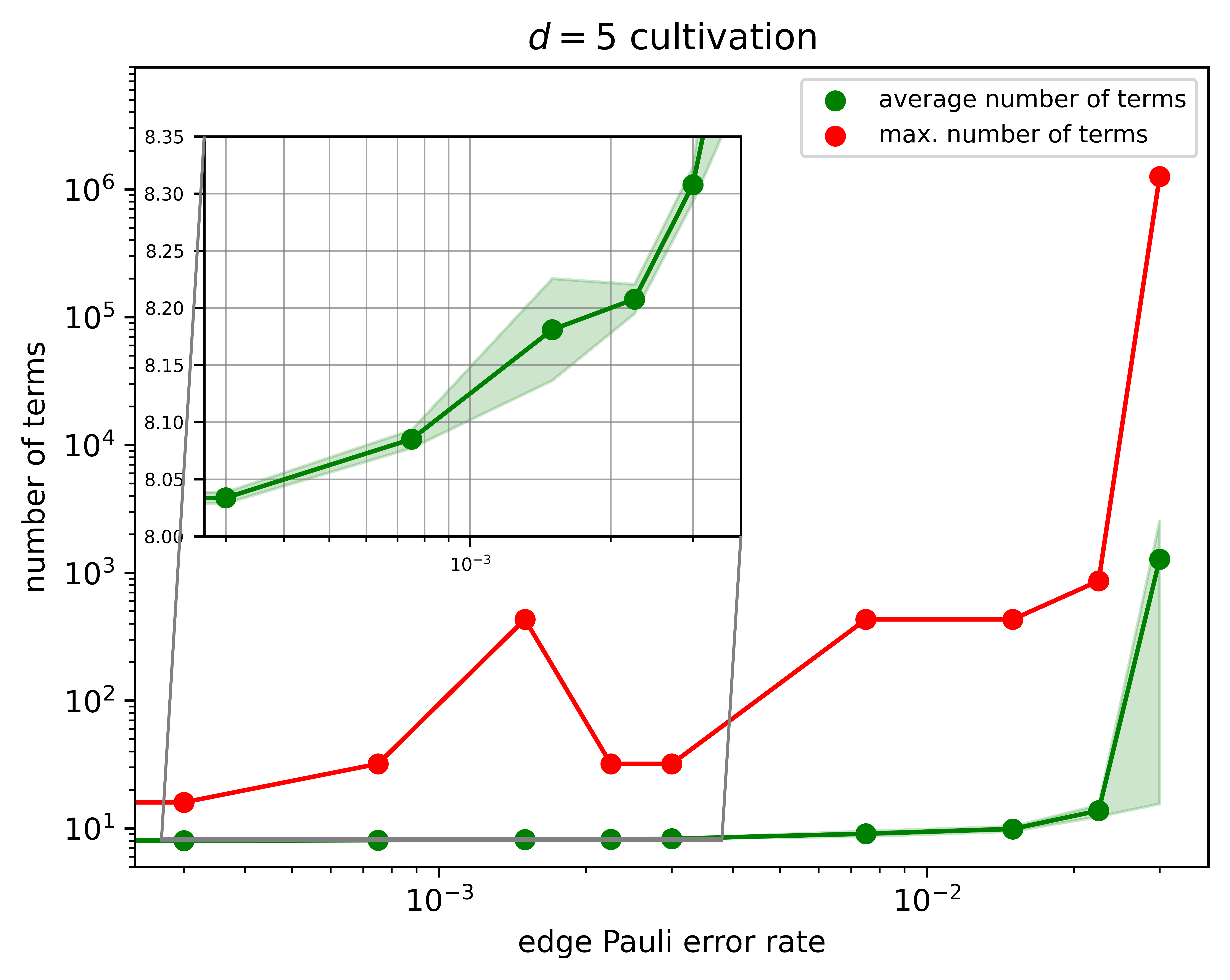}
    \caption{Average and maximum number of terms in the cutting stabiliser decomposition for the $d=5$ circuit. There is a single event containing $432$ terms at edge error rate of $0.0015$.}
    \label{fig:numerics_main_d5}
\end{figure}

\begin{figure}[h]
    \centering
    \includegraphics[width=0.8\linewidth]{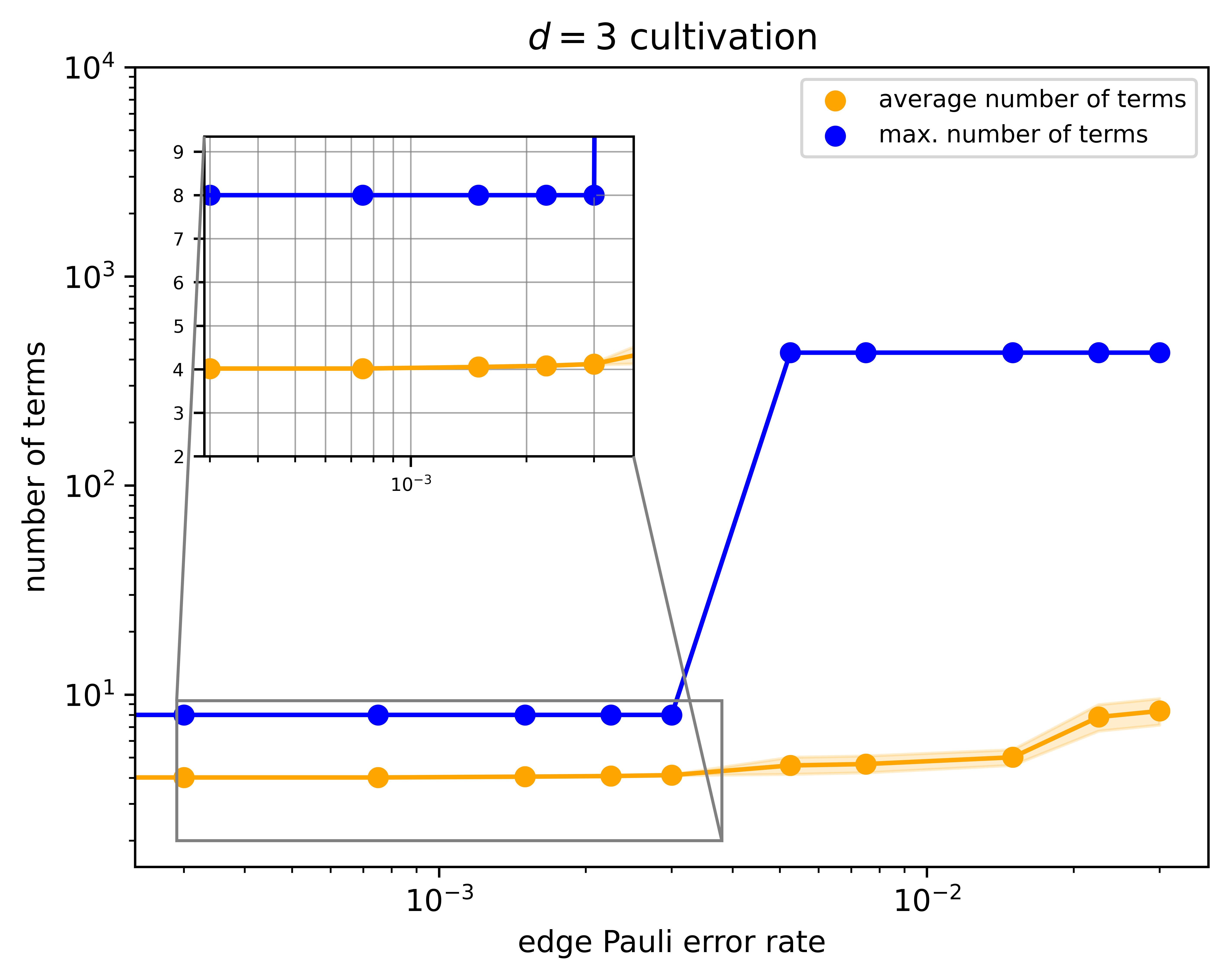}
    \caption{Average and maximum number of terms in the cutting stabiliser decomposition for the $d=3$ circuit.}
    \label{fig:numerics_main_d3}
\end{figure}

We then applied the same cutting decomposition outlined in \cite{wan2025cuttingstabiliserdecompositionsmagic} to the Pauli-errored $d=3$ and $d=5$ cultivation ZX-diagrams. At the end of the cutting decomposition, if we failed to decompose terms into low $T$-count ZX-diagrams (whereby each term contains only $1$ or fewer non-Clifford spiders), we further apply \textbf{secondary} magic cat state decomposition (outlined previously) to those terms. We noticed that in the operationally relevant error regimes of magic state cultivation - $\mathcal{O}(10^{-4})$ to $10^{-3}$ edge error rate, the number of terms is still very reasonable. On average, $\sim8$ ($\sim4$) terms are needed for the $d=5$ ($d=3$) circuits in those error ranges (see figures \ref{fig:numerics_main_d5} and \ref{fig:numerics_main_d3}). To account for any potential large deviation from the mean number of terms, albeit with low probability, we show the maximum number of terms in the same plots. In our numerical experiment, we observed no more than $432$ ($8$) maximum number of terms in the same error intervals over $10^4$ shots. There is a single shot containing $432$ terms at $0.0015$ error rate for the $d=5$ circuit. We don't observe this in the two subsequent higher error data points; this is likely due to our sample size. Nonetheless, the average number of terms is still very low.

Based on these numerical results, we suggest that one can simulate the magic state cultivation circuits, post-selected upon all $+1$ measurements with the cutting stabiliser decomposition \cite{wan2025cuttingstabiliserdecompositionsmagic}. On average, this appears to add only a minor overhead on top of the stabiliser simulations. Next, we will consider the case where $-1$ parity measurement outcomes are permitted.

\section{Randomised measurement spiders}
Post-selecting upon $+1$ outcome exclusively is unrealistic in the context of magic state cultivation, since some parity measurement results may yield a $-1$ value and still correspond to an accepted shot provided the syndrome (products of parities) in the associated detecting region \cite{rodatz2025faulttoleranceconstruction} returns a $+1$ value. To address this issue\footnote{On Scirate, the discussion platform for arXiv preprints (especially in quantum information sciences), user KdV raised concerns about the first version of our arXiv submission, in particular, our reliance on post-selection of $+1$ parity measurement results. See the full discussion at \url{https://scirate.com/arxiv/2509.08658}.}, we present additional simulations in this section. Our simulation will focus on the $d=5$ circuit, although the same reasoning can be applied to the lower $T$-count $d=3$ circuit.

As an initial step, we repeated the simulations from the previous section for the $d=5$ circuit, this time allowing each measurement spider to be randomly flipped with uniform probability. In the simulation, we effectively change all $\scalebox{1}{
\begin{tikzpicture}
    \begin{pgfonlayer}{nodelayer}
        \node [style=X dot] (0) at (0.00, -0.00) {};
        \node [style=none] (1) at (-1.00, -0.00) {};
    \end{pgfonlayer}
    \begin{pgfonlayer}{edgelayer}
        \draw (0) to (1);
    \end{pgfonlayer}
\end{tikzpicture}
}\rightarrow\scalebox{1}{
\begin{tikzpicture}
    \begin{pgfonlayer}{nodelayer}
        \node [style=X phase dot] (0) at (0.00, -0.00) {$a_i\pi$};
        \node [style=none] (1) at (-1.00, -0.00) {};
    \end{pgfonlayer}
    \begin{pgfonlayer}{edgelayer}
        \draw (0) to (1);
    \end{pgfonlayer}
\end{tikzpicture}
}$ and $\scalebox{1}{\begin{tikzpicture}
    \begin{pgfonlayer}{nodelayer}
        \node [style=Z dot] (0) at (0.00, -0.00) {};
        \node [style=none] (1) at (-1.00, -0.00) {};
    \end{pgfonlayer}
    \begin{pgfonlayer}{edgelayer}
        \draw (0) to (1);
    \end{pgfonlayer}
\end{tikzpicture}
}\rightarrow\scalebox{1}{\begin{tikzpicture}
    \begin{pgfonlayer}{nodelayer}
        \node [style=Z phase dot] (0) at (0.00, -0.00) {$a_j\pi$};
        \node [style=none] (1) at (-1.00, -0.00) {};
    \end{pgfonlayer}
    \begin{pgfonlayer}{edgelayer}
        \draw (0) to (1);
    \end{pgfonlayer}
\end{tikzpicture}}$, where $a_k \in \{0,1\}$ are uniform random numbers representing potential measurement results. Our approach to relax the constraint of post-selecting on all $+1$ parities is to emulate it by introducing random flips on the measurement spiders.

\begin{figure}[H]
    \centering
    \includegraphics[width=0.8\linewidth]{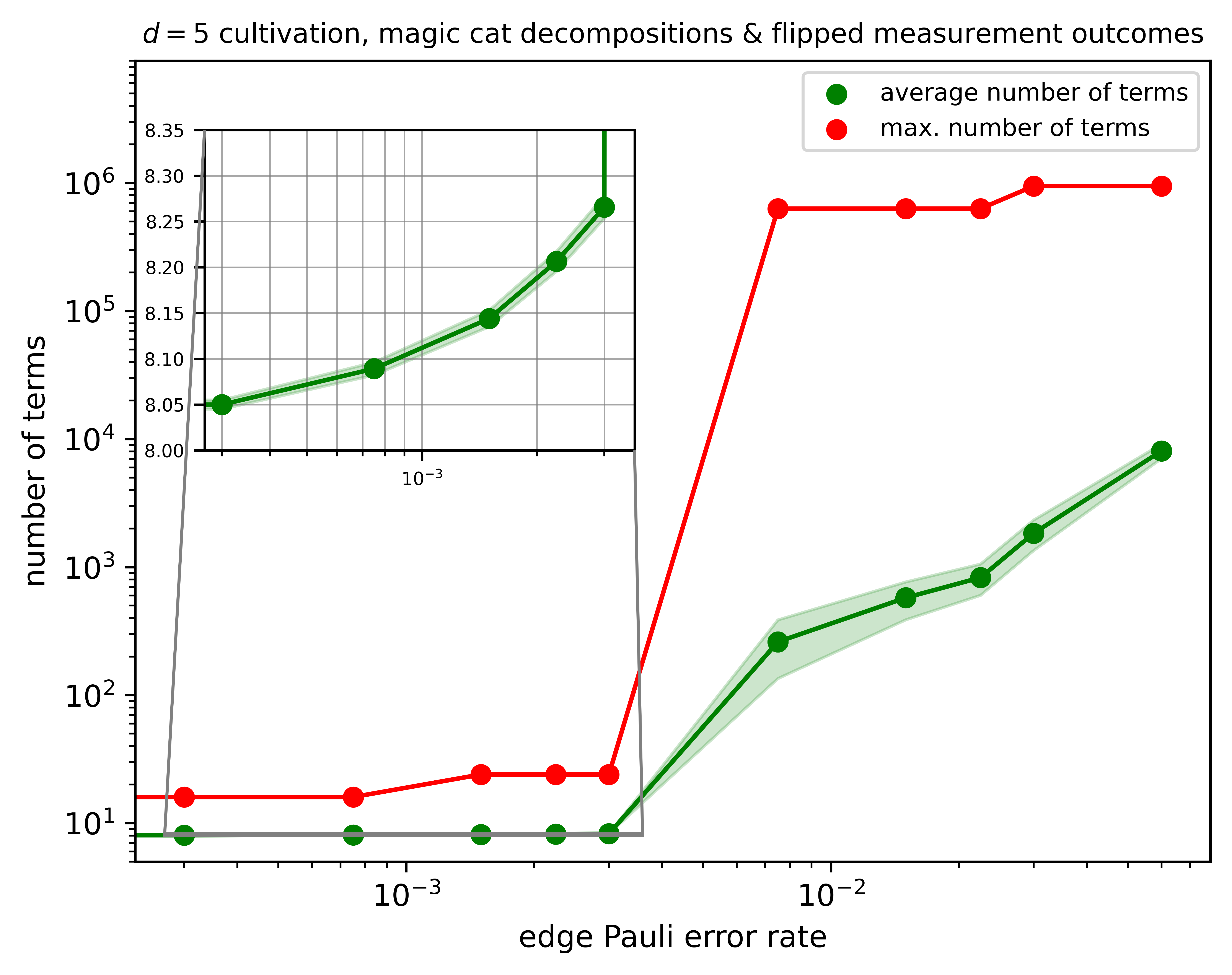}
    \caption{Average and maximum number of terms in the cutting stabiliser decomposition followed by magic cat secondary decompositions for the $d=5$ circuit. Measurement flips are applied, flipping each measurement single leg spiders to $a\pi$, $a\in\{0,1\}$ uniformly.}
    \label{fig:d5_meas_flipped_magic_cat}
\end{figure}

We noticed a sharp increase in the mean and maximum number of terms required as shown in figure \ref{fig:d5_meas_flipped_magic_cat}. This simulation proceeds by decomposing terms via magic cat state decomposition if the cutting decomposition failed to produce terms containing $1$ or fewer $\scalebox{1}{
\begin{tikzpicture}
    \begin{pgfonlayer}{nodelayer}
        \node [style=Z phase dot] (0) at (-2.00, -0.00) {$\frac{\pi}{4}$};
        \node [style=none] (1) at (-1.00, -0.00) {};
    \end{pgfonlayer}
    \begin{pgfonlayer}{edgelayer}
        \draw (0) to (1);
    \end{pgfonlayer}
\end{tikzpicture}
}/\scalebox{1}{
\begin{tikzpicture}
    \begin{pgfonlayer}{nodelayer}
        \node [style=Z phase dot] (0) at (-2.00, -0.00) {$\frac{7\pi}{4}$};
        \node [style=none] (1) at (-1.00, -0.00) {};
    \end{pgfonlayer}
    \begin{pgfonlayer}{edgelayer}
        \draw (0) to (1);
    \end{pgfonlayer}
\end{tikzpicture}
}$ spider. The results from figure \ref{fig:d5_meas_flipped_magic_cat} are concerning as early fault-tolerant quantum devices are expected to exhibit error rates of $\mathcal{O}(10^{-3})$ \cite{rodriguez2024experimentaldemonstrationlogicalmagic}. Confirming cultivation on such devices would require $\sim 10^{2}$ to $10^{3}$ more terms per shot in a Monte-Carlo simulation, potentially making this stabiliser decomposition approach impractical. The sharp increase around $p \sim 3 \times 10^{-3}$ when using the magic cat secondary decomposition (figure~\ref{fig:d5_meas_flipped_magic_cat}) appears to coincide with the regime where the cutting decomposition begins to leave behind a larger number of surviving non-Clifford spiders. We do not currently have a detailed theoretical explanation for this behaviour, this is left for future work.

\begin{figure}[H]
    \centering
    \includegraphics[width=0.8\linewidth]{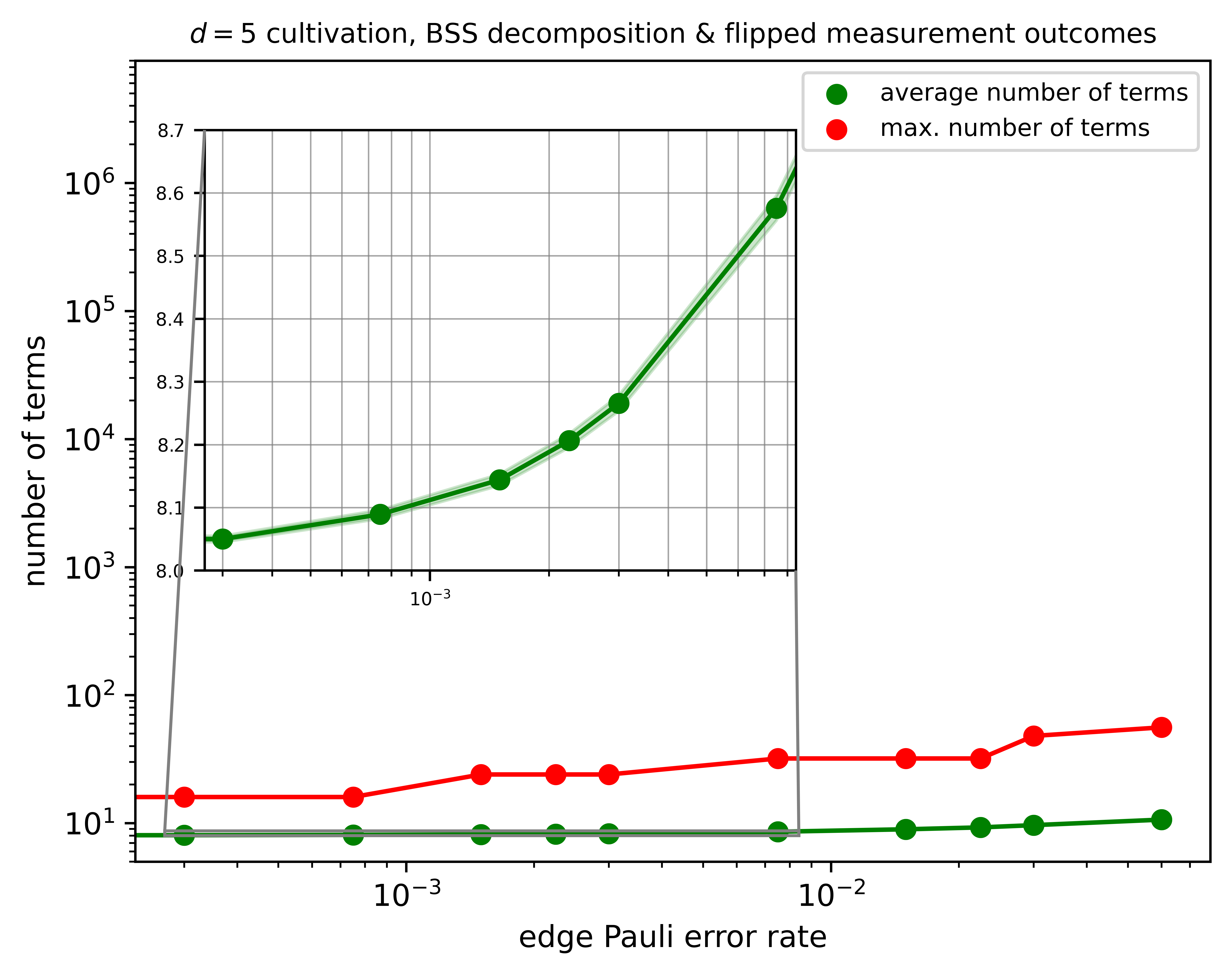}
    \caption{Average and maximum number of terms in the cutting stabiliser decomposition followed by BSS secondary decompositions for the $d=5$ circuit (with measurement flips).}
    \label{fig:d5_meas_flipped_BSS}
\end{figure}

Instead of using magic cat states in the secondary decomposition, we can:
\begin{enumerate}
    \item perform a BSS decomposition\footnote{With $\mathtt{pyzx.simulate.find\_stabilizer\_decomp(g)}$.} \cite{BSS2016, kissinger2022reduced} iteratively on each surviving ZX-diagram stemming from the cutting decompositions, and
    \item perform more cuts in the first stage of cutting decompositions.
\end{enumerate}
By combining these two strategies, we obtain significantly improved scaling of both average and maximum number of terms with respect to edge Pauli error rate, as shown in figure \ref{fig:d5_meas_flipped_BSS}. The average number of terms stays near $\approx8$, even at higher error rates once intractable. The cutting decomposition appears to struggle when the ZX-diagram being decomposed contains a large number of spiders or errors. Hence it must be supplemented with additional cuts and/or BSS as a secondary step. The BSS decomposition was chosen arbitrarily as it was readily available in $\mathtt{pyzx}$~\cite{kissinger2020Pyzx}, a more systematic approach to using all possible stabiliser decomposition in ZX-diagram is currently being explored in \cite{QuEraComputing_tsim_2026,haenel2026tsimfastuniversalsimulator}, see Appendix \ref{appendix:dc} and \ref{appendix:an_obs} for further intuitions and preliminary ideas on other stabiliser decompositions.

The results presented here are preliminary simulations. There are certainly opportunities to further optimise these decompositions and how to combine them appropriately. Potentially, one can incorporate the star edge decompositions from \cite{koch2023speedycontractionzxdiagrams,vollmeier2025graphicalstabilizerdecompositionsmulticontrol} (see appendix \ref{appendix:an_obs}). See section \ref{sec:numerical-methods} for an end-to-end $d=3$ circuit simulation via integration of cutting with \texttt{tsim} \cite{QuEraComputing_tsim_2026}. 

\section{Post-selected closed Pauli webs}

Additional optimisation can be implemented such that stabiliser decomposition is applied to only a subset of all measurement outcomes. This will further reduce runtimes. Magic state cultivation is a post-selected protocol, one can determine which measurement outcomes are considered acceptable shots and should therefore be retained during simulation. This effectively corresponds to performing stabiliser decomposition on error realisations for which all detecting regions \cite{Bombin_2024,rodatz2025faulttoleranceconstruction} remain un-violated \cite{rodatz2025faulttoleranceconstruction}. Six examples ($3$ {\color{red}red}, $3$ {\color{green}green}) of these detecting regions/closed Pauli webs are illustrated in \eqref{eq:d3_with_web} for the $d=3$ circuit. 

It is not necessary to perform stabiliser decomposition to all possible shots as certain measurement patterns are rejected due to post-selection. For the $d=3$ and $d=5$ circuits with degenerate injection, we have translated all the detectors from \cite{gidneyy2024cultivationdata} into closed Pauli webs. These are available in appendix \ref{appendix:d3_webs_det} and \ref{appendix:d5_webs_det} as figures and under the folders $\mathtt{d3\_det\_webs}$ and $\mathtt{d5\_det\_webs}$ as supplementary tikz files.

\begin{widetext}
\begin{equation}
\centering
    \label{eq:d3_with_web}
    \scalebox{0.55}{\includegraphics{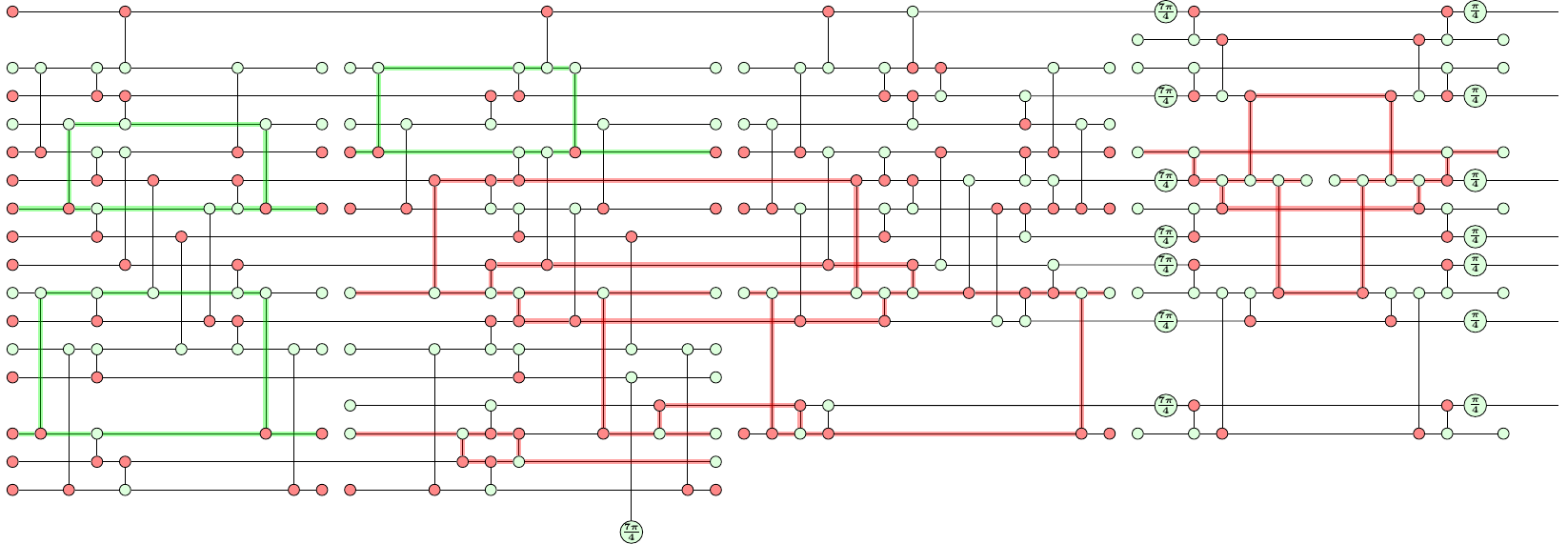}}
\end{equation}
\end{widetext}

\begin{figure}[H]
    \centering
    \includegraphics[width=0.8\linewidth]{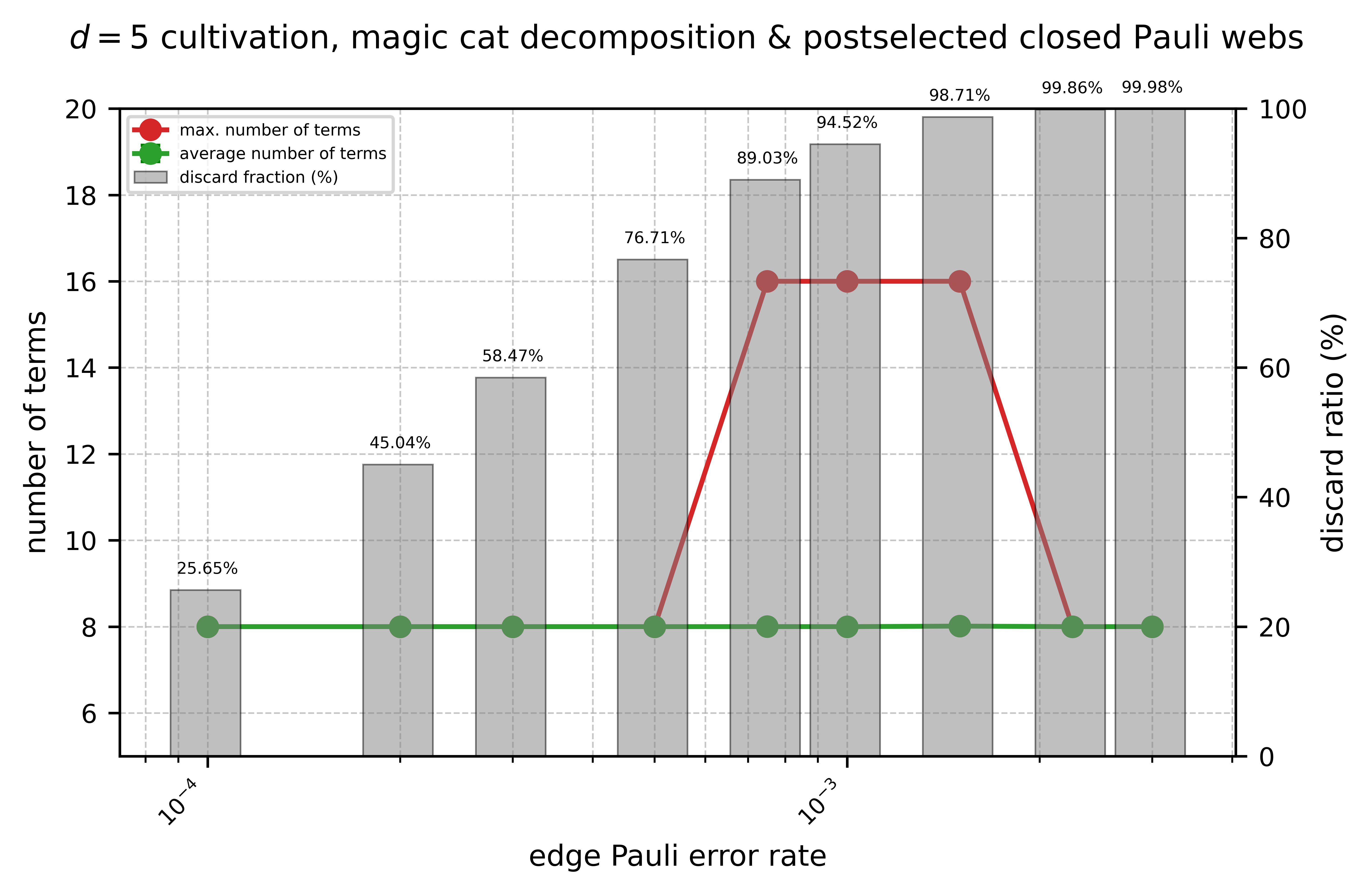}
    \caption{Average and maximum number of terms in the cutting stabiliser decomposition followed by magic cat secondary decompositions, for the $d=5$ circuit. The error realisation were post-selected such that all detecting regions returned $+1$ parity. We also included the discard ratio at each error rate.}
\label{fig:d5_post_selected_pauliweb}
\end{figure}

We have repeated the simulations from the previous sections, this time post-selecting on all the detecting regions and ensuring all the closed Pauli webs \cite{Bombin_2024} (equivalent to detectors in \cite{gidneyy2024cultivationdata} or detecting regions in \cite{rodatz2025faulttoleranceconstruction}) remain un-violated, for the $d=5$ circuit (see figure \ref{fig:d5_post_selected_pauliweb}).
We observed a near constant average number of terms ($\approx 8$) across the relevant error regime of $\mathcal{O}(10^{-4})$ to $\mathcal{O}(10^{-3})$ when using only the magic cat state secondary decomposition. However, single occurrences of error realisations led to $16$ terms at edge Pauli error rate of $0.00075$, $0.001$ and $0.0015$ with $10^5$ samples\footnote{Likely due to small sample size.}. We also included the discard ratio for various error rates under our error model in the same figure (\ref{fig:d5_post_selected_pauliweb}). Note that our error model has a higher error density, so the discard ratio is expected to be higher compared to \cite{gidney2024magicstatecultivationgrowing} as these correspond to different error models\footnote{Note: $p_{\text{edge Pauli error}}$ here $\neq$ $p_{\text{circuit-level noise}}$ from \cite{gidney2024magicstatecultivationgrowing}.}. Figure \ref{fig:d5_post_selected_pauliweb} provides early indications that the realistic number of terms in simulations with post-selected detectors/closed Pauli webs may still remain very low. 

\begin{figure}[H]
    \centering
    \includegraphics[width=0.8\linewidth]{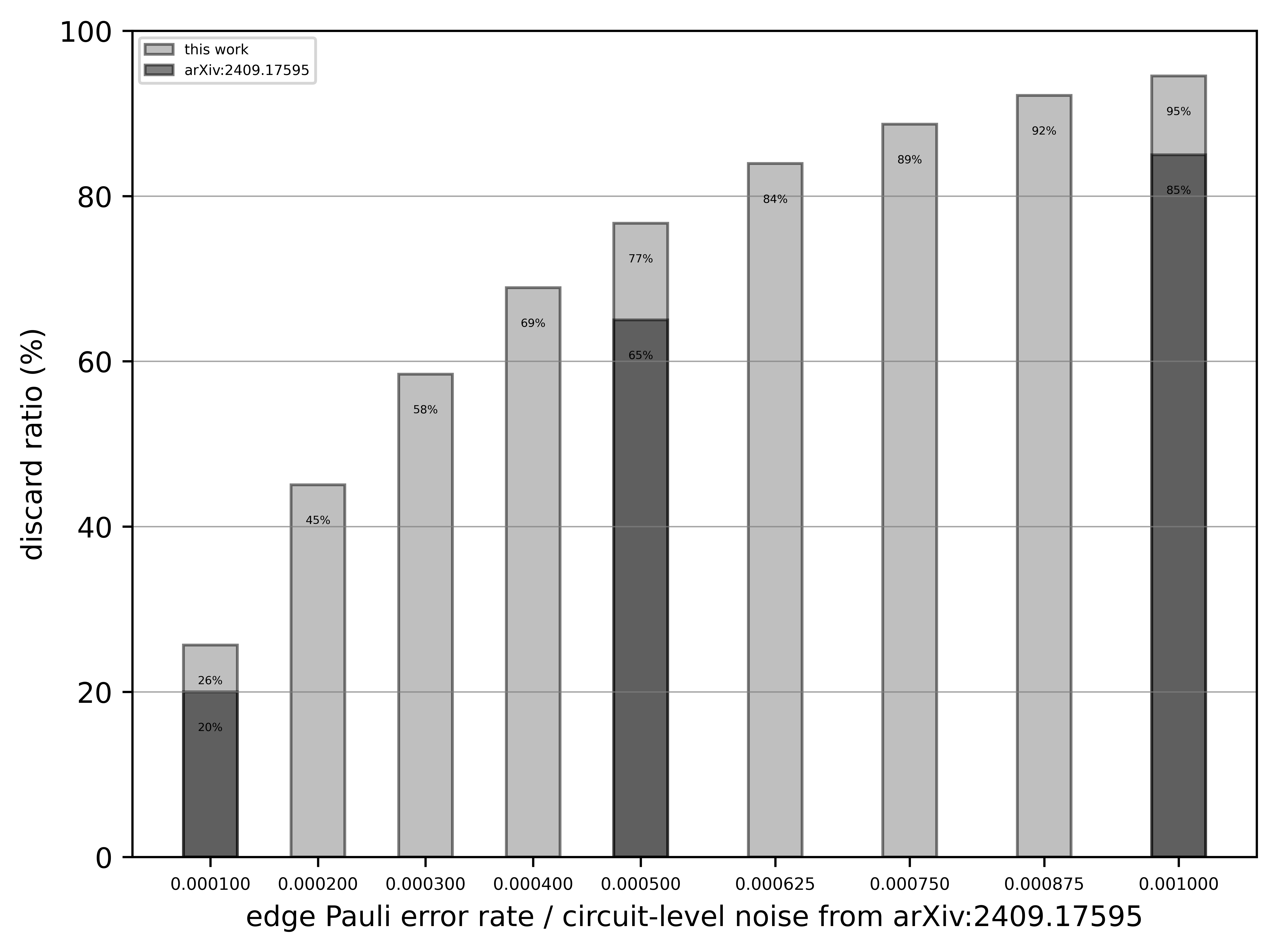}
    \caption{Comparison of discard ratios with results from \cite{gidney2024magicstatecultivationgrowing}.}
\label{fig:compare_gsj}
\end{figure}

To compare with the existing literature, figure \ref{fig:compare_gsj} shows the discard ratio for the $d=5$ circuit over a range of edge Pauli error rates, alongside the discard ratios reported in figure 2 of \cite{gidney2024magicstatecultivationgrowing}. As anticipated, our error led to slightly higher discard ratio, in line with our initial expectations.

Overall, the numerical evidences from the past three sections indicate that, for simulations in the $\mathcal{O}(10^{-3})$ error regime, the average number of terms remains at $\approx 8$ regardless of the chosen secondary decomposition, or if we post-select upon detector values.

\section{A sketch on an end-to-end logical error rate simulation}

In order to perform an and end-to-end representation of the resultant state from magic state cultivation (by which we mean simulating the injection and double-checking stages of the protocol), we present two approaches, which we shall name the \textit{sample measurements} approach and \textit{closed Pauli web post-selection} approach. See section \ref{sec:numerical-methods} for an end-to-end simulation of the magic state cultivation circuit sampled under uniform circuit level noise.

\subsection{Sample measurements approach}
In this first approach, we sample the measurement results explicitly. We can start by contracting/simplifying the errored\footnote{With known error locations and types.} ZX-diagram up to a measurement layer, excluding the measurement spiders. Then perform a stabiliser decomposition on that sub-diagram up to the measurement time slice with measurement spiders removed, hence more open legs. A sum of stabiliser terms will be returned from the decomposition. Sample those terms (they are all Clifford) and combine them into actual measurement realisations. Then input the sampled measurement outcomes into the original sub-diagram, included as $n\pi$ ($n\in \{0,1\}$) phases in the corresponding measurement spiders: $\scalebox{1}{
\begin{tikzpicture}
    \begin{pgfonlayer}{nodelayer}
        \node [style=Z phase dot] (0) at (-2.00, -0.00) {$n\pi$};
        \node [style=none] (1) at (-1.00, -0.00) {};
    \end{pgfonlayer}
    \begin{pgfonlayer}{edgelayer}
        \draw (0) to (1);
    \end{pgfonlayer}
\end{tikzpicture}
}/\scalebox{1}{
\begin{tikzpicture}
    \begin{pgfonlayer}{nodelayer}
        \node [style=X phase dot] (0) at (-2.00, -0.00) {$n\pi$};
        \node [style=none] (1) at (-1.00, -0.00) {};
    \end{pgfonlayer}
    \begin{pgfonlayer}{edgelayer}
        \draw (0) to (1);
    \end{pgfonlayer}
\end{tikzpicture}
}$. Finally, conditioned on all un-violated detecting regions inferred from measured parities, proceeding to the next measurement stage and repeat this procedure. Please see an example of the sketch outlined applied to the initial stages of the $d=3$ circuits in appendix \ref{appendix:sketch_example} and the following pseudocode in algorithm \ref{algo:sample} in appendix \ref{appendix:sketch_state_sample}. A drawback to this approach is that each shot may require multiple rounds of stabiliser decompositions proportional to the number of measurement layers. This will increase the actual run time.

\subsection{Closed Pauli web post-selection approach}
In this next (better) approach, we can post-select on closed Pauli webs without ever generating measurement values. To run and simulate a numerical experiment with errors, Pauli errors must first be initialised in the ZX-diagram, with their error types and locations stored. Using this information about the errors, one can form parity values of closed Pauli webs. This allows us to determine, before performing any stabiliser decompositions, whether a particular error realisation results in a rejected shot. If any of the closed Pauli webs (appendix \ref{appendix:d3_webs_det} and \ref{appendix:d5_webs_det}) are violated (returning $-1$ parity), the shot is discarded and we proceed onto the next shot. If none of the closed Pauli webs are violated, only then do we perform a cutting followed by any necessary further secondary decomposition(s). 

\subsection{Porting into large near-Clifford circuits}
For both approaches, a series of stabiliser tableaus corresponding to terms in the stabiliser decomposition is obtained at the end. With a stabiliser decomposition of $\chi$ terms, 
\begin{equation*}
    \ket{\psi} = \sum_{j=1}^{\chi} c_j \begin{pmatrix}
        \scalebox{0.7}{\input{clifford_zx_diag_2.tikz}}
    \end{pmatrix}_j
\end{equation*}
one would have to simulate $\chi$ different numerical experiments at the escape stage of cultivation (per shot) for an end-to-end simulation. Even if one has to port those states into a much larger Clifford circuit with thousands of qubits, the mean number of numerical experiments required is still proportional to the mean number of terms in the stabiliser decomposition. Effectively, we can also simulate any Clifford circuit applied subsequent to the cultivation circuit, such as tomography. Or even a single $T$-count noiseless projection on to a $\bra{\bar{T}}$ with marginal ($2 \times $) overhead. In an unlikely scenario where $\chi$ is very large for a particular error realisation, we can always count that particular shot as a logical failure to obtain a conservative bound on the logical error rate. 

We shall move on to explicit numerical experiments implemented via optimisations around $\mathtt{tsim}$ \cite{QuEraComputing_tsim_2026} next.

\section{Numerical Methods}
\label{sec:numerical-methods}
In this section and the next (section \ref{sec:accelerated}), we present numerical simulation methods and results for the $d=3$ $\ket{T}$ state cultivation circuit, based around our previously discussed cutting decomposition. The full pipeline achieves ${\sim}4$ million exact shots per second on a laptop at $p = 5 \times 10^{-4}$.

\subsection{Parametric Cutting and \texttt{tsim} Integration}
\label{sec:parametric-cutting}

Since the $d=3$ cultivation circuit contains non-Clifford gates, standard stabiliser simulation is insufficient. Instead, we evaluate stabiliser decompositions of parametric ZX-diagrams, whose scalar amplitudes depend on binary variables not fixed at compile time, enabling exact Monte Carlo sampling of detector and observable outcomes.

Unlike \cite{gidney2024magicstatecultivationgrowing}, which applies uniform circuit-level noise throughout the entire protocol, we apply the same circuit-level noise model (with physical noise parameter $p$) to the injection and cultivation stages only. A noiseless projection circuit is used to calculate logical error rates at the end of the circuit. Within the noisy injection and cultivation stages, two sources of runtime uncertainty exist in the simulation. \textit{f-parameters} represent the outcomes of depolarising noise channels and the particular Pauli error (or identity) realised is encoded as binary f-parameters sampled freshly for each shot. \textit{m-parameters} represent unmeasured output qubits, essentially the open legs of the ZX-diagram after all detectors and observables have been plugged in, whose values must be drawn from the correct Born-rule distribution conditioned on the noise realisation. The noisy projection onto a particular detector outcome pattern is thus determined jointly by the f- and m-parameters: each shot corresponds to a specific noise realisation (f) and a sampled set of unmeasured outputs (m), and the full detector sample is read off from the resulting plugged ZX-diagram.

The ZX-diagram evaluated by the sampler is not the circuit itself but the
\textit{sampling graph} constructed by \texttt{tsim} \cite{QuEraComputing_tsim_2026} (denoted by $G$), which is equivalent to the circuit's ZX-representation composed with its adjoint, with data qubits traced out. Detector and observable outcomes remain as open legs (m-parameters) of this doubled graph (figure \ref{fig:pyzx-param}). Plugging each leg with a binary value $m_i \in \{0,1\}$ closes the diagram into a scalar proportional to the Born-rule probability of that outcome pattern. For a given noise realisation $\mathbf{f}$, the sampler enumerates all $M = 2^N$ pluggings and draws the detector outcome from the resulting categorical distribution (section \ref{sec:enumeration}). Henceforth, $G$ denotes this sampling graph (still a ZX-diagram) and $\tilde{G}_j$ its stabiliser decomposition terms unless stated otherwise.

Rather than treating each parameter assignment as an independent simulation, we implement a custom sampler pipeline within the
\texttt{tsim} \cite{QuEraComputing_tsim_2026} framework that evaluates the parametric amplitude $\mathcal{S}(\mathbf{f}, \mathbf{m})$ for many assignments simultaneously, achieving up to ${\sim}132\text{k}$ shots/s on a laptop. The pipeline is compiled end-to-end using JAX \cite{jax2018github}.

\subsection{Stabiliser Rank Decomposition via Spider Cutting}
\label{sec:zx-decomposition}

The primary decomposition method is spider cutting, as described previously. The cutting decomposition expresses a parametric ZX-graph $G$ containing non-Clifford ($T$ gate) phases as a sum of $L$ Clifford graphs:
\begin{equation}
\label{eq:cutting-sum}
    G = \sum_{i=1}^{L} \tilde{G}_i, \qquad
    \mathrm{tcount}(\tilde{G}_i) = 0 \;\;\forall\, i\, \ ,
\end{equation}
where $\mathrm{tcount}(\tilde{G}_i)$ is the T-count (number of spiders with phase $\in\{\pi/4, 3\pi/4,5\pi/4,7\pi/4\}$) in ZX-graph $\tilde{G}_i$.
\subsubsection{Cutting Identity}
\label{sec:cutting-identity}

For a Z-spider (or X-spider, reverse the colourings) with phase $\alpha$ and $n$ legs connected to
neighbours $\{v_1, \ldots, v_n\}$, the cutting identity reads \cite{Sutcliffe2025thesis}:
\begin{equation}
\label{eq:cut}
        \overbrace{\scalebox{0.85}{\input{spider_decomp_abs_L.tikz}}}^{\mathrm{Spider}_\alpha^{(n)}} = \Bigg(\frac{1}{\sqrt{2}}\Bigg)^n \Bigg(\underbrace{\scalebox{0.85}{\input{spider_decomp_abs_R_1.tikz}}}_{G_{\text{left}}} + \ e^{i\alpha} \underbrace{\scalebox{0.85}{\input{spider_decomp_abs_R_2.tikz}}}_{G_{\text{right}}}\Bigg) \ ,
\end{equation}
where $G_{\text{left}}$ removes the spider and replaces each leg with a new opposite-colour spider of phase $0$ connected to the corresponding neighbour, and $G_{\text{right}}$ is the same but each new spider has phase $\pi$. For Hadamard edges, the created spider is same-colour (since $H$ commutes through as a colour change).

This identity extends to the \textit{parametric} setting: when the
cut spider carries parametric phase variables $\{p_1, \ldots, p_k\}$ from noise channels, the right branch acquires an additional sign factor $(-1)^{\bigoplus_j p_j}$, tracked via the scalar's \texttt{phasevars\_pi} field. This ensures that the noise dependence is faithfully preserved through the decomposition, leaving the noisy projection onto any particular error pattern exact.

\subsubsection{Cutting Algorithm}
\label{sec:cutting-algorithm}

\begin{figure}
    \centering
    \scalebox{0.7}{\input{tikz_largest_component.tikz}}
    \caption{Sampling graph of the main component of the $d=3$ $T$ gate
    cultivation circuit, as represented in \texttt{pyzx\_param}. Boundary
    vertices (open legs) correspond to detector and observable outcomes
    ($m$-parameters); plugging each with $m_i \in \{0,1\}$ yields a scalar
    proportional to the Born-rule probability of that outcome pattern.
    Non-Clifford phases ($\pi/4$, $7\pi/4$) arise from the $T$ and
    $T^\dagger$ gates.}
    \label{fig:pyzx-param}
\end{figure}

Given a ZX-graph represented as a \texttt{pyzx\_param} graph object
(figure \ref{fig:pyzx-param}) \cite{pyzx_param}, the algorithm proceeds as follows:
\begin{enumerate}
    \item Apply \texttt{full\_reduce(paramSafe=True)} to simplify
          (fuses spiders, removes identities, preserves parametric phases).
    \item Select a non-Clifford spider to cut
          ($\alpha \in \{\pi/4,\, 3\pi/4,\, 5\pi/4,\, 7\pi/4\}$) using the
          \textit{fewest-neighbours} heuristic, which minimises the
          $(1/\sqrt{2})^n$ scalar penalty; cutting a low-degree spider minimally perturbs the surrounding graph, preserving locality for subsequent \texttt{full\_reduce} simplification, and keeps each Clifford term's per-cut scalar factor closer to unity. The spider must not be adjacent
          to boundary vertices.
    \item Apply the cutting identity \eqref{eq:cut}$\;\to\;$ two new graphs,
          each with one fewer $T$ gate.
    \item Apply \texttt{full\_reduce(paramSafe=True)} to each new graph
          (\textit{inter-cut reduction}).
    \item Recurse until all terms are Clifford ($T$-count $= 0$).
\end{enumerate}
The inter-cut reduction at step 4 is critical: by simplifying after each cut, the algorithm exploits cancellations between spider fusions and identity removals that are invisible to methods operating on the unreduced graph.

\subsubsection{Relation to BSS Decomposition}
\label{sec:bss-relation}

After cutting reduces the $T$-count as far as possible, any remaining non-Clifford terms are finished with a secondary BSS stabiliser rank decomposition~\cite{BSS2016, kissinger2022reduced, sutcliffe2024fast}
(\texttt{find\_stab} from \texttt{tsim}). In practice, cutting with inter-cut \texttt{full\_reduce} eliminates all $T$ gates for the cultivation circuit without needing the BSS fallback. The hybrid approach: cutting first then BSS second produces significantly fewer Clifford terms than pure BSS, because intermediate simplification exploits cancellations that BSS misses. As a reference, the native BSS only stabiliser decomposition of the sampling graph, $G$, in $\mathtt{tsim}$ produces $16{,}627$ graphs (compared to $120$ graphs from the cutting then BSS decomposition, see next subsection). This makes the BSS only decomposition far too impractical for any realistic computation.

Before compilation, \texttt{phasevars\_pi} terms created by cutting (encoding $(-1)^{\bigoplus \mathcal{S}}$ for a set $\mathcal{S}$ of parameter variables) are
converted to \texttt{phasevars\_pi\_pair} format: $(-1)^{(\bigoplus \mathcal{S})\,\cdot\, 1}$, which the compiler handles as C-terms (section \ref{sec:term-types}).

\subsection{Sub-Component Product Factorisation}
\label{sec:factorisation}

After stabiliser decomposition and \texttt{full\_reduce}, the fully-plugged ZX-graph of the $d=3$ cultivation circuit disconnects into $K$ independent sub-components (connected components of the ZX-graph), enabling a factored evaluation:
\begin{equation}
\label{eq:factored}
    \mathcal{S}(\mathbf{f}, \mathbf{m})
    = \prod_{k=1}^{K} \mathcal{S}_k(\mathbf{f}_k, \mathbf{m}_k),
\end{equation}
where each sub-component amplitude is a sum over the $G_k$ Clifford graphs belonging to that sub-component:
\begin{equation}
    \mathcal{S}_k(\mathbf{f}_k, \mathbf{m}_k)
    = \sum_{g=1}^{G_k} s_{k,g}(\mathbf{f}_k, \mathbf{m}_k)\,.
\end{equation}
For the $d=3$ cultivation circuit: $K = 2$, with $G_1 = G_2 = 32$ Clifford graphs per sub-component ($64$ total). The remaining $56$ graphs belong to $28$ single-output components evaluated autoregressively, giving $120$ graphs in total across the full circuit.
\begin{figure}
    \centering
    \scalebox{0.7}{\input{tikz_plugged_paper.tikz}}
    \caption{The fully-plugged sampling graph for the largest connected component, after \texttt{full\_reduce} with parametric-safe rewriting. The graph disconnects into two sub-components $G^{(1)}$ and $G^{(2)}$, each containing 16 $T$ gates (green nodes with $\frac{\pi}{4}$ or $\frac{7\pi}{4}$ phases). The five free outputs have been plugged with measurement-outcome parameters $m_{19}, m_{26}, m_{28}, m_{30}, m_{32}$, which appear as parametric phases on internal Z-spiders. Each sub-component is independently decomposed via spider cutting and BSS into 32 Clifford terms. Sub-component factorisation reduces the evaluation cost from $32 \times 32 = 1024$ cross-term pairs to $32 + 32 = 64$ independent term evaluations per measurement-outcome combination. All measurement-outcome parameters $m_k$ carry an implicit factor of $\pi$, for example, a label $m_{19}$ on a Z-spider denotes a phase of $m_{19}\pi \in \{0, \pi\}$.
}
    \label{fig:zx-diagram}
\end{figure}

\subsection{Compiled Scalar Graph Evaluation}
\label{sec:compiled-eval}

\subsubsection{Term Types}
\label{sec:term-types}

Each fully reduced Clifford graph $g$ is compiled into a
\texttt{CompiledScalarGraphs}
structure~\cite{sutcliffe2024fast}, which
decomposes the scalar amplitude into a product of four algebraically distinct term types; for more detail see the appendices of \cite{sutcliffe2024fast}. These arise naturally from the structure of a simplified ZX-diagram: after \texttt{full\_reduce}, the remaining graph features such as individual spiders, phase gadgets, and parity constraints, each contribute a characteristic functional form to the scalar. The amplitude factorises as
\begin{equation}
\label{eq:scalar-amplitude}
    s_g(\mathbf{p})
    = \omega^{\phi_g} \cdot \lambda_g \cdot 2^{r_g}
      \cdot \prod_{t=1}^{N_A^{(g)}} \!A_t
      \cdot B
      \cdot C
      \cdot \prod_{t=1}^{N_D^{(g)}} \!D_t\,,
\end{equation}
where $\omega = e^{i\pi/4}$, $\phi_g$ is a static phase index, $\lambda_g = a + b\omega + ci + d\bar\omega$ is an exact scalar factor (stored as four integer coefficients), and $r_g$ is a power-of-two scaling exponent. All term types depend on binary row sums of the parameter vector $\mathbf{p} = [\mathbf{f},\, \mathbf{m}]$: linear functions $r_t = \bigoplus_j B_{t,j}\, p_j$ over $\mathbb{F}_2$.

\paragraph{A-terms (node terms).}
Each surviving interior spider in the reduced graph contributes a factor
$A_t = 1 + \omega^{\alpha_t}$, where
$\alpha_t = (4\, r_t + c_t) \bmod 8$. The row sum $r_t$ encodes how the
spider's phase depends on runtime parameters, while $c_t$ captures its
fixed Clifford phase.

\paragraph{B-terms (half-pi terms).}
Phase gadgets that reduce to $\pi/4$-multiples contribute a collective
phase $B = \omega^{\beta}$, where
$\beta = \sum_t (r_t \cdot \tau_t) \bmod 8$ and
$\tau_t \in \{0,\ldots,7\}$ are compile-time type constants encoding
each gadget's base phase.

\paragraph{C-terms (pi-pair terms).}
Parity constraints between pairs of $\pi$-phase contributions yield a
collective sign $C = (-1)^{\sum_t (\psi_t \cdot \phi_t) \bmod 2}$, where
$\psi_t$ and $\phi_t$ are binary row sums each with an additive constant
bit. In our pipeline, these arise specifically from the
\texttt{phasevars\_pi\_pair} conversion of cutting-generated sign factors
(Sec. \ref{sec:bss-relation}).

\paragraph{D-terms (phase-pair terms).}
Pairs of phase gadgets that resist further simplification contribute
$D_t = 1 + \omega^{\alpha_t} + \omega^{\beta_t}
     - \omega^{(\alpha_t + \beta_t) \bmod 8}$,
with two independent row sums. These are the dominant cost:
$N_D \approx 1408$ total across both sub-components, compared with
${\sim}28$ A-terms in the $d=3$ cultivation circuit. The total amplitude is: $\displaystyle\mathcal{S}(\mathbf{p}) = \sum_{g=1}^{G} s_g(\mathbf{p})$.

\subsubsection{Enumeration-Based Sampling}
\label{sec:enumeration}

The $d=3$ cultivation circuit's sampling graph has $N = 5$ unmeasured output bits, giving $M = 2^N = 32$ possible m-parameter combinations. For each noise realisation $\mathbf{f}$, we enumerate all $M$ combinations and compute the noisy
projection probability:
\begin{equation}
\label{eq:sampling-prob}
    P(\mathbf{m}_c \mid \mathbf{f})
    = \frac{\displaystyle\prod_{k=1}^{K} |\mathcal{S}_k(\mathbf{f}_k,\, \mathbf{m}_{c,k})|}
           {\displaystyle \sum_{c'=1}^{M} \prod_{k=1}^{K}
            |\mathcal{S}_k(\mathbf{f}_k,\, \mathbf{m}_{c',k})|}\, \ .
\end{equation}
This eliminates the autoregressive sampling chain, which would require $N+1$ sequential evaluation levels, in favour of a single parallel evaluation of all $M$ combinations at the fully-plugged level. The detector outcomes for each shot are then determined by the sampled $(\mathbf{f}, \mathbf{m})$ pair.

\begin{table*}[t]
\centering
\small
\caption{Sampling throughput for the $d=3$ cultivation circuit ($N\!=\!5$,
$M\!=\!32$, $K\!=\!2$, batch size $65{,}536$) on Apple M4 Pro series CPU at
$p = 0.001$ with $2^{26}$ shots. All methods are exact.}
\label{tab:throughput}
\scalebox{0.7}{
\begin{tabular}{lccl}
\toprule
\textbf{Optimisation} & \textbf{shots/s} & \textbf{Speedup} & \textbf{Key change} \\
\midrule
Autoregressive baseline
    & ${\sim}20\text{k}$ & $1.0\times$
    & Sequential $N\!+\!1$ evaluation levels \\
$+$ Enumeration $+$ BLAS matmul
    & ${\sim}38\text{k}$ & $1.9\times$
    & Single-level eval of all $M\!=\!32$ combos \\
$+$ Complex64 LUT
    & ${\sim}42\text{k}$ & $2.1\times$
    & Float32 lookup replaces exact integer arith. \\
$+$ Full-program JIT $+$ noiseless cache
    & ${\sim}54\text{k}$ & $2.7\times$
    & No Python overhead; cache ${\sim}64\%$ noiseless shots \\
$+$ Split f/m decomposition
    & ${\sim}94\text{k}$ & $4.7\times$
    & $32\times$ matmul reduction via row-sum linearity \\
\bottomrule
\end{tabular}
}
\end{table*}

\subsubsection{BLAS-Accelerated Binary Row Sums}
\label{sec:blas}

All term types require binary row sums $\displaystyle r_{g,t} = \bigoplus_{j=1}^{P} B_{g,t,j}\, p_j$, where $B \in \{0,1\}^{G \times T \times P}$ is a compile-time binary tensor and $\mathbf{p} \in \{0,1\}^P$ are runtime parameters. For a batch of $n$ parameter vectors, we reshape to a matrix multiply and reduce modulo 2:
\begin{equation}
    \underbrace{(G \!\cdot\! T,\; P)}_{\text{binary matrix}}
    \times
    \underbrace{(P,\; n)}_{\text{param batch}}
    \xrightarrow{\;\bmod 2\;}
    (n, G, T)\,.
\end{equation}
This activates multi-threaded BLAS (SIMD) on CPU via JAX/XLA. There are six such matrix multiplications per sub-component: two for D-term row sums ($\alpha$ and $\beta$), and one each for A, B, C-$\psi$, and C-$\phi$.

\subsubsection{Complex64 Lookup Table Arithmetic}
\label{sec:lut}

The eighth roots of unity $\omega^k$ for $k = 0,\ldots,7$ are precomputed as a length-8 \texttt{complex64} array. Term values are computed via integer index arithmetic on the row sums followed by table lookup, avoiding all trigonometric calls at runtime. The full evaluation uses \texttt{complex64} ($2\times\texttt{float32}$) throughout, replacing the four-component exact integer arithmetic (\texttt{ExactScalarArray}) of the reference implementation.

\subsubsection{Split f/m Row Sum Decomposition}
\label{sec:split-fm}

In the na\"ive enumeration pipeline, evaluating all $n \times M$ (shot, combo) pairs requires expanding f-parameters by a factor of $M$ to form an $(nM,\; P)$ input matrix. We exploit the linearity of the binary row sum to avoid this expansion.

Since $\mathbf{p} = [\mathbf{f},\, \mathbf{m}]$ and $B = [B_f \mid B_m]$, the row sum splits additively:
\begin{equation}
    \label{eq:split}
\begin{split}
    r & = \Bigl(\sum_j B_{f,j}\, f_j + \sum_j B_{m,j}\, m_j\Bigr) \bmod 2 \\
      & = (r_f + r_m) \bmod 2\, \ .
\end{split}
\end{equation}
The f-contribution $r_f$ is computed via a $(G \!\cdot\! T,\; n_f) \times (n_f,\; n)$ matrix multiply, executed once per batch (six small matmuls). The m-contribution $r_m$ is computed via a $(G \!\cdot\! T,\; n_m) \times (n_m,\; M)$ matrix multiply that depends only on the $M = 32$ fixed m-parameter combinations and is precomputed once at compile time, stored as an $(M, G, T)$ tensor.

Per-combo evaluation then reduces to the element-wise operation
\begin{equation}
    r^{(c)} = \bigl(r_f + r_m^{(c)}\bigr) \bmod 2\,,
\end{equation}
followed by lookup-table evaluation and product accumulation. This reduces the dominant matrix-multiply dimension from $nM$ to $n$ columns, a factor of $M = 32$ for the cultivation circuit. The identity $(a + b) \bmod 2 = ((a \bmod 2) + (b \bmod 2)) \bmod 2$ ensures the decomposition is exact.

After computing f-only row sums, we iterate over combos $c = 1, \ldots, M$. Each iteration combines precomputed m-row sums with f-row sums, evaluates all four term types via the LUT, computes per-graph products, and sums over graphs to obtain $\mathcal{S}_k(\mathbf{f}, \mathbf{m}_c)$ for each sub-component. The combo loop is unrolled at JIT trace time into a static XLA computation graph. The memory footprint is bounded by $(n, G, T_{\max})$ tensors per combo iteration (${\sim}367$\,MB at $n = 65{,}536$), rather than the infeasible $(n, M, G, T_{\max})$ that full materialisation would require (${\sim}12$\,GB).

\subsubsection{Noiseless Caching}
\label{sec:noiseless-cache}

Under circuit-level, uniform depolarising noise at strength $p$, a fraction $\eta$ of shots have all-zero f-parameters. For these noiseless shots, the noisy projection reduces to the noise-free case and the sampling distribution
$P(\mathbf{m} \mid \mathbf{f} = \mathbf{0})$ is identical across all such
shots, so it can be precomputed once. The sampler operates in two phases:
\begin{enumerate}
    \item \textbf{Phase 1.} Sample noise channel outcomes. Noiseless shots
          ($\mathbf{f} = \mathbf{0}$) are served from the cached distribution
          via a single categorical draw. Noisy shots are collected into a
          separate buffer.
    \item \textbf{Phase 2.} All noisy shots are consolidated and processed
          through the full split f/m pipeline in fixed-size batches (avoiding
          JIT recompilation).
\end{enumerate}
At $p = 0.001$, approximately $\eta \approx 63.9\%$ of shots are noiseless, so only ${\sim}36\%$ require full amplitude evaluation.

\subsubsection{Full-Program JIT Compilation}
\label{sec:jit}

The entire sampling function, including 
\begin{enumerate}
    \item noise channel categorical draws, 
    \item per-component enumeration, 
    \item per-sub-component split f/m evaluation, and 
    \item categorical output sampling, 
\end{enumerate}
is compiled into a single \texttt{jax.jit}-compiled function. All Python control flow (loops over components, sub-components, and combos) is resolved at trace time, producing a static XLA computation graph with no Python dispatch overhead at runtime.

\subsection{Performance}
\label{sec:performance}

The split f/m sampler achieves $94.8$ thousand shots/s with exact sampling at $p = 0.001$. All Monte Carlo estimates are computed with $2^{26} \approx 67$ million shots (except at $p=0.0005$, which uses $2^{28}$ shots). Every shot evaluates the exact non-Clifford amplitude of the $T$ gate circuit via a noiseless projection, a departure from \cite{gidney2024magicstatecultivationgrowing}, which replaces $T(T^\dagger) \to S(S^\dagger)$ to enable Clifford-only simulation via \texttt{stim} \cite{Gidney_2021}. See figure \ref{fig:throughput} for a shots per second comparison. Our approach requires no such gate substitution, the stabiliser decomposition (and BSS fall back per term) handles the full non-Clifford circuit directly.
\begin{figure}[h]
    \centering
    \includegraphics[width=0.95\columnwidth]{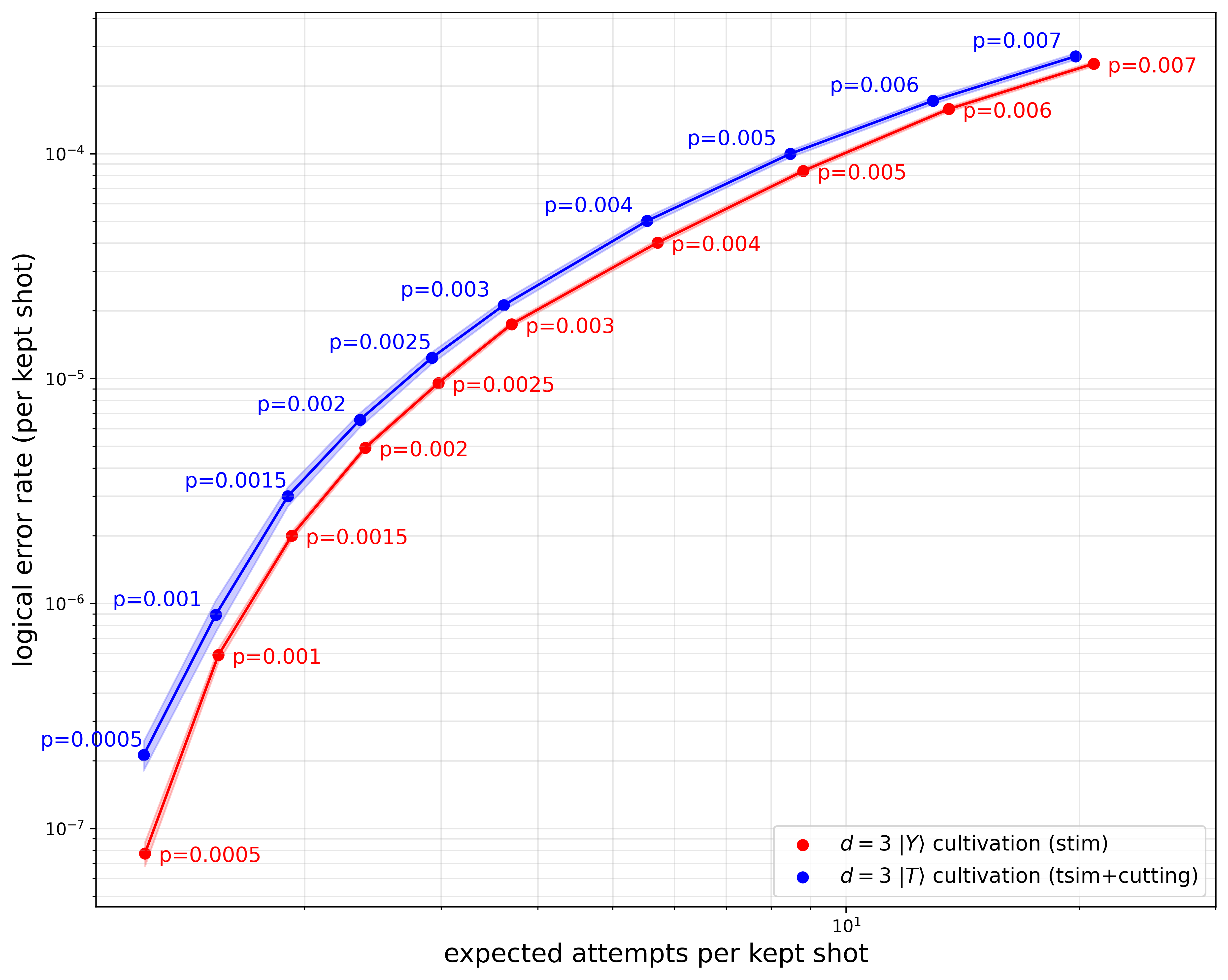}
    \caption{Logical error rate (LER) of the $d=3$ $T$ gate cultivation
    circuit as a function of physical noise strength $p$, sampled with
    $2^{26}$ shots per data point via exact noiseless projection (in {\color{blue}blue}). For comparison, the $S$ gate proxy of the equivalent noiseless projection circuit simulated with $\mathtt{stim}$ is shown in {\color{red}red}. The $T$ gate LER is approximately $2\times$ higher than the $S$ gate proxy across the plotted noise range, consistent with the
    proxy underestimating detector sensitivity by a factor of two as
    noted in figure 13 of \cite{gidney2024magicstatecultivationgrowing}.}
    \label{fig:ler}
\end{figure}

At $p = 0.001$, the full run produces a post-selection rate of
$0.651$, with 39 errors in 43.7 million post-selected shots,
corresponding to logical error rate of $\approx 8.9 \times 10^{-7}$. This is, to our knowledge, the first exact (non-Clifford-proxy) Monte Carlo estimate of the logical error rate for the $d=3$ $T$ gate cultivation circuit at operationally
relevant noise strengths, made feasible by the combination of cutting
decomposition and the optimised sampling pipeline. At $p=0.0005$, the sampler is only ${\sim}34\times$ slower (${\sim}132{,}400$ shots a second) than the fully Clifford proxy simulation via \texttt{stim} \cite{Gidney_2021} using $S$ gates ($\sim 4.47$ million shots a second). Figure \ref{fig:ler} shows the logical error rate as a function of physical noise strength for the exact $T$ gate simulation (with noiseless projection) alongside its equivalent $S$ gate Clifford proxy. Across the operationally relevant regime $p \in [10^{-4},\, 10^{-3}]$, the $T$ gate logical error rate tracks approximately within a factor of two relative to the $S$ gate proxy LER value. Full pseudocode for this decomposition, compilation, and sampling pipeline is provided in Appendix \ref{appendix:pseudocode}.
\begin{figure}[h]
    \centering
    \includegraphics[width=0.95\columnwidth]{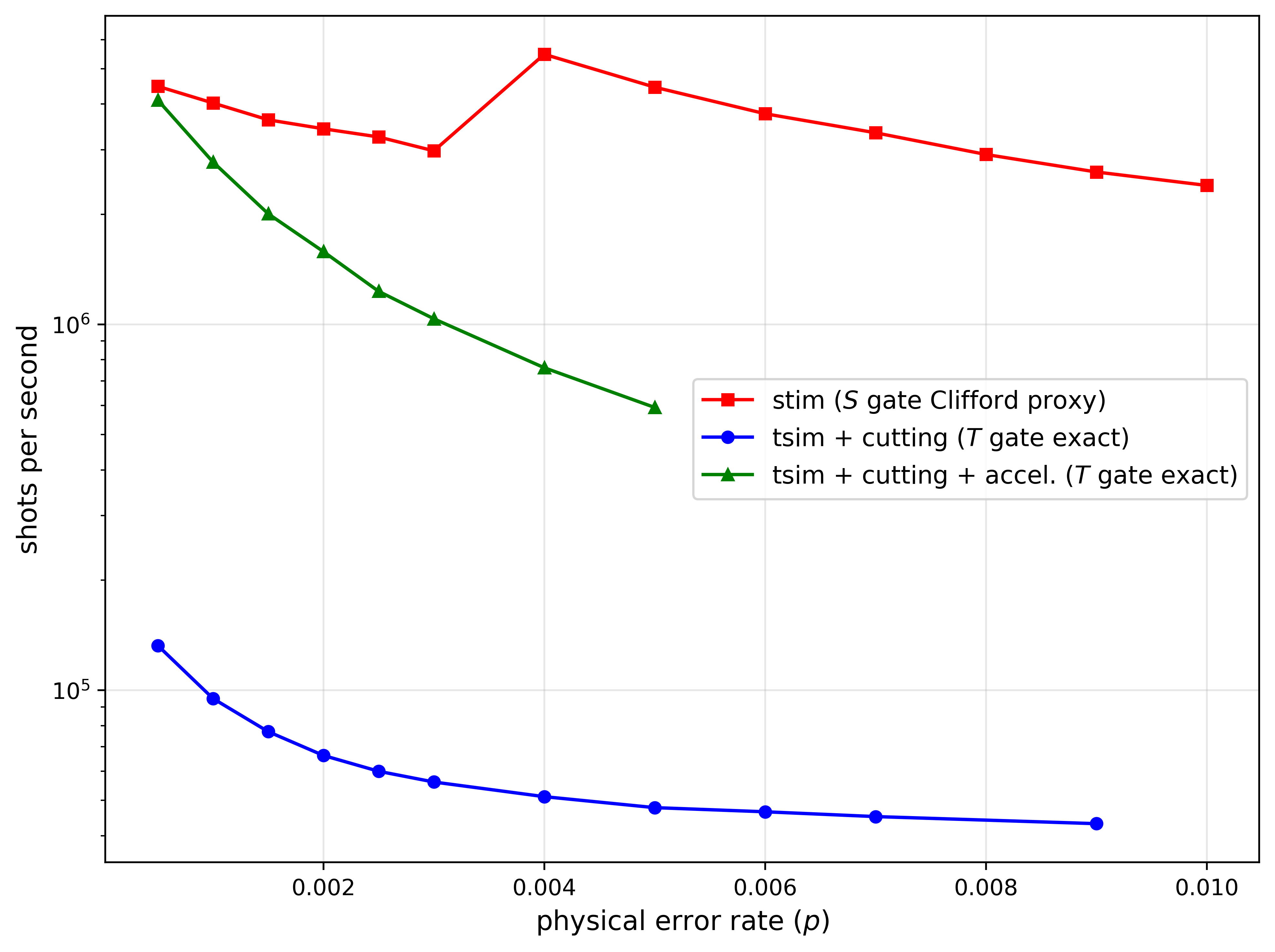}
    \caption{Sampling throughput (shots per second) as a function of physical
    circuit-level noise strength $p$ for the $d=3$ cultivation circuit. Throughput is measured as total shots divided by wall-clock time (excluding JIT warmup) over $2^{29}$--$2^{32}$ shots; run-to-run variance is negligible as the pipeline is deterministic after compilation. The $S$ gate Clifford proxy via \texttt{stim} ({\color{red}red}) achieves ${\sim}\,4.5$ million shots/s. The baseline exact $T$ gate simulation via \texttt{tsim}+cutting ({\color{blue}blue}) reaches ${\sim}\,132{,}000$ shots/s at $p = 5\times10^{-4}$. The accelerated pipeline ({\color{green}green}), combining sparse geometric channel sampling, cross-batch deduplication, and scan-based evaluation, achieves ${\sim}4.1$ million shots/s at $p = 5\times10^{-4}$. This is a ${\sim}31\times$ speed-up over the baseline and only ${\sim}\,1.1\times$ slower than \texttt{stim}. The accelerated sampler's advantage diminishes as the dedup compression ratio falls with increasing $p$.}
    \label{fig:throughput}
\end{figure}

We shall now move onto an even faster implementation of the $d=3$ circuits, reaching ${\sim}4$ million shots per second on a laptop at $p=0.0005$.

\section{Accelerated Sampling Pipeline}
\label{sec:accelerated}

Several further optimisations target the remaining bottlenecks in the sampling pipeline identified by profiling: random number generation in the channel sampler (Phase~1), redundant tensor network evaluations across shots sharing identical fault patterns (Phase~2), and JIT compilation overhead in the combo evaluation loop. Combined, these yield a ${\sim}30\times$ additional throughput improvement over the split~f/m baseline of section \ref{sec:split-fm} at $p = 10^{-3}$, while preserving exact sampling.

\subsection{Sparse Geometric Channel Sampling}
\label{sec:sparse-geometric}

The JIT-compiled channel sampler of section \ref{sec:jit} uses the Gumbel-Max trick \cite{maddison2015asampling, gumbel1954}: for each channel with $K$ outcomes, it draws $K$ i.i.d.\ Gumbel random variables, adds them to the log-probabilities, and returns the argmax. For the $d{=}3$ cultivation circuit with $N_{\mathrm{ch}} = 124$ channels averaging $\bar{K} \approx 4$ outcomes
each, this generates $N_{\mathrm{ch}} \cdot \bar{K} \cdot B \approx 10^9$ random numbers per batch of $B = 2{,}097{,}152$ shots. An intermediate step replaces the Gumbel-Max sampler with inverse CDF sampling via \texttt{searchsorted}, reducing random number generation by a factor of $\bar{K}$ to $N_{\mathrm{ch}} \cdot B$ draws per batch. However, at low noise, even one uniform per channel per shot is wasteful: the identity outcome dominates with probability $P(\text{identity}) \approx 1 - 3p \approx 0.9985$ at $p = 5 \times 10^{-4}$, so the vast majority of draws produce no effect.

We exploit this sparsity via geometric skip sampling: instead of asking ``what happened at each shot?'' for every channel, we ask ``which shot fires next?'' The geometric distribution $\mathrm{Geom}(p_{\mathrm{fire}})$ gives the gap between consecutive non-identity (``fire'') events, where $p_{\mathrm{fire}} = 1 - P(\text{identity})$ is precomputed per channel. Fire positions are obtained by cumulative summation of geometric draws, giving sorted (and therefore cache-friendly) write positions. For channels with multiple non-identity outcomes, a single conditional uniform per fire event selects among them via \texttt{searchsorted} on a precomputed conditional CDF. The selected outcome's XOR pattern is applied in-place to the result array.

The total random number budget is ${\sim}\,2 N_{\mathrm{ch}} B \bar{p}_{\mathrm{fire}}$: one geometric draw per skip plus one conditional uniform per fire event. At $p = 5 \times 10^{-4}$, this amounts to ${\sim}500\text{K}$ draws instead of ${\sim}260\text{M}$ (inverse CDF) or ${\sim}10^9$ (Gumbel-Max) per batch. The entire sampler runs in NumPy, eliminating JAX dispatch overhead for the $N_{\mathrm{ch}} = 124$ per-channel operations.

\subsection{Unique-Pattern Deduplication with Persistent Cache}
\label{sec:dedup}

At low physical error rates, the channel sampler produces highly repetitive fault patterns: among the ${\sim}36\%$ of shots that are noisy at $p = 0.001$, the vast majority share one of a small number of distinct f-parameter vectors (those with exactly one or two channels in a non-identity state). Since the enumerated sub-component evaluation of section \ref{sec:enumeration} depends on $\mathbf{f}$ only through the $n_f = 37$ f-parameters selected by the enum component, we can eliminate this redundancy by evaluating only the $U \ll B$ unique f-patterns and scattering the results back.

The deduplication proceeds in three steps:
\begin{enumerate}
    \item \textbf{Hash.} Each shot's $n_f$-bit f-pattern is packed into
          a single \texttt{int64} scalar via the inner product
          $h_i = \mathbf{f}_i \cdot (2^0, 2^1, \ldots, 2^{n_f - 1})$,
          computed as a BLAS matrix-vector product. This is exact for
          $n_f \leq 63$.
    \item \textbf{Unique.} \texttt{numpy.unique} on the scalar hashes
          returns $U$ unique values and an inverse index~$\sigma$ mapping
          each shot to its unique pattern. Operating on \texttt{int64}
          scalars, this runs in $O(B \log B)$ time, which is vastly faster than
          row-wise unique on $(B, n_f)$ boolean arrays.
    \item \textbf{Decode.} The $U$ unique hashes are decoded back to
          $n_f$-bit patterns via bitwise extraction, padded to the next
          power of two for JIT shape stability, and passed to the split
          f/m evaluator.
\end{enumerate}
The per-sub-component magnitudes $|\mathcal{S}_k(\mathbf{f}_u, \mathbf{m}_c)|$ are evaluated for each unique pattern $u = 1, \ldots, U$ and each combo $c = 1, \ldots, M$. The joint magnitudes are then scattered back to all $B$ shots via the inverse index: $P_i = \prod_k |\mathcal{S}_k|[\sigma(i)]$, after which each shot draws independently from its own categorical distribution.

The deduplication is performed in NumPy \textbf{outside} the JIT boundary, avoiding JAX's 64-bit integer limitations, while the tensor network evaluation remains inside a JIT-compiled function operating on the (much smaller) unique pattern array. At $p = 5 \times 10^{-4}$, within-batch deduplication reduces the number of split~f/m evaluations per batch from ${\sim}400{,}000$ noisy shots to ${\sim}500$ unique patterns---an $800\times$ reduction.

\paragraph{Cross-batch persistent cache.} The within-batch dedup finds $U \ll B$ unique patterns per batch, but many patterns recur across batches, especially at low noise where most noisy shots have only one or two channels firing. A persistent dictionary mapping \texttt{int64} hashes to magnitude vectors is maintained across all batches. After dedup, each unique hash is checked against this cache; only genuinely new patterns trigger the split~f/m evaluation. At $p = 5 \times 10^{-4}$, approximately $U \approx 500$ unique patterns appear per batch, but the total number of distinct patterns across all $n_{\mathrm{batches}} = 512$ batches is only ${\sim}1{,}000$. The persistent cache therefore reduces total evaluations from $U \times n_{\mathrm{batches}} \approx 256{,}000$ to ${\sim}1{,}000$---a further ${\sim}250\times$ reduction beyond within-batch dedup alone.

\subsection{Scan-Based Combo Evaluation}
\label{sec:scan-combo}

The split f/m evaluator of section \ref{sec:split-fm} iterates over the $M = 32$ measurement-outcome combos in a Python \texttt{for}-loop that is unrolled at JIT trace time into $M$ copies of the loop body. This increases compilation time and instruction-cache pressure. We replace this with \texttt{jax.lax.scan}, which compiles the loop body once and iterates at runtime. Additionally, the D-term values (section \ref{sec:term-types}) are precomputed into a 4-entry lookup table per $(G, T_D)$ position, indexed by the two parity bits of the f- and m-row sums, eliminating redundant complex-exponential computation inside the loop body. Together, these changes halve JIT warmup time and reduce per-batch evaluation cost.

\begin{table*}[t]
\centering
\small
\caption{Accelerated sampling optimisations for the $d=3$ cultivation circuit ($N\!=\!5$, $M\!=\!32$, $K\!=\!2$, batch size $2{,}097{,}152$) on Apple M4 Pro
CPU at $p = 10^{-3}$ with $2^{29}$ shots. Speed-ups are relative to the split f/m baseline of table \ref{tab:throughput}. All methods are exact.}
\label{tab:throughput-accel}
\scalebox{0.7}{
\begin{tabular}{lccl}
\toprule
\textbf{Optimisation} & \textbf{shots/s} & \textbf{Speedup} & \textbf{Key change} \\
\midrule
Split f/m baseline (table \ref{tab:throughput})
    & ${\sim}94\text{k}$ & $1.0\times$
    & Batch size $65{,}536$ \\
$+$ Inverse CDF + deduplication
    & ${\sim}974\text{k}$ & $10.4\times$
    & $\bar{K}{\times}$ fewer draws; eval only $U \!\ll\! B$ unique f-patterns \\
$+$ Geometric skip + persistent cache + scan
    & ${\sim}2{,}778\text{k}$ & ${\sim}30\times$
    & Sparse fire events; cross-batch cache; single-body scan \\
\bottomrule
\end{tabular}
}
\end{table*}

\subsection{Results}

\begin{figure}[h]
    \centering
    \includegraphics[width=0.95\columnwidth]{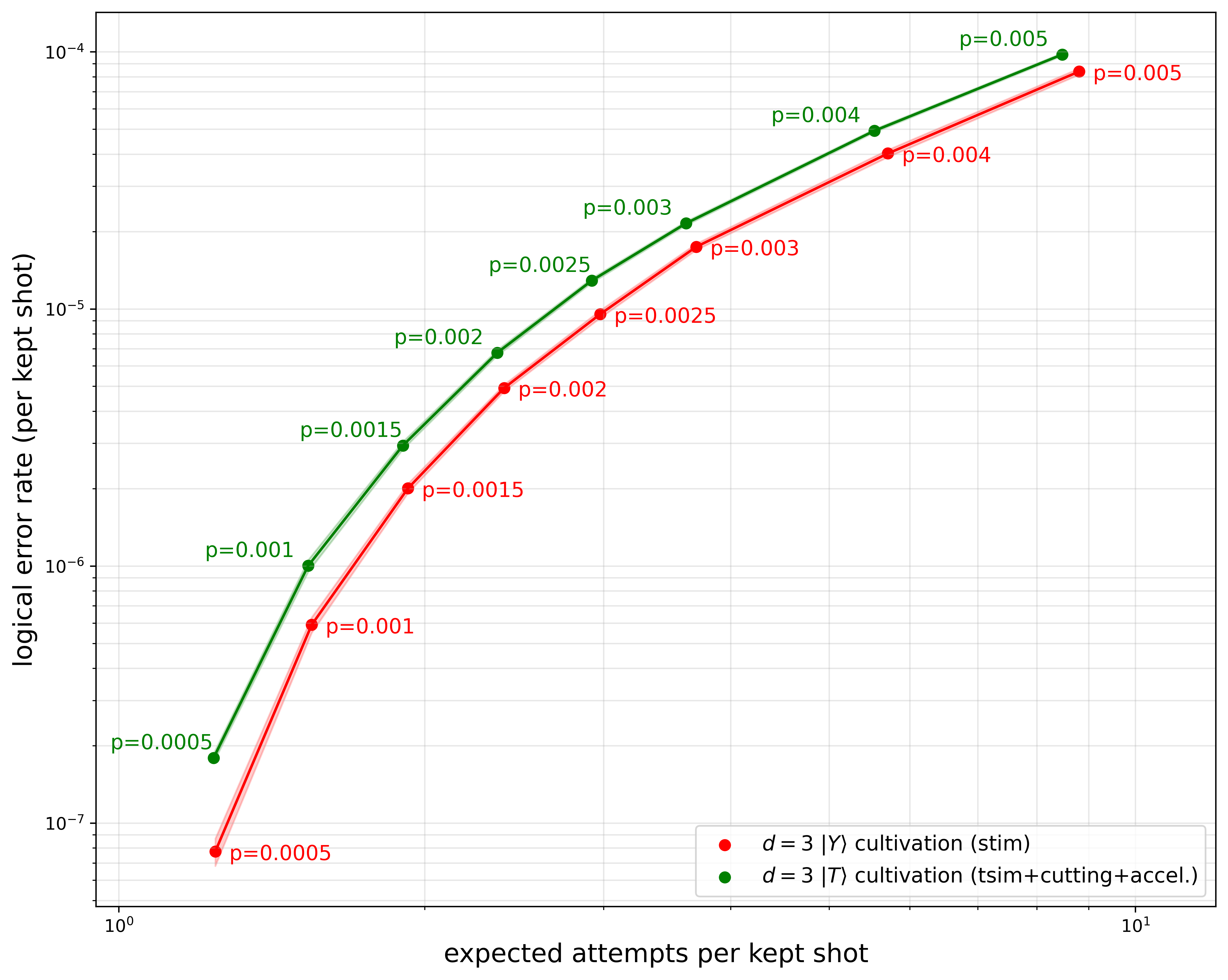}
    \caption{Logical error rate versus expected attempts per kept shot for the accelerated $|T\rangle$ simulation ({\color{green}green}, $2^{29}$ to $2^{32}$ shots) and the $|Y\rangle$ ($S$ gate) Clifford proxy via \texttt{stim} ({\color{red}red}). Same format as figure \ref{fig:ler}, shaded bands show binomial standard error.}
    \label{fig:ler_accel}
\end{figure}

\begin{figure*}
    \centering
    \includegraphics[width=0.9\linewidth]{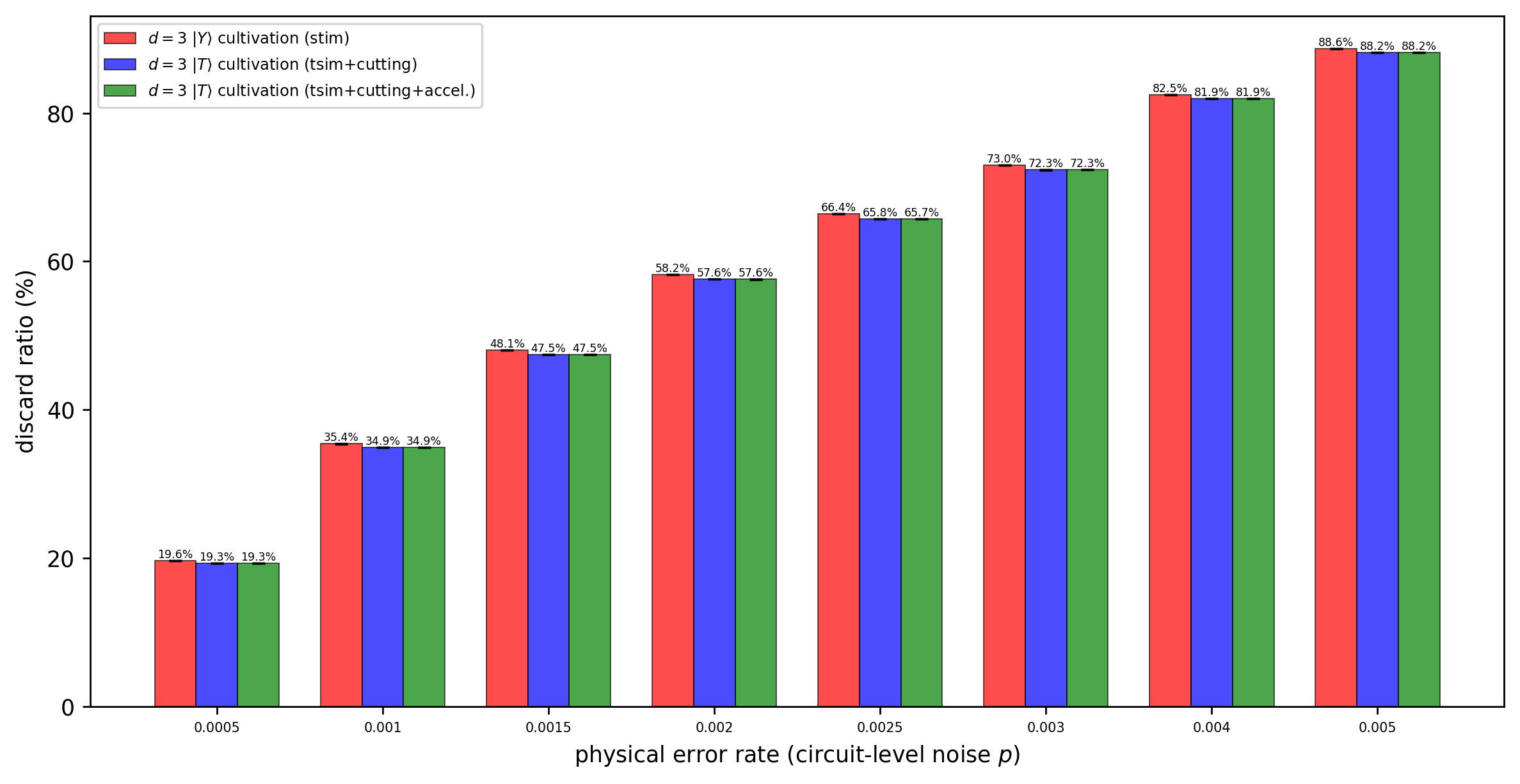}
    \caption{Discard ratio ($1-$PSR) across physical error rates $p \in \{0.0005, \ldots, 0.005\}$ for three configurations: stabiliser $S$ gate proxy via \texttt{stim} ({\color{red}red}), unaccelerated $T$ gate via \texttt{tsim} ({\color{blue}blue}), and accelerated $T$ gate via \texttt{tsim} with the optimisations of section \ref{sec:accelerated} ({\color{green}green}). All three circuits share the same noise model and post-selection criteria. The $S$ and $T$ gate PSR values are nearly indistinguishable, confirming that the non-Clifford gate does not affect the discard rate. Each bar is generated from at least $2^{26}$ shots.}
    \label{fig:psr-bar}
\end{figure*}

Table \ref{tab:v3-results} (and figure \ref{fig:ler_accel}) reports throughput and logical error rates for the combined pipeline at $2^{29}$ to $2^{32}$ shots per
noise strength on Apple M4 Pro CPU with batch size $B = 2{,}097{,}152$. At $p = 5 \times 10^{-4}$, the accelerated pipeline achieves ${\sim}4.1$ million~shots/s, within ${\sim}1.1\times$ of the Clifford-proxy \texttt{stim} throughput. The \texttt{stim} reference throughput is measured at $2^{29}$ shots in a single call; unlike the exact sampler, which loops over fixed $2$M-shot batches, \texttt{stim} allocates all shots at once and its throughput is sensitive to call size. We did not explore batching strategies for \texttt{stim} that could improve its throughput. At $p = 10^{-3}$, this represents a ${\sim}30\times$ improvement over the split~f/m baseline (see table \ref{tab:throughput-accel}). All optimisations preserve exact sampling: the output distribution is identical to the baseline pipeline. The logical error rate scales as ${\sim}p^3$, consistent with a distance-3 code. Pseudocode for the full
accelerated pipeline is provided in Appendix~\ref{appendix:pseudocode_fast}. The pipeline's throughput degrades at higher physical error rates because the persistent cache's compression ratio decreases and the sparse channel sampler generates more fire events, increasing the fraction of shots that require full ZX/tensor-network evaluation.
In addition, post-selection rates are consistent across all three configurations and decrease with increasing physical error rate, as shown in figure \ref{fig:psr-bar} as discard ratios ($1-$PSR).

\begin{table*}[t]
\centering
\small
\caption{Accelerated sampling results for the $d{=}3$ cultivation
circuit (batch size $2{,}097{,}152$, Apple M4 Pro CPU).
All methods are exact (non-Clifford).}
\label{tab:v3-results}
\begin{tabular}{ccccccc}
\toprule
$p$ & \textbf{Shots} & \textbf{shots/s} & \textbf{PSR} & \textbf{Errors} & $n_{\mathrm{kept}}$ & \textbf{LER} \\
\midrule
$5 \!\times\! 10^{-4}$  & $2^{32}$ & $4{,}102{,}367$ & $0.807$ & $622$       & $3{,}465{,}069{,}335$ & $1.8 \!\times\! 10^{-7}$ \\
$1 \!\times\! 10^{-3}$  & $2^{29}$ & $2{,}777{,}295$ & $0.651$ & $351$      & $349{,}529{,}061$    & $1.0 \!\times\! 10^{-6}$ \\
$1.5\!\times\! 10^{-3}$ & $2^{29}$ & $2{,}005{,}014$ & $0.525$ & $831$      & $282{,}110{,}746$    & $2.9 \!\times\! 10^{-6}$ \\
$2 \!\times\! 10^{-3}$  & $2^{29}$ & $1{,}580{,}499$ & $0.424$ & $1{,}540$  & $227{,}748{,}263$    & $6.8 \!\times\! 10^{-6}$ \\
$2.5\!\times\! 10^{-3}$ & $2^{29}$ & $1{,}230{,}763$ & $0.343$ & $2{,}373$  & $183{,}908{,}903$    & $1.3 \!\times\! 10^{-5}$ \\
$3 \!\times\! 10^{-3}$  & $2^{29}$ & $1{,}035{,}449$ & $0.277$ & $3{,}201$  & $148{,}548{,}040$    & $2.2 \!\times\! 10^{-5}$ \\
$4 \!\times\! 10^{-3}$  & $2^{29}$ & $760{,}769$     & $0.181$ & $4{,}785$  & $96{,}970{,}177$     & $4.9 \!\times\! 10^{-5}$ \\
$5 \!\times\! 10^{-3}$  & $2^{29}$ & $592{,}605$     & $0.118$ & $6{,}193$  & $63{,}357{,}051$     & $9.8 \!\times\! 10^{-5}$ \\
\bottomrule
\end{tabular}
\end{table*}

\section{Summary}
Magic state cultivation circuits with moderate $T$-count (in the $\sim 50$ range) can be simulated exactly via stabiliser decomposition with manageable overhead per shot relative to pure stabiliser circuits. We provide an optimised numerical pipeline based on $\mathtt{tsim}$ ($\mathtt{bloqade\text{-}tsim}$ v0.1.0) \cite{QuEraComputing_tsim_2026}, incorporating spider cutting, BLAS-accelerated evaluation, enumeration-based sampling, and an accelerated two-phase architecture combining sparse geometric channel sampling, persistent cross-batch deduplication, and scan-based evaluation (section \ref{sec:accelerated}), achieving ${\sim}4$ million exact shots per second for the $d{=}3$ circuit at $p = 5 \times 10^{-4}$. This is within a factor of ${\sim}1.1$ of the fully Clifford proxy via $\mathtt{stim}$. We note that our pipeline is compiled and optimised for the specific $d{=}3$ cultivation circuit, whereas $\mathtt{stim}$ is a general-purpose Clifford simulator. Many of the optimisations described in this work were developed on top of the $\mathtt{tsim}$ framework, and $\mathtt{tsim}$ is currently in the process of incorporating these optimisations into its main codebase \cite{haenel2026tsimfastuniversalsimulator}. Extending these simulations to the $d=5$ cultivation circuit, potentially with GPU-accelerated $\mathtt{tsim}$ \cite{haenel2026tsimfastuniversalsimulator,QuEraComputing_tsim_2026}, is left for future work.

For the $d=5$ case, our analytical results show that $\sim 8$ Clifford terms suffice on average to represent an error-free state via the cutting decomposition. The main challenges for a full numerical pipeline are (i) the larger graph size after compilation in the ABCD formalism, increasing per-term evaluation cost, and (ii) the need for a more sophisticated treatment of the escape stage. Whether the cutting decomposition continues to produce a manageable number of terms at $d=5$ or $d=7$ remains an open question. Simulating the escape stage is in principle feasible---one ports each stabiliser term into the subsequent Clifford circuit independently---but has not yet been demonstrated numerically. We leave these extensions to future work in the $\mathtt{tsim}$ community \cite{QuEraComputing_tsim_2026, haenel2026tsimfastuniversalsimulator}.

We note that several concurrent works on magic state cultivation have recently appeared \cite{sahay2025foldtransversalsurfacecodecultivation,claes2025cultivatingtstatessurface,
surti2025efficientsimulationlogicalmagic} or been updated \cite{vaknin2025efficientmagicstatecultivation} on the arXiv. It would be interesting to investigate whether our simulation method extends to other cultivation circuit variants \cite{chen2025efficientmagicstatecultivation}. We note that the throughput in our numerical experiments is several orders of magnitude higher than that of the competing approach \cite{surti2025efficientsimulationlogicalmagic} and out-performs \cite{li2025softhighperformancesimulatoruniversal} in terms of average shots per second by a factor of $11.7$, without the aid of a pricey H800 Nvidia GPU\footnote{Googled price of Nvidia H800 as of Feb 2026, $\sim$\$$30{,}000$ (USD). The Macbook Pro used for our computation cost $\sim$\$$2{,}000$ (USD).}. 

See also Appendix~\ref{appendix:an_obs} for a potential approach to further reducing the total number of stabiliser terms required for simulating the double-checking circuits.

\section{Acknowledgements}
We acknowledge useful feedback from Yves Vollmeier, Austin Fowler, Isaac Kim, Beatriz Dias, Rafael Haenel and our journal submission editors/reviewers. Z.Z was supported by the ERC Consolidator Grant \# 864828 ``Algebraic Foundations of Supersymmetric Quantum Field Theory'' (SCFTAlg). Packages such as $\mathtt{zxlive}$, $\mathtt{pyzx}$ \cite{kissinger2020Pyzx}, $\mathtt{tsim}$ \cite{QuEraComputing_tsim_2026}, magic state cultivation simulation python code \cite{gidneyy2024cultivationdata} and cutting stabiliser decomposition python code \cite{Sutcliffe_2024} have been extensively used in the preparation of this manuscript. All of the numerics were performed on an Apple M4 Macbook Pro, or devices of less computational power. Minimally reproducible code available at \cite{Wan2026accel}. We invite readers to open the supplementary tikz files in $\mathtt{zxlive}$ or $\mathtt{pyzx}$ \cite{kissinger2020Pyzx} for interactive exploration of the ZX-diagrams presented in this work. The wording in certain sections of this manuscript had been refined using LLMs.

\bibliography{main}

\clearpage

\appendix

\onecolumn

\section{Sampling Pipeline Pseudocode}
\label{appendix:pseudocode}

We present the full sampling pipeline in three stages: stabiliser
decomposition via cutting (Algorithm \ref{algo:decompose}), compilation and
precomputation (Algorithm \ref{algo:compile}), and per-batch sampling
(Algorithm \ref{algo:sample-batch}). Throughout, $\omega = e^{i\pi/4}$,
$\oplus$ denotes XOR, and $M = 2^N$ is the number of m-parameter
combinations.

\begin{algorithm}[H]
\small
\caption{Cutting-based stabiliser decomposition.}
\label{algo:decompose}
\begin{algorithmic}[1]
\Input ZX-graph $G$ with $\mathrm{tcount}(G) > 0$, max iterations $I_{\max}$
\Output Set of Clifford ZX-graphs $\mathcal{T}_{\mathrm{Cliff}}$

\Function{Decompose}{$G, I_{\max}$}
    \State $G \gets \texttt{full\_reduce}(G,\; \texttt{paramSafe}=\textsc{True})$
    \State $\mathcal{T} \gets \{G\}$;\quad $\mathcal{T}_{\mathrm{Cliff}} \gets \emptyset$
    \For{$i = 1$ \To $I_{\max}$}
        \State $\mathcal{T}_{\mathrm{new}} \gets \emptyset$
        \For{each $T \in \mathcal{T}$}
            \State $T \gets \texttt{full\_reduce}(T,\; \texttt{paramSafe}=\textsc{True})$
            \If{$\mathrm{scalar}(T) = 0$} \textbf{continue} \EndIf
            \If{$\mathrm{tcount}(T) = 0$}
                $\mathcal{T}_{\mathrm{Cliff}} \gets \mathcal{T}_{\mathrm{Cliff}} \cup \{T\}$;\quad \textbf{continue}
            \EndIf
            \State $v^* \gets \arg\min_{v \,\in\, \mathrm{cuttable}(T)} |\mathrm{neighbours}(v)|$
                \Comment{fewest-neighbours heuristic}
            \If{$v^* = \textsc{None}$}
                \Comment{BSS fallback}
                \State $\mathcal{T}_{\mathrm{Cliff}} \gets \mathcal{T}_{\mathrm{Cliff}} \cup \textsc{BSS}(T)$;\quad \textbf{continue}
            \EndIf
            \State $(G_L,\, G_R) \gets \textsc{CutSpider}(T,\, v^*)$
                \Comment{Eq. \eqref{eq:cut}}
            \State $\mathcal{T}_{\mathrm{new}} \gets \mathcal{T}_{\mathrm{new}} \cup \{G_L,\, G_R\}$
        \EndFor
        \State $\mathcal{T} \gets \mathcal{T}_{\mathrm{new}}$
        \If{$\mathcal{T} = \emptyset$} \textbf{break} \EndIf
    \EndFor
    \For{each $T \in \mathcal{T}$ with $\mathrm{tcount}(T) > 0$}
        \Comment{BSS cleanup}
        \State $\mathcal{T}_{\mathrm{Cliff}} \gets \mathcal{T}_{\mathrm{Cliff}} \cup \textsc{BSS}(T)$
    \EndFor
    \State Convert all \texttt{phasevars\_pi} $\to$ \texttt{phasevars\_pi\_pair}
        \Comment{$(-1)^{\bigoplus S} \to (-1)^{(\bigoplus S)\cdot 1}$}
    \State \Return $\mathcal{T}_{\mathrm{Cliff}}$
\EndFunction

\Statex
\Statex \hrulefill
\Statex \textbf{Subroutine} \textsc{CutSpider}($G$, $v$):
    spider $v$ has phase $\alpha$, $n$ legs, parametric variables $\{p_1,\ldots,p_k\}$.
\Statex \quad $G_L$: remove $v$; attach $n$ new opposite-colour phase-$0$ spiders;
    scalar $\times (1/\sqrt{2})^n$.
\Statex \quad $G_R$: remove $v$; attach $n$ new opposite-colour phase-$\pi$ spiders;
    scalar $\times (1/\sqrt{2})^n \cdot e^{i\pi\alpha} \cdot (-1)^{\bigoplus_j p_j}$.
\Statex \quad For Hadamard edges: created spider is same-colour.
\end{algorithmic}
\end{algorithm}

\clearpage

\begin{algorithm}[H]
\small
\caption{Compile circuit and precompute split f/m row sums.}
\label{algo:compile}
\begin{algorithmic}[1]
\Input Noisy circuit $\mathcal{C}$, noise strength $p$, max iterations $I_{\max}$
\Output Compiled sampling program (JIT-compiled)

\Function{Compile}{$\mathcal{C},\, p,\, I_{\max}$}
    \State $G_{\mathrm{ZX}} \gets \texttt{tsim.circuit\_to\_zx}(\mathcal{C})$
    \State $\{\mathcal{K}_1, \ldots\} \gets \mathrm{connected\_components}(G_{\mathrm{ZX}})$
    \For{each component $\mathcal{K}$ with $N$ output bits}
        \State $M \gets 2^N$
        \State $G_{\mathrm{plug}} \gets \texttt{plug\_all\_outputs}(\mathcal{K})$
        \State $\texttt{full\_reduce}(G_{\mathrm{plug}},\; \texttt{paramSafe}=\textsc{True})$
        \State $\{G^{(1)},\ldots,G^{(K)}\} \gets \mathrm{connected\_components}(G_{\mathrm{plug}})$
            \Comment{$K$ sub-components}
        \Statex
        \For{$k = 1$ \To $K$}
            \Comment{\textbf{Decompose and compile}}
            \State $\mathcal{T}_k \gets \textsc{Decompose}(G^{(k)},\, I_{\max})$
                \Comment{Alg. \ref{algo:decompose}}
            \State $\mathrm{CSG}_k \gets \texttt{compile\_scalar\_graphs}(\mathcal{T}_k)$
                \Comment{$G_k$ graphs, A/B/C/D terms}
        \EndFor
        \Statex
        \State $\mathbf{m}_{\mathrm{combos}} \gets \{0,1\}^{M \times N}$
            \Comment{\textbf{Enumerate} all $M$ output combos}
        \Statex
        \For{$k = 1$ \To $K$}
            \Comment{\textbf{Precompute m-contribution row sums}}
            \For{$\tau \in \{D_\alpha,\, D_\beta,\, A,\, B,\, C_\psi,\, C_\phi\}$}
                \State $B_m^{(\tau)} \gets \mathrm{CSG}_k.\mathrm{param\_bits}_\tau[\,:,\, :,\, \mathrm{m\_cols}]$
                    \Comment{$(G_k, T_\tau, n_{m,k})$}
                \State $\mathbf{v}_m \gets \mathbf{m}_{\mathrm{combos}}[\,:,\, \mathrm{m\_indices}_k]$
                    \Comment{$(M, n_{m,k})$}
                \State $R_m^{(\tau,k)} \gets \bigl(\mathrm{reshape}(B_m^{(\tau)}) \;\times\; \mathbf{v}_m^\top\bigr) \bmod 2$
                    \Comment{$(G_k \!\cdot\! T_\tau,\, M) \to (M,\, G_k,\, T_\tau)$}
            \EndFor
        \EndFor
        \Statex
        \State $P(\mathbf{m} \mid \mathbf{f}\!=\!\mathbf{0}) \gets$ evaluate all $M$ combos at $\mathbf{f}=\mathbf{0}$
            \Comment{\textbf{Noiseless cache}}
    \EndFor
    \State \Return $\texttt{jax.jit}(\textsc{SampleBatch})$
        \Comment{\textbf{JIT-compile} full sampling function}
\EndFunction
\end{algorithmic}
\end{algorithm}

\begin{algorithm}[H]
\small
\caption{Per-batch sampling with split f/m evaluation.}
\label{algo:sample-batch}
\begin{algorithmic}[1]
\Input Compiled program, batch size $n$, PRNG key
\Output Sampled detector outcomes for $n$ shots

\Function{SampleBatch}{$\mathrm{program},\, n,\, \mathrm{key}$}
    \Statex \Comment{\textbf{Phase 1: Channel sampling + noiseless cache}}
    \State $\mathbf{F} \gets \texttt{channel\_sample}(n)$
        \Comment{$(n, n_f)$ binary matrix from depolarising noise}
    \State $\mathcal{I}_0 \gets \{i : \mathbf{F}[i,:] = \mathbf{0}\}$
        \Comment{noiseless shots (${\sim}63.9\%$ at $p=0.001$)}
    \State $\mathrm{results}[\mathcal{I}_0] \gets \texttt{categorical}(P(\mathbf{m} \mid \mathbf{f}\!=\!\mathbf{0}))$
        \Comment{single cached draw}
    \State $\mathbf{F}' \gets \mathbf{F}[\overline{\mathcal{I}_0},\, :]$
        \Comment{$(n', n_f)$ noisy shots, $n' \approx 0.36\, n$}
    \Statex
    \Statex \Comment{\textbf{Phase 2: Full evaluation for noisy shots}}
    \For{each component with sub-components $k = 1,\ldots,K$}
        \State $W \gets \mathbf{1}_{n' \times M}$
            \Comment{joint magnitudes, initialised to $1$}
        \For{$k = 1$ \To $K$}
            \Statex \hspace{2em} \Comment{\textbf{Step A: f-only row sums} (6 matmuls, computed once)}
            \For{$\tau \in \{D_\alpha,\, D_\beta,\, A,\, B,\, C_\psi,\, C_\phi\}$}
                \State $B_f^{(\tau)} \gets \mathrm{CSG}_k.\mathrm{param\_bits}_\tau[\,:,\, :,\, \mathrm{f\_cols}_k]$
                    \Comment{$(G_k, T_\tau, n_{f,k})$}
                \State $R_f^{(\tau)} \gets \bigl(\mathrm{reshape}(B_f^{(\tau)}) \;\times\; \mathbf{F}'_k{}^\top \bigr) \bmod 2$
                    \Comment{$(G_k \!\cdot\! T_\tau,\; n') \to (n',\; G_k,\; T_\tau)$}
            \EndFor
            \Statex \hspace{2em} \Comment{\textbf{Step B: Loop over combos} (unrolled at JIT trace time)}
            \For{$c = 1$ \To $M$}
                \For{$\tau \in \{D_\alpha,\, D_\beta,\, A,\, B,\, C_\psi,\, C_\phi\}$}
                    \State $R^{(\tau)} \gets (R_f^{(\tau)} + R_m^{(\tau,k)}[c]) \bmod 2$
                        \Comment{element-wise, $(n', G_k, T_\tau)$}
                \EndFor
                \Statex
                \State \Comment{D-terms: LUT $\to$ product over $t$}
                \State $\boldsymbol{\alpha} \gets (4\, R^{(D_\alpha)} + \mathbf{c}_{D_\alpha}) \bmod 8$;\quad
                       $\boldsymbol{\beta} \gets (4\, R^{(D_\beta)} + \mathbf{c}_{D_\beta}) \bmod 8$
                \State $P_D \gets \textstyle\prod_t \bigl(1 + \omega[\boldsymbol{\alpha}_t] + \omega[\boldsymbol{\beta}_t] - \omega[(\boldsymbol{\alpha}_t\!+\!\boldsymbol{\beta}_t) \bmod 8]\bigr)$
                    \Comment{$(n', G_k)$}
                \Statex
                \State \Comment{A-terms: LUT $\to$ product over $t$}
                \State $P_A \gets \textstyle\prod_t \bigl(1 + \omega[(4\, R^{(A)}_t + \mathbf{c}_A) \bmod 8]\bigr)$
                    \Comment{$(n', G_k)$}
                \Statex
                \State \Comment{B-term: collective phase;\quad C-term: collective sign}
                \State $V_B \gets \omega\bigl[\textstyle\sum_t (R^{(B)}_t \cdot \boldsymbol{\tau}_t) \bmod 8\bigr]$;\quad
                        $V_C \gets (-1)^{\sum_t ((R^{(C_\psi)}_t + c_{\psi,t}) \cdot (R^{(C_\phi)}_t + c_{\phi,t})) \bmod 2}$
                \Statex
                \State \Comment{Sum over graphs}
                \State $S_k^{(c)} \gets \textstyle\sum_{g=1}^{G_k}
                       \omega^{\phi_g} \cdot \lambda_g \cdot 2^{r_g}
                       \cdot P_A[:,g] \cdot V_B[:,g] \cdot V_C[:,g] \cdot P_D[:,g]$
                    \Comment{$(n',)$}
                \State $W[:,\, c] \gets W[:,\, c] \;\times\; |S_k^{(c)}|$
            \EndFor
        \EndFor
        \Statex
        \State \Comment{Categorical sampling from noisy projection distribution}
        \State $\mathbf{p} \gets W \;/\; \mathrm{rowsum}(W)$
            \Comment{$P(\mathbf{m}_c \mid \mathbf{f})$ for each shot}
        \State $\mathrm{results}[\overline{\mathcal{I}_0}] \gets \texttt{categorical}(\log \mathbf{p},\; \mathrm{key})$
    \EndFor
    \State \Return results
\EndFunction
\end{algorithmic}
\end{algorithm}
\vspace{-1cm}
\paragraph{Complexity.}
The baseline pipeline requires six matrix multiplications of size
$(G \cdot T,\, P) \times (P,\, nM)$ per sub-component per batch, which is the
dominant cost. The split f/m decomposition (see section \ref{sec:split-fm})
replaces these with six multiplications of size
$(G \cdot T,\, n_f) \times (n_f,\, n)$, reducing the matmul cost by a factor
of $M = 32$. The m-contributions are six multiplications of size
$(G \cdot T,\, n_m) \times (n_m,\, M)$, precomputed once at compile time.
The per-combo element-wise operations cost
$\mathcal{O}(n \cdot M \cdot G \cdot T_{\max})$, bounded in memory by
$(n,\, G,\, T_{\max})$ tensors per combo iteration. The total per-batch cost
is $\mathcal{O}(n \cdot G \cdot T \cdot n_f + n \cdot M \cdot G \cdot T)$
versus the baseline $\mathcal{O}(n \cdot M \cdot G \cdot T \cdot P)$.

\clearpage
\section{Accelerated Sampling Pipeline Pseudocode}
\label{appendix:pseudocode_fast}
Algorithm~\ref{algo:sample-batch-fast} extends the per-batch sampling of algorithm \ref{algo:sample-batch} with sparse geometric channel sampling (section \ref{sec:sparse-geometric}), unique-pattern deduplication with a persistent cross-batch evaluation cache (section \ref{sec:dedup}), and scan-based combo evaluation (section \ref{sec:scan-combo}). The compile-time setup (Algorithm~\ref{algo:compile}) is unchanged except for precomputing per-channel fire probabilities and conditional CDFs, the noiseless cache distributions, and D-term lookup tables.

\begin{algorithm}[H]
\tiny
\caption{Accelerated per-batch sampling with sparse geometric channel sampling, cross-batch deduplication, and scan-based evaluation.}
\label{algo:sample-batch-fast}
\begin{algorithmic}[1]
\Input Compiled program, batch size $B$, PRNG key, per-channel
    $(p_{\mathrm{fire},j},\, \mathrm{cond\_cdf}_j,\, \mathrm{xor\_pattern}_j)$,
    noiseless cache $\hat{\mu}$, persistent eval cache $\mathcal{C}$
\Output Sampled detector outcomes for $B$ shots

\Function{SampleBatchFast}{$\mathrm{program},\, B,\, \mathrm{key}$}
    \Statex \Comment{\textbf{Phase 1: Sparse geometric channel sampling + noiseless cache}}
    \State $\mathbf{F} \gets \mathbf{0}^{B \times n_f}$;\quad
        $\mathrm{noisy\_mask} \gets \mathbf{0}^B$
    \For{$j = 1$ \To $N_{\mathrm{ch}}$}
        \Comment{geometric skip: only visit fire events}
        \State $\mathrm{skips} \gets \textsc{Geometric}(p_{\mathrm{fire},j})$
            \Comment{draw skip distances}
        \State $\mathrm{pos} \gets \textsc{CumSum}(\mathrm{skips}) - 1$;\quad
            keep $\mathrm{pos} < B$
        \If{$K_j = 2$}
            \Comment{binary channel: single non-identity outcome}
            \State $\mathbf{F}[\mathrm{pos}] \mathrel{\oplus}= \mathrm{xor\_pattern}_j$
        \Else
            \Comment{multi-outcome: conditional sampling}
            \State $\mathbf{u} \gets \textsc{Uniform}(0,1)^{|\mathrm{pos}|}$
            \State $\mathrm{idx} \gets \textsc{SearchSorted}(\mathrm{cond\_cdf}_j,\, \mathbf{u})$
            \State $\mathbf{F}[\mathrm{pos}] \mathrel{\oplus}= \mathrm{xor\_pattern}_j[\mathrm{idx}]$
        \EndIf
        \State $\mathrm{noisy\_mask}[\mathrm{pos}] \gets 1$
    \EndFor
    \Statex
    \State $\mathrm{results} \gets \textsc{SampleNoiseless}(\hat{\mu},\, B)$
        \Comment{NumPy PCG64, column-by-column}
    \State $\mathcal{I}_{\mathrm{noisy}} \gets \{i : \mathrm{noisy\_mask}[i] = 1\}$
    \State $\mathbf{F}' \gets \mathbf{F}[\mathcal{I}_{\mathrm{noisy}},\, :]$
        \Comment{$(n', n_f)$, $n' = |\mathcal{I}_{\mathrm{noisy}}|$}
    \Statex
    \Statex \Comment{\textbf{Phase 2: Deduplication + persistent cache + scan evaluation}}
    \Statex \Comment{\textit{Step 1: Hash f-patterns to int64 scalars (NumPy, outside JIT)}}
    \State $\mathrm{sel} \gets$ f-parameter indices for enum component
        \Comment{$|\mathrm{sel}| = n_f = 37$}
    \State $\mathbf{h} \gets \mathbf{F}'[\,:,\, \mathrm{sel}] \;\cdot\; (2^0,\, 2^1,\, \ldots,\, 2^{n_f - 1})$
        \Comment{$(n',)$ int64 via BLAS matmul}
    \Statex
    \Statex \Comment{\textit{Step 2: Find unique patterns + cross-batch cache lookup}}
    \State $(\mathbf{h}_{\mathrm{uniq}},\, \sigma) \gets \textsc{NumpyUnique}(\mathbf{h})$
        \Comment{$U$ unique hashes + inverse map}
    \State $\mathrm{new} \gets \{i : \mathbf{h}_{\mathrm{uniq}}[i] \notin \mathcal{C}\}$
        \Comment{only genuinely new patterns}
    \Statex
    \Statex \Comment{\textit{Step 3: Decode new patterns, pad, and evaluate}}
    \State $\tilde{\mathbf{F}} \gets \bigl((\mathbf{h}_{\mathrm{uniq}}[\mathrm{new}] \gg [0,\ldots,n_f{-}1]) \;\&\; 1\bigr)$
    \State $\hat{U} \gets 2^{\lceil\log_2 |\mathrm{new}|\rceil}$;\quad
        pad $\tilde{\mathbf{F}}$ to $(\hat{U}, n_f)$
    \Statex
    \Statex \Comment{\textit{Step 4: Split f/m evaluation via \texttt{jax.lax.scan} (JAX JIT)}}
    \State $\mathbf{J} \gets \mathbf{1}^{\hat{U} \times M}$
        \Comment{joint magnitudes for new patterns}
    \For{$k = 1$ \To $K$}
        \Comment{each sub-component}
        \For{$\tau \in \{D_\alpha,\, D_\beta,\, A,\, B,\, C_\psi,\, C_\phi\}$}
            \State $R_f^{(\tau)} \gets \bigl(\mathrm{reshape}(B_f^{(\tau)}) \;\times\; \tilde{\mathbf{F}}_k{}^\top\bigr) \bmod 2$
                \Comment{f-only row sums, once per $k$}
        \EndFor
        \State D-term lookup: precompute $d_{00}, d_{01}, d_{10}, d_{11}$ per $(G_k, T_D)$ position
        \Statex \Comment{combo loop via \texttt{jax.lax.scan} (single compiled body)}
        \For{$c = 1$ \To $M$}
            \State $R^{(\tau)} \gets (R_f^{(\tau)} + R_m^{(\tau,k)}[c]) \bmod 2$ for each $\tau$
            \State D-terms: select from $\{d_{00}, d_{01}, d_{10}, d_{11}\}$ by parity bits
            \State $S_k^{(c)} \gets \textsc{EvalTerms}(\ldots)$;\quad
                $\mathbf{J}[\,:,\, c] \gets \mathbf{J}[\,:,\, c] \times |S_k^{(c)}|$
        \EndFor
    \EndFor
    \State Store $\mathbf{J}$ in persistent cache $\mathcal{C}$ for new hashes
    \Statex
    \Statex \Comment{\textit{Step 5: Assemble all magnitudes from cache + categorical sampling}}
    \State $\mathbf{J}_{\mathrm{all}} \gets$ look up $\mathcal{C}[\mathbf{h}_{\mathrm{uniq}}]$ for all $U$ patterns
    \State $\mathbf{P}_i \gets \mathbf{J}_{\mathrm{all}}[\sigma(i),\, :]$ for each noisy shot $i$
    \State $\mathbf{p}_i \gets \mathbf{P}_i \;/\; \mathrm{rowsum}(\mathbf{P}_i)$
    \State $\mathbf{c} \gets \texttt{categorical}(\log \mathbf{p},\; \mathrm{key})$
    \State $\mathrm{results}[\mathcal{I}_{\mathrm{noisy}}] \gets \mathbf{m}_{\mathbf{c}}$
    \Statex
    \State \Return results
\EndFunction
\end{algorithmic}
\end{algorithm}

\paragraph{Complexity.}
Let $\eta$ denote the noiseless fraction and $U$ the number of unique fault patterns among the $(1{-}\eta)B$ noisy shots per batch. Phase~1 costs $O(N_{\mathrm{ch}} \cdot B \cdot \bar{p}_{\mathrm{fire}})$ per batch via geometric skip: only fire events incur random draws, compared with $O(N_{\mathrm{ch}} \cdot B)$ for inverse CDF or $O(N_{\mathrm{ch}} \cdot \bar{K} \cdot B)$ for Gumbel-Max. Phase~2 costs $O(B \log B)$ for hashing and deduplication, plus $O(\hat{U}_{\mathrm{new}} \cdot K \cdot M \cdot G \cdot T_{\max})$ for the split~f/m evaluation on genuinely new patterns only, where $\hat{U}_{\mathrm{new}}$ is the number of patterns not found in the persistent cache. The total number of evaluations across all batches is bounded by the number of distinct patterns in the entire run (${\sim}1{,}000$ at $p = 5 \times 10^{-4}$), rather than $U \times n_{\mathrm{batches}}$ as in within-batch dedup alone.

\paragraph{Random number budget.}
The sparse geometric sampler generates ${\sim}\,2 N_{\mathrm{ch}} B \bar{p}_{\mathrm{fire}}$ random numbers per batch (one geometric draw per skip plus one conditional uniform per fire event). At $p = 5 \times 10^{-4}$ with $\bar{p}_{\mathrm{fire}} \approx 0.001$, this amounts to ${\sim}500\text{K}$ draws versus $N_{\mathrm{ch}} \cdot B \approx 2.6 \times 10^8$ for inverse CDF or $N_{\mathrm{ch}} \cdot \bar{K} \cdot B \approx 1.0 \times 10^9$ for Gumbel-Max. The noiseless cache adds $n_{\mathrm{binary}} + n_{\mathrm{categorical}}$ draws per batch ($28$ Bernoulli $+ 1$ categorical for this circuit, generated via NumPy PCG64). Phase~2 adds one categorical draw per noisy shot.

\section{Cutting the basic flag-free double checking circuit}
\label{appendix:dc}
The double checking sub-circuit dominates the $T$-count in cultivation circuits. Essentially, they are a series of $n$ controlled-$(H_{XY})$ gates applied to qubits on a $n$-GHZ state as control and qubits on the injected colour code as target. Taking the double checking circuit from \cite{gidney2024magicstatecultivationgrowing}, removing all the flag like checks and allow $n \geq 2$ to be an arbitrary integer, we arrive at the following circuit (with $n$ total inputs/outputs): 
\begin{equation}
\centering
        \scalebox{1}{\input{check_free_DC.tikz}}  \ .
\end{equation}
If we were to make two cuts at the two Z-spiders inside the cyan dotted circles via the cutting decompositions \cite{Codsi2022Masters,Sutcliffe2025thesis,wan2025cuttingstabiliserdecompositionsmagic}, we will have the following $2^2 = 4$ terms stabiliser decomposition:
\begin{equation}
        \scalebox{0.65}{\input{cfdcLHS.tikz}}  \propto  \scalebox{0.65}{\input{cfdct0.tikz}}   +   \scalebox{0.65}{\input{cfdct1.tikz}}   +   \scalebox{0.65}{\input{cfdct2.tikz}}   +  \scalebox{0.65}{\input{cfdct3.tikz}}  \ .
\end{equation}
With a bit of simplifications and collecting terms, we arrive at:
\begin{equation}
        \scalebox{0.65}{\input{cfdcLHS.tikz}}  \propto  \scalebox{0.65}{\input{cfdc_f_t0.tikz}}   +   \scalebox{0.65}{\input{cfdc_f_t1.tikz}} \ ,
\end{equation}
this implies
\begin{equation}
        \scalebox{0.65}{\input{cfdcLHS.tikz}}  \propto  I^{\otimes n}   +   (XS^{\dagger})^{\otimes n} \ .
\end{equation}
The basic flag-free double checking circuit can be written as a sum of $2$ Clifford ZX-diagrams. Similar arguments can be made about the double checking circuit with Pauli errors and different measurement values. This suggest that the difficulty in finding stabiliser decomposition for such circuits from \cite{gidney2024magicstatecultivationgrowing} maybe due to the presence of additional flag-like structures, leading to higher number of terms ($4$ and $8$ terms) as we see in the main text and from \cite{wan2025cuttingstabiliserdecompositionsmagic}. 

Another potentially useless observation, $I+XS^{\dagger} = \begin{pmatrix}
    1 & -i \\
    1 & 1
\end{pmatrix}$, which can be related to the Hadamard box:
\[\scalebox{1}{
\begin{tikzpicture}
    \begin{pgfonlayer}{nodelayer}
        \node [style=hadamard] (0) at (-1.00, 0.00) {$a$};
        \node [style=none] (1) at (0.00, 0.00) {};
        \node [style=none] (2) at (-2.00, 0.00) {};
    \end{pgfonlayer}
    \begin{pgfonlayer}{edgelayer}
        \draw (0) to (2);
        \draw (0) to (1);
    \end{pgfonlayer}
\end{tikzpicture}
}=\begin{pmatrix}
    1 & 1 \\
    1 & a
\end{pmatrix} \ , \]
\cite{KissingerWetering2024Book} from ZH-calculus via 
\[I+XS^{\dagger} = \scalebox{1}{\input{ihbox.tikz}} \ . \] 
We wonder if $(I+XS^{\dagger})^{\otimes n} = I^{\otimes n}+(XS^{\dagger})^{\otimes n} - \sum_{j}(\text{cross terms})_j$ admits a nice interpretation in ZH-calculus. Probably not useful.

\section{An observation}
\label{appendix:an_obs}
We noticed that the orange dotted (star) edge \cite{koch2023speedycontractionzxdiagrams} defined as:
\begin{equation}
        \begin{tikzpicture}
	\begin{pgfonlayer}{nodelayer}
		\node [style=none] (0) at (0, 1) {};
		\node [style=none] (1) at (0, -1) {};
	\end{pgfonlayer}
	\begin{pgfonlayer}{edgelayer}
		\draw [style=star edge] (0.center) to (1.center);
	\end{pgfonlayer}
\end{tikzpicture}
  =    \sqrt{2}        \scalebox{0.7}{\input{orange_edge_r.tikz}} \ ,
\end{equation}
bears resemblance to the double checking circuits. This ZX-diagram has some very nice stabiliser decompositions\footnote{Tikz figures taken from \cite{koch2023speedycontractionzxdiagrams}.}.
\begin{equation} \label{decomp/star-1}
	  =  \textstyle{
	\sqrt 2   \begin{tikzpicture}
	\begin{pgfonlayer}{nodelayer}
		\node [style=Z dot] (0) at (0, 0.5) {};
		\node [style=Z dot] (1) at (0, -0.5) {};
		\node [style=none] (2) at (0, 1.25) {};
		\node [style=none] (3) at (0, -1.25) {};
	\end{pgfonlayer}
	\begin{pgfonlayer}{edgelayer}
		\draw (3.center) to (1);
		\draw [style=hadamard edge] (0) to (2.center);
	\end{pgfonlayer}
\end{tikzpicture}
   +   
	2   \begin{tikzpicture}
	\begin{pgfonlayer}{nodelayer}
		\node [style=Z dot] (0) at (0, -0.5) {};
		\node [style=Z phase dot] (1) at (0, 0.5) {$\pi$};
		\node [style=none] (2) at (0, 1.25) {};
		\node [style=none] (3) at (0, -1.25) {};
	\end{pgfonlayer}
	\begin{pgfonlayer}{edgelayer}
		\draw [style=hadamard edge] (3.center) to (0);
		\draw [style=hadamard edge] (1) to (2.center);
	\end{pgfonlayer}
\end{tikzpicture}
}
\end{equation}

\begin{align}
	\label{decomp/star-2} 
	\begin{tikzpicture}
	\begin{pgfonlayer}{nodelayer}
		\node [style=none] (0) at (0, 1) {};
		\node [style=none] (1) at (0, -1) {};
		\node [style=none] (3) at (0.75, 1) {};
		\node [style=none] (4) at (0.75, -1) {};
	\end{pgfonlayer}
	\begin{pgfonlayer}{edgelayer}
		\draw [style=star edge] (0.center) to (1.center);
		\draw [style=star edge] (3.center) to (4.center);
	\end{pgfonlayer}
\end{tikzpicture}
  &=  \textstyle{
		\frac{1}{\sqrt 2}  \begin{tikzpicture}
	\begin{pgfonlayer}{nodelayer}
		\node [style=Z dot] (0) at (0.75, 0.5) {};
		\node [style=Z dot] (1) at (0.75, -0.5) {};
		\node [style=none] (2) at (0.75, 1.25) {};
		\node [style=none] (3) at (0.75, -1.25) {};
		\node [style=none] (4) at (0, 1.25) {};
		\node [style=none] (5) at (0, -1.25) {};
	\end{pgfonlayer}
	\begin{pgfonlayer}{edgelayer}
		\draw (3.center) to (1);
		\draw (0) to (2.center);
		\draw [style=hadamard edge] (5.center) to (4.center);
	\end{pgfonlayer}
\end{tikzpicture}
   +  
		\frac{1}{\sqrt 2}  \begin{tikzpicture}
	\begin{pgfonlayer}{nodelayer}
		\node [style=Z dot] (0) at (0, 0.5) {};
		\node [style=Z dot] (1) at (0, -0.5) {};
		\node [style=none] (2) at (0, 1.25) {};
		\node [style=none] (3) at (0, -1.25) {};
		\node [style=none] (4) at (0.75, 1.25) {};
		\node [style=none] (5) at (0.75, -1.25) {};
	\end{pgfonlayer}
	\begin{pgfonlayer}{edgelayer}
		\draw (3.center) to (1);
		\draw (0) to (2.center);
		\draw [style=hadamard edge] (5.center) to (4.center);
	\end{pgfonlayer}
\end{tikzpicture}
   +   
		4   \begin{tikzpicture}
	\begin{pgfonlayer}{nodelayer}
		\node [style=Z phase dot] (0) at (0, 0.5) {$\pi$};
		\node [style=Z phase dot] (1) at (1, 0.5) {$\pi$};
		\node [style=Z phase dot] (2) at (0, -0.5) {$\pi$};
		\node [style=Z phase dot] (3) at (1, -0.5) {$\pi$};
		\node [style=none] (4) at (0, 1.25) {};
		\node [style=none] (5) at (1, 1.25) {};
		\node [style=none] (6) at (1, -1.25) {};
		\node [style=none] (7) at (0, -1.25) {};
	\end{pgfonlayer}
	\begin{pgfonlayer}{edgelayer}
		\draw [style=hadamard edge] (7.center) to (2);
		\draw [style=hadamard edge] (0) to (4.center);
		\draw [style=hadamard edge] (5.center) to (1);
		\draw [style=hadamard edge] (6.center) to (3);
	\end{pgfonlayer}
\end{tikzpicture}
}
	\\[5pt]
	\label{decomp/star-3} 
	\input{star-3/star.tikz}  &=  \textstyle{
		\frac{1}{2\sqrt 2}   \input{star-3/1.tikz}   +  
		\frac{1}{2\sqrt 2}   \input{star-3/2.tikz}   +  
		\frac{1}{2\sqrt 2}   \input{star-3/3.tikz}  +  
		\frac{1}{\sqrt 2}   \input{star-3/4.tikz}   +  
		8   \input{star-3/5.tikz}}
\end{align}

\begin{align}
	\label{decomp/star-3-state/0} 
	\input{star-3-state/star-0.tikz}  &=  \textstyle{
	3   \input{star-3-state/1.tikz}   -  
	\input{star-3-state/2.tikz}   +  
	\frac{3}{\sqrt{2}}   \input{star-3-state/3.tikz}   -  
	\frac{3}{2\sqrt{2}}   \input{star-3-state/4.tikz}}
	\\[5pt]
	\label{decomp/star-3-state/pi2} 
	\input{star-3-state/star-pi2.tikz}  &=  \textstyle{
	\frac{1 \pm 3i}{2}   \input{star-3-state/1.tikz}   +  
	\frac{1 \mp i}{2}   \input{star-3-state/2.tikz}   -  
	\frac{3-i}{2\sqrt{2}}   \input{star-3-state/3.tikz}   +  
	\frac{1 \mp i}{2\sqrt{2}}   \input{star-3-state/4.tikz}}
\end{align} 
We wonder if these can be adapted to reduce the number of terms needed to simulate magic state cultivation even further via re-write etc. To any potential ZX-calculus experts, please do let us know.

\section{State sampling}
\label{appendix:sketch_state_sample}
\begin{algorithm}[H]
    \scriptsize    
    \begin{algorithmic}
      
        \Input{Original errored ZX-diagram $g = \scalebox{0.25}{
\begin{tikzpicture}[yshift=6cm]
    \begin{pgfonlayer}{nodelayer}
        \node [style=Z dot] (0) at (1.00, -7.00) {};
        \node [style=X dot] (1) at (2.00, -5.00) {};
        \node [style=Z dot] (2) at (-1.00, -4.00) {};
        \node [style=Z dot] (3) at (6.00, -4.00) {};
        \node [style=Z dot] (4) at (4.00, -6.00) {};
        \node [style=Z dot] (5) at (4.00, -4.00) {};
        \node [style=X dot] (6) at (0.00, -10.50) {};
        \node [style=X dot] (7) at (-1.00, -1.50) {};
        \node [style=Z dot] (8) at (2.00, -6.00) {};
        \node [style=X dot] (9) at (6.00, -7.00) {};
        \node [style=X dot] (10) at (6.00, -10.50) {};
        \node [style=X dot] (11) at (-1.00, -10.50) {};
        \node [style=Z dot] (12) at (0.00, -6.00) {};
        \node [style=Z dot] (13) at (2.00, -7.00) {};
        \node [style=Z dot] (14) at (2.00, -4.00) {};
        \node [style=X dot] (15) at (3.00, -8.00) {};
        \node [style=X dot] (16) at (-1.00, -8.00) {};
        \node [style=X dot] (17) at (3.00, -7.00) {};
        \node [style=Z dot] (18) at (6.00, -6.00) {};
        \node [style=X dot] (20) at (2.00, -1.50) {};
        \node [style=X dot] (21) at (1.00, -5.00) {};
        \node [style=Z dot] (22) at (0.00, -4.00) {};
        \node [style=Z dot] (23) at (3.00, -4.00) {};
        \node [style=Z dot] (24) at (2.00, -10.50) {};
        \node [style=Z dot] (25) at (-1.00, -6.00) {};
        \node [style=X dot] (26) at (-1.00, -7.00) {};
        \node [style=X dot] (27) at (0.00, -7.00) {};
        \node [style=X dot] (28) at (1.00, -8.00) {};
        \node [style=Z dot] (29) at (1.00, -4.00) {};
        \node [style=X dot] (30) at (4.00, -10.50) {};
        \node [style=X phase dot] (31) at (1.00, -12.00) {$\frac{\pi}{4}$};
        \node [style=Z dot] (32) at (3.00, -10.50) {};
        \node [style=X dot] (33) at (-1.00, -5.00) {};
        \node [style=none] (34) at (7.95, -1.50) {};
        \node [style=none] (36) at (7.95, -5.00) {};
        \node [style=none] (37) at (7.95, -8.00) {};
        \node [style=Z phase dot] (38) at (4.00, -0.00) {$\frac{\pi}{4}$};
        \node [style=none] (39) at (-0.50, -1.00) {};
        \node [style=none] (40) at (5.00, -11.00) {};
        \node [style=X dot] (41) at (8.00, -7.00) {};
        \node [style=Z dot] (42) at (8.00, -4.00) {};
        \node [style=X dot] (43) at (9.00, -7.00) {};
        \node [style=X dot] (44) at (11.00, -5.00) {};
        \node [style=Z dot] (45) at (15.00, -4.00) {};
        \node [style=Z dot] (46) at (13.00, -4.00) {};
        \node [style=none] (47) at (16.00, -1.50) {};
        \node [style=Z dot] (48) at (11.00, -7.00) {};
        \node [style=X dot] (49) at (15.00, -10.50) {};
        \node [style=X dot] (50) at (10.00, -8.00) {};
        \node [style=X dot] (51) at (10.00, -5.00) {};
        \node [style=Z dot] (52) at (9.00, -4.00) {};
        \node [style=none] (53) at (14.00, -11.00) {};
        \node [style=Z dot] (54) at (13.00, -6.00) {};
        \node [style=X dot] (55) at (11.00, -1.50) {};
        \node [style=X dot] (56) at (12.00, -7.00) {};
        \node [style=Z dot] (57) at (9.00, -6.00) {};
        \node [style=Z dot] (58) at (10.00, -7.00) {};
        \node [style=X dot] (59) at (9.00, -10.50) {};
        \node [style=none] (60) at (16.00, -5.00) {};
        \node [style=none] (61) at (16.00, -8.00) {};
        \node [style=none] (62) at (7.50, -8.00) {};
        \node [style=X dot] (63) at (12.00, -8.00) {};
        \node [style=Z dot] (64) at (11.00, -6.00) {};
        \node [style=Z dot] (65) at (12.00, -10.50) {};
        \node [style=Z phase dot] (66) at (10.00, -12.00) {$\frac{\pi}{4}$};
        \node [style=X dot] (67) at (8.00, -6.00) {};
        \node [style=none] (68) at (8.50, -1.00) {};
        \node [style=Z dot] (69) at (12.00, -4.00) {};
        \node [style=Z dot] (70) at (8.00, -10.50) {};
        \node [style=none] (71) at (7.50, -5.00) {};
        \node [style=Z dot] (72) at (15.00, -6.00) {};
        \node [style=X dot] (73) at (15.00, -7.00) {};
        \node [style=none] (74) at (7.50, -1.50) {};
        \node [style=none] (75a) at (11.075, -12.00) {$\dots$};
        \node [style=Z phase dot] (75) at (12.00, -12.00) {$\frac{7\pi}{4}$};
        \node [style=X dot] (76) at (13.00, -10.50) {};
        \node [style=Z dot] (77) at (10.00, -4.00) {};
        \node [style=Z dot] (78) at (11.00, -4.00) {};
        \node [style=none] (79) at (6.50, -2.75) {$\vdots$};
        \node [style=none] (80) at (6.50, -9.25) {$\vdots$};
        \node [style=none] (81) at (15.50, -9.25) {$\vdots$};
        \node [style=none] (82) at (15.50, -2.75) {$\vdots$};
        \node [style=none] (83) at (7.25, .75) {$t=\tau$};
    \end{pgfonlayer}
    \begin{pgfonlayer}{edgelayer}
        \draw (0) to (28);
        \draw (0) to (27);
        \draw (0) to (13);
        \draw (1) to (8);
        \draw (1) to (21);
        \draw (1) to (36);
        \draw (2) to (22);
        \draw (3) to (5);
        \draw (4) to (18);
        \draw (4) to (30);
        \draw (4) to (8);
        \draw (5) to (23);
        \draw (5) to (38);
        \draw (6) to (11);
        \draw (6) to (12);
        \draw (6) to (24);
        \draw (7) to (20);
        \draw (8) to (12);
        \draw (9) to (17);
        \draw (10) to (30);
        \draw (12) to (25);
        \draw (13) to (17);
        \draw (14) to (20);
        \draw (14) to (23);
        \draw (14) to (29);
        \draw (15) to (32);
        \draw (15) to (37);
        \draw (16) to (28);
        \draw (17) to (23);
        \draw (20) to (34);
        \draw (21) to (29);
        \draw (21) to (33);
        \draw (22) to (29);
        \draw (22) to (27);
        \draw (24) to (32);
        \draw (26) to (27);
        \draw (28) to (31);
        \draw (30) to (32);
        \draw (41) to (43);
        \draw (42) to (52);
        \draw (43) to (52);
        \draw (43) to (58);
        \draw (44) to (64);
        \draw (44) to (51);
        \draw (44) to (60);
        \draw (45) to (46);
        \draw (46) to (69);
        \draw (47) to (55);
        \draw (48) to (56);
        \draw (48) to (58);
        \draw (49) to (76);
        \draw (50) to (66);
        \draw (50) to (58);
        \draw (50) to (62);
        \draw (51) to (77);
        \draw (51) to (71);
        \draw (52) to (77);
        \draw (54) to (72);
        \draw (54) to (76);
        \draw (54) to (64);
        \draw (55) to (74);
        \draw (55) to (78);
        \draw (56) to (69);
        \draw (56) to (73);
        \draw (57) to (67);
        \draw (57) to (59);
        \draw (57) to (64);
        \draw (59) to (70);
        \draw (61) to (63);
        \draw (63) to (65);
        \draw (65) to (75);
        \draw (65) to (76);
        \draw (69) to (78);
        \draw (77) to (78);  
        \draw[dashed] (7.25, 0.25) -- (7.25, -12.25);

    \end{pgfonlayer}
    \draw [fill = white] (53) rectangle (68);
	\draw [fill = white] (39) rectangle (40);
    \node [] (84) at (2.50, -6.00) {$\mathcal{G}$};
    \node [] (85) at (11.50, -6.00) {$\mathcal{H}$};

\end{tikzpicture}
}$.} 
        \Output{$g_{\text{meas}}$, ZX-diagram with sampled measurements.}
        \\
        \Function{SubDiagram}{$g,\tau$}
            \State Obtain sub-diagram up to the first measurement time slice $(t = \tau)$: $h$. 
            \State \Return $h= \scalebox{0.24}{
\begin{tikzpicture}[yshift=6.5cm]
    \begin{pgfonlayer}{nodelayer}
        \node [style=Z dot] (2) at (4.50, -4.50) {};
        \node [style=X dot] (3) at (6.50, -11.00) {};
        \node [style=X dot] (4) at (1.50, -8.50) {};
        \node [style=X dot] (6) at (1.50, -5.50) {};
        \node [style=none] (8) at (5.50, -11.50) {};
        \node [style=X dot] (9) at (2.50, -2.00) {};
        \node [style=Z dot] (10) at (0.50, -6.50) {};
        \node [style=Z dot] (11) at (1.50, -7.50) {};
        \node [style=Z phase dot] (14) at (4.50, -0.50) {$\frac{\pi}{4}$};
        \node [style=Z dot] (16) at (2.50, -6.50) {};
        \node [style=X dot] (17) at (-0.50, -11.00) {};
        \node [style=Z dot] (18) at (1.50, -4.50) {};
        \node [style=Z dot] (20) at (2.50, -4.50) {};
        \node [style=X dot] (23) at (0.50, -7.50) {};
        \node [style=Z dot] (24) at (6.50, -4.50) {};
        \node [style=Z dot] (29) at (0.50, -4.50) {};
        \node [style=X dot] (30) at (3.50, -7.50) {};
        \node [style=Z dot] (36) at (-0.50, -6.50) {};
        \node [style=Z dot] (37) at (3.50, -4.50) {};
        \node [style=Z dot] (39) at (6.50, -6.50) {};
        \node [style=Z dot] (40) at (2.50, -11.00) {};
        \node [style=X dot] (43) at (2.50, -5.50) {};
        \node [style=Z dot] (48) at (2.50, -7.50) {};
        \node [style=Z dot] (50) at (4.50, -6.50) {};
        \node [style=none] (51) at (7.25, -5.50) {};
        \node [style=none] (52) at (7.25, -8.50) {};
        \node [style=X dot] (54) at (3.50, -8.50) {};
        \node [style=Z dot] (59) at (3.50, -11.00) {};
        \node [style=X phase dot] (60) at (1.50, -12.50) {$\frac{\pi}{4}$};
        \node [style=X dot] (62) at (6.50, -7.50) {};
        \node [style=X dot] (66) at (-0.50, -7.50) {};
        \node [style=Z dot] (67) at (-0.50, -4.50) {};
        \node [style=none] (72) at (7.25, -2.00) {};
        \node [style=none] (73) at (7.00, -3.25) {$\vdots$};
        \node [style=X dot] (74) at (0.50, -11.00) {};
        \node [style=X dot] (75) at (-0.50, -8.50) {};
        \node [style=none] (79) at (7.00, -9.75) {$\vdots$};
        \node [style=none] (80) at (0.00, -1.50) {};
        \node [style=X dot] (81) at (-0.50, -5.50) {};
        \node [style=X dot] (82) at (-0.50, -2.00) {};
        \node [style=X dot] (83) at (4.50, -11.00) {};
    \end{pgfonlayer}
    \begin{pgfonlayer}{edgelayer}
        \draw (2) to (37);
        \draw (2) to (14);
        \draw (2) to (24);
        \draw (3) to (83);
        \draw (4) to (60);
        \draw (4) to (11);
        \draw (4) to (54);
        \draw (4) to (75);
        \draw (6) to (18);
        \draw (6) to (81);
        \draw (6) to (43);
        \draw (9) to (72);
        \draw (9) to (20);
        \draw (9) to (82);
        \draw (10) to (36);
        \draw (10) to (16);
        \draw (10) to (74);
        \draw (11) to (23);
        \draw (11) to (48);
        \draw (16) to (43);
        \draw (16) to (50);
        \draw (17) to (74);
        \draw (18) to (20);
        \draw (18) to (29);
        \draw (20) to (37);
        \draw (23) to (29);
        \draw (23) to (66);
        \draw (29) to (67);
        \draw (30) to (37);
        \draw (30) to (48);
        \draw (30) to (62);
        \draw (39) to (50);
        \draw (40) to (59);
        \draw (40) to (74);
        \draw (43) to (51);
        \draw (50) to (83);
        \draw (52) to (54);
        \draw (54) to (59);
        \draw (59) to (83);
        \draw[dashed] (7.25, 0.25) -- (7.25, -12.25);

    \end{pgfonlayer}
	\draw [fill = white] (8) rectangle (80);
    \node [] (84) at (3.00, -6.500) {$\mathcal{G}$};
    \node [style=none] (7) at (7.25, 0.85) {$t=\tau$};
\end{tikzpicture}
}$
        \EndFunction
        \\
        \Function{StripMeasSpiders}{$h,\tau$}
            \State Strip measurement spiders in ZX-diagram $h$ at timeslice $t=\tau$.
            \State \Return $\displaystyle h = \scalebox{0.25}{
\begin{tikzpicture}[yshift=6.5cm]
    \begin{pgfonlayer}{nodelayer}
        \node [style=Z dot] (2) at (4.50, -4.50) {};
        \node [style=none] (3) at (6.50, -11.00) {};
        \node [style=X dot] (4) at (1.50, -8.50) {};
        \node [style=X dot] (6) at (1.50, -5.50) {};
        \node [style=none] (8) at (5.50, -11.50) {};
        \node [style=X dot] (9) at (2.50, -2.00) {};
        \node [style=Z dot] (10) at (0.50, -6.50) {};
        \node [style=Z dot] (11) at (1.50, -7.50) {};
        \node [style=Z phase dot] (14) at (4.50, -0.50) {$\frac{\pi}{4}$};
        \node [style=Z dot] (16) at (2.50, -6.50) {};
        \node [style=X dot] (17) at (-0.50, -11.00) {};
        \node [style=Z dot] (18) at (1.50, -4.50) {};
        \node [style=Z dot] (20) at (2.50, -4.50) {};
        \node [style=X dot] (23) at (0.50, -7.50) {};
        \node [style=none] (24) at (6.50, -4.50) {};
        \node [style=Z dot] (29) at (0.50, -4.50) {};
        \node [style=X dot] (30) at (3.50, -7.50) {};
        \node [style=Z dot] (36) at (-0.50, -6.50) {};
        \node [style=Z dot] (37) at (3.50, -4.50) {};
        \node [style=none] (39) at (6.50, -6.50) {};
        \node [style=Z dot] (40) at (2.50, -11.00) {};
        \node [style=X dot] (43) at (2.50, -5.50) {};
        \node [style=Z dot] (48) at (2.50, -7.50) {};
        \node [style=Z dot] (50) at (4.50, -6.50) {};
        \node [style=none] (51) at (7.25, -5.50) {};
        \node [style=none] (52) at (7.25, -8.50) {};
        \node [style=X dot] (54) at (3.50, -8.50) {};
        \node [style=Z dot] (59) at (3.50, -11.00) {};
        \node [style=X phase dot] (60) at (1.50, -12.50) {$\frac{\pi}{4}$};
        \node [style=none] (62) at (6.50, -7.50) {};
        \node [style=X dot] (66) at (-0.50, -7.50) {};
        \node [style=Z dot] (67) at (-0.50, -4.50) {};
        \node [style=none] (72) at (7.25, -2.00) {};
        \node [style=none] (73) at (7.00, -3.25) {$\vdots$};
        \node [style=X dot] (74) at (0.50, -11.00) {};
        \node [style=X dot] (75) at (-0.50, -8.50) {};
        \node [style=none] (79) at (7.00, -9.75) {$\vdots$};
        \node [style=none] (80) at (0.00, -1.50) {};
        \node [style=X dot] (81) at (-0.50, -5.50) {};
        \node [style=X dot] (82) at (-0.50, -2.00) {};
        \node [style=X dot] (83) at (4.50, -11.00) {};
    \end{pgfonlayer}
    \begin{pgfonlayer}{edgelayer}
        \draw (2) to (37);
        \draw (2) to (14);
        \draw (2) to (24);
        \draw (3) to (83);
        \draw (4) to (60);
        \draw (4) to (11);
        \draw (4) to (54);
        \draw (4) to (75);
        \draw (6) to (18);
        \draw (6) to (81);
        \draw (6) to (43);
        \draw (9) to (72);
        \draw (9) to (20);
        \draw (9) to (82);
        \draw (10) to (36);
        \draw (10) to (16);
        \draw (10) to (74);
        \draw (11) to (23);
        \draw (11) to (48);
        \draw (16) to (43);
        \draw (16) to (50);
        \draw (17) to (74);
        \draw (18) to (20);
        \draw (18) to (29);
        \draw (20) to (37);
        \draw (23) to (29);
        \draw (23) to (66);
        \draw (29) to (67);
        \draw (30) to (37);
        \draw (30) to (48);
        \draw (30) to (62);
        \draw (39) to (50);
        \draw (40) to (59);
        \draw (40) to (74);
        \draw (43) to (51);
        \draw (50) to (83);
        \draw (52) to (54);
        \draw (54) to (59);
        \draw (59) to (83);
        \draw[dashed] (7.25, 0.25) -- (7.25, -12.25);

    \end{pgfonlayer}
	\draw [fill = white] (8) rectangle (80);
    \node [] (84) at (3.00, -6.500) {$\mathcal{G}$};
    \node [style=none] (7) at (7.25, 0.85) {$t=\tau$};
\end{tikzpicture}
}$
        \EndFunction
        \\
        \Function{StabiliserDecompose}{$h$}
            \State Perform stabiliser decomposition (cutting and further secondary decompositions) on ZX-diagram $h$.
            \State \Return $\displaystyle s = \sum_{j=1}^{\chi} a_j \begin{pmatrix}
        \scalebox{0.25}{\input{cc_meas.tikz}}
        \end{pmatrix}_j$
        \EndFunction
        \\
        \Function{SampleAndReplaceMeasSpiders}{$s,g$}
            \State Sample decomposed sum of Clifford ZX-diagrams, only at the stripped measurement spider legs. Obtaining measurement sample: $\vec{M} = (\pi, 0, 0, \pi, \dots)$ for example. Replace original ZX-diagram $g$ with measurement results.
            
            \State \Return $g_{\text{meas}} = \scalebox{0.25}{
\begin{tikzpicture}[yshift=6cm]
    \begin{pgfonlayer}{nodelayer}
        \node [style=Z dot] (0) at (1.00, -7.00) {};
        \node [style=X dot] (1) at (2.00, -5.00) {};
        \node [style=Z dot] (2) at (-1.00, -4.00) {};
        \node [style=Z phase dot] (3) at (6.00, -4.00) {$\pi$};
        \node [style=Z dot] (4) at (4.00, -6.00) {$\pi$};
        \node [style=Z dot] (5) at (4.00, -4.00) {};
        \node [style=X dot] (6) at (0.00, -10.50) {};
        \node [style=X dot] (7) at (-1.00, -1.50) {};
        \node [style=Z dot] (8) at (2.00, -6.00) {};
        \node [style=X dot] (9) at (6.00, -7.00) {};
        \node [style=X dot] (10) at (6.00, -10.50) {};
        \node [style=X dot] (11) at (-1.00, -10.50) {};
        \node [style=Z dot] (12) at (0.00, -6.00) {};
        \node [style=Z dot] (13) at (2.00, -7.00) {};
        \node [style=Z dot] (14) at (2.00, -4.00) {};
        \node [style=X dot] (15) at (3.00, -8.00) {};
        \node [style=X dot] (16) at (-1.00, -8.00) {};
        \node [style=X dot] (17) at (3.00, -7.00) {};
        \node [style=Z dot] (18) at (6.00, -6.00) {};
        \node [style=X dot] (20) at (2.00, -1.50) {};
        \node [style=X dot] (21) at (1.00, -5.00) {};
        \node [style=Z dot] (22) at (0.00, -4.00) {};
        \node [style=Z dot] (23) at (3.00, -4.00) {};
        \node [style=Z dot] (24) at (2.00, -10.50) {};
        \node [style=Z dot] (25) at (-1.00, -6.00) {};
        \node [style=X dot] (26) at (-1.00, -7.00) {};
        \node [style=X dot] (27) at (0.00, -7.00) {};
        \node [style=X dot] (28) at (1.00, -8.00) {};
        \node [style=Z dot] (29) at (1.00, -4.00) {};
        \node [style=X dot] (30) at (4.00, -10.50) {};
        \node [style=X phase dot] (31) at (1.00, -12.00) {$\frac{\pi}{4}$};
        \node [style=Z dot] (32) at (3.00, -10.50) {};
        \node [style=X dot] (33) at (-1.00, -5.00) {};
        \node [style=none] (34) at (7.95, -1.50) {};
        \node [style=none] (36) at (7.95, -5.00) {};
        \node [style=none] (37) at (7.95, -8.00) {};
        \node [style=Z phase dot] (38) at (4.00, -0.00) {$\frac{\pi}{4}$};
        \node [style=none] (39) at (-0.50, -1.00) {};
        \node [style=none] (40) at (5.00, -11.00) {};
        \node [style=X dot] (41) at (8.00, -7.00) {};
        \node [style=Z dot] (42) at (8.00, -4.00) {};
        \node [style=X dot] (43) at (9.00, -7.00) {};
        \node [style=X dot] (44) at (11.00, -5.00) {};
        \node [style=Z dot] (45) at (15.00, -4.00) {};
        \node [style=Z dot] (46) at (13.00, -4.00) {};
        \node [style=none] (47) at (16.00, -1.50) {};
        \node [style=Z dot] (48) at (11.00, -7.00) {};
        \node [style=X dot] (49) at (15.00, -10.50) {};
        \node [style=X dot] (50) at (10.00, -8.00) {};
        \node [style=X dot] (51) at (10.00, -5.00) {};
        \node [style=Z dot] (52) at (9.00, -4.00) {};
        \node [style=none] (53) at (14.00, -11.00) {};
        \node [style=Z dot] (54) at (13.00, -6.00) {};
        \node [style=X dot] (55) at (11.00, -1.50) {};
        \node [style=X dot] (56) at (12.00, -7.00) {};
        \node [style=Z dot] (57) at (9.00, -6.00) {};
        \node [style=Z dot] (58) at (10.00, -7.00) {};
        \node [style=X dot] (59) at (9.00, -10.50) {};
        \node [style=none] (60) at (16.00, -5.00) {};
        \node [style=none] (61) at (16.00, -8.00) {};
        \node [style=none] (62) at (7.50, -8.00) {};
        \node [style=X dot] (63) at (12.00, -8.00) {};
        \node [style=Z dot] (64) at (11.00, -6.00) {};
        \node [style=Z dot] (65) at (12.00, -10.50) {};
        \node [style=Z phase dot] (66) at (10.00, -12.00) {$\frac{\pi}{4}$};
        \node [style=X dot] (67) at (8.00, -6.00) {};
        \node [style=none] (68) at (8.50, -1.00) {};
        \node [style=Z dot] (69) at (12.00, -4.00) {};
        \node [style=Z dot] (70) at (8.00, -10.50) {};
        \node [style=none] (71) at (7.50, -5.00) {};
        \node [style=Z dot] (72) at (15.00, -6.00) {};
        \node [style=X dot] (73) at (15.00, -7.00) {};
        \node [style=none] (74) at (7.50, -1.50) {};
        \node [style=none] (75a) at (11.075, -12.00) {$\dots$};
        \node [style=Z phase dot] (75) at (12.00, -12.00) {$\frac{7\pi}{4}$};
        \node [style=X dot] (76) at (13.00, -10.50) {};
        \node [style=Z dot] (77) at (10.00, -4.00) {};
        \node [style=Z dot] (78) at (11.00, -4.00) {};
        \node [style=none] (79) at (6.50, -2.75) {$\vdots$};
        \node [style=none] (80) at (6.50, -9.25) {$\vdots$};
        \node [style=none] (81) at (15.50, -9.25) {$\vdots$};
        \node [style=none] (82) at (15.50, -2.75) {$\vdots$};
        \node [style=none] (83) at (7.25, .75) {$t=\tau$};
    \end{pgfonlayer}
    \begin{pgfonlayer}{edgelayer}
        \draw (0) to (28);
        \draw (0) to (27);
        \draw (0) to (13);
        \draw (1) to (8);
        \draw (1) to (21);
        \draw (1) to (36);
        \draw (2) to (22);
        \draw (3) to (5);
        \draw (4) to (18);
        \draw (4) to (30);
        \draw (4) to (8);
        \draw (5) to (23);
        \draw (5) to (38);
        \draw (6) to (11);
        \draw (6) to (12);
        \draw (6) to (24);
        \draw (7) to (20);
        \draw (8) to (12);
        \draw (9) to (17);
        \draw (10) to (30);
        \draw (12) to (25);
        \draw (13) to (17);
        \draw (14) to (20);
        \draw (14) to (23);
        \draw (14) to (29);
        \draw (15) to (32);
        \draw (15) to (37);
        \draw (16) to (28);
        \draw (17) to (23);
        \draw (20) to (34);
        \draw (21) to (29);
        \draw (21) to (33);
        \draw (22) to (29);
        \draw (22) to (27);
        \draw (24) to (32);
        \draw (26) to (27);
        \draw (28) to (31);
        \draw (30) to (32);
        \draw (41) to (43);
        \draw (42) to (52);
        \draw (43) to (52);
        \draw (43) to (58);
        \draw (44) to (64);
        \draw (44) to (51);
        \draw (44) to (60);
        \draw (45) to (46);
        \draw (46) to (69);
        \draw (47) to (55);
        \draw (48) to (56);
        \draw (48) to (58);
        \draw (49) to (76);
        \draw (50) to (66);
        \draw (50) to (58);
        \draw (50) to (62);
        \draw (51) to (77);
        \draw (51) to (71);
        \draw (52) to (77);
        \draw (54) to (72);
        \draw (54) to (76);
        \draw (54) to (64);
        \draw (55) to (74);
        \draw (55) to (78);
        \draw (56) to (69);
        \draw (56) to (73);
        \draw (57) to (67);
        \draw (57) to (59);
        \draw (57) to (64);
        \draw (59) to (70);
        \draw (61) to (63);
        \draw (63) to (65);
        \draw (65) to (75);
        \draw (65) to (76);
        \draw (69) to (78);
        \draw (77) to (78);  
        \draw[dashed] (7.25, 0.25) -- (7.25, -12.25);

    \end{pgfonlayer}
    \draw [fill = white] (53) rectangle (68);
	\draw [fill = white] (39) rectangle (40);
    \node [] (84) at (2.50, -6.00) {$\mathcal{G}$};
    \node [] (85) at (11.50, -6.00) {$\mathcal{H}$};

\end{tikzpicture}
}$
        \EndFunction
        \\
        \Function{Main}{$g$}
            \For {$\tau \in \{1,2,3,...,\text{End}\}$ }
                \State $h \gets$ \Call{SubDiagram}{$g,\tau$}
                \State $h \gets$ \Call{StripMeasSpiders}{$h,\tau$}
                \State $s \gets$ \Call{StabiliserDecompose}{$h$}
                \State $g \gets$ \Call{SampleAndReplaceMeasSpiders}{$s,g$}
            \EndFor
          
            \State \Return $g$ (with measurement spider legs sampled)
        \EndFunction
        \\
        \State $g_{\text{meas}} \gets$ \Call{Main}{$g$} 
        \\
        \end{algorithmic}   
    \caption{Sample measurements.}
    \label{algo:sample}
\end{algorithm}

\section{A few detecting regions for the $d=3$ circuit with degenerate injection}
\label{appendix:d3_webs_det}
Following the same detector numbering from \cite{gidneyy2024cultivationdata}, here are a few detectors for the $d=3$ circuits with degenerate injection.
\foreach \i in {0,2,18} {%
  \begin{figure}[h!]
    \centering
    \includegraphics[width=0.8\linewidth]{d3_det_webs/det_\i.pdf}
    \caption{Detecting region/closed Pauli web/detector $\i$.}
  \end{figure}
}

\section{A few detecting regions for the $d=5$ circuit with degenerate injection}
\label{appendix:d5_webs_det}
Again, following the same detector numbering from \cite{gidneyy2024cultivationdata}, here are $6$ detectors for the $d=5$ circuit with degenerate injection. The rest are available in the supplementary tikz folder $\mathtt{d5\_det\_webs}$.

\foreach \i in {81,48,55} {%
  \begin{figure}[h!]
    \centering
    \includegraphics[width=0.8\linewidth]{d5_det_webs/det_\i.pdf}
    \caption{Detector $\i$.}
  \end{figure}
}

\clearpage

\section{Sketch example}
\label{appendix:sketch_example}
The figure illustrates the sample measurements approach applied to the initial stages of the $d=3$ cultivation circuit, showing the contraction up to the first measurement layer and subsequent stabiliser decomposition.
\begin{equation*}
\centering
    \label{eq:sketch}
    \scalebox{0.35}{\includegraphics{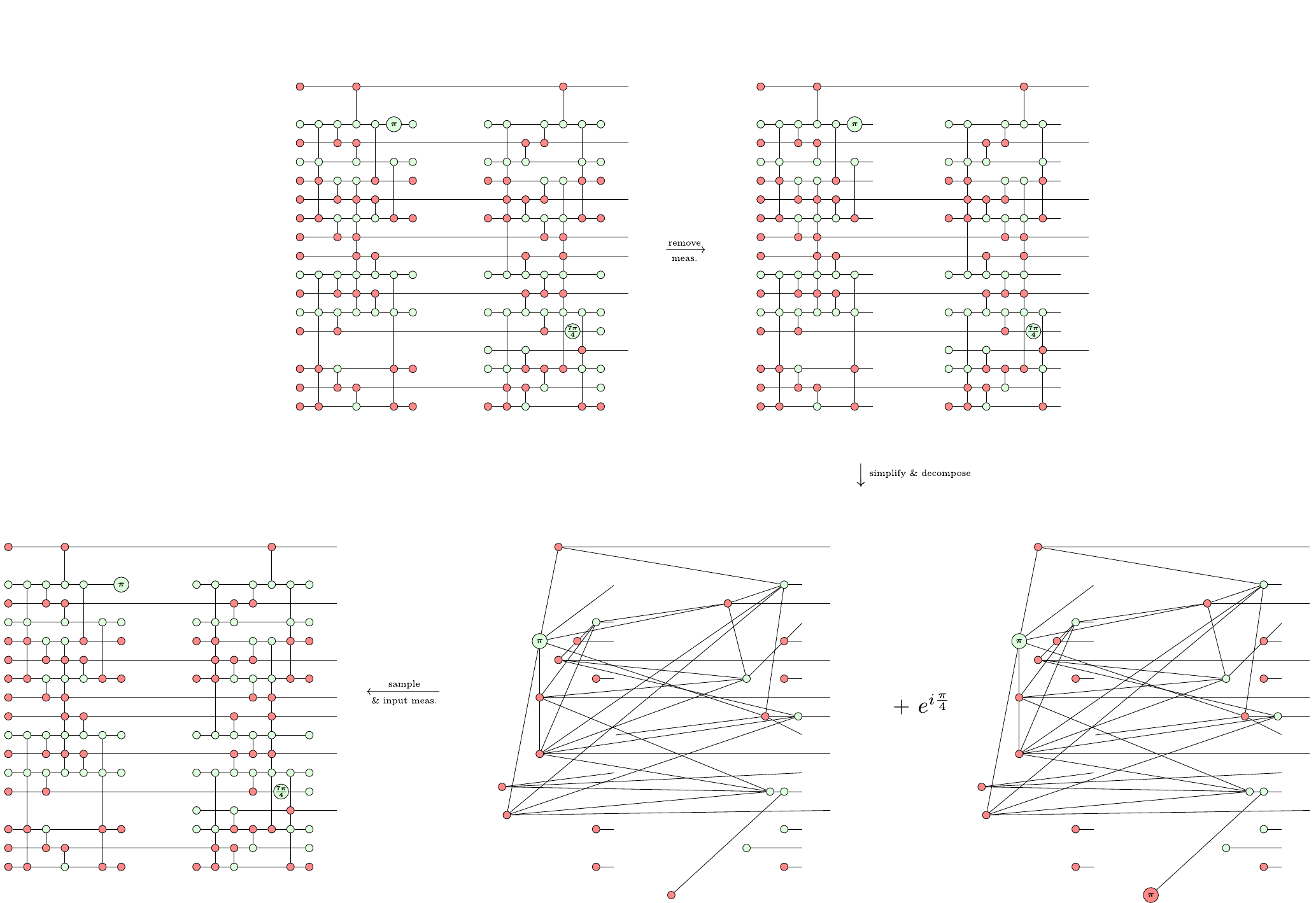}}
\end{equation*}

\section{Error model}
\label{appendix:error_model}

This paper uses two noise models in different sections. In the analytical sections preceding the numerical methods (Sections~6--9), we apply a depolarising channel to \textit{every edge} of the ZX-diagram:
\begin{equation*}
    \mathcal{E}(\rho) = (1-p)\rho + \frac{p}{3}(X\rho X + Y\rho Y + Z\rho Z),
\end{equation*}
where $p$ is the edge error rate. Each Pauli error (${\color{red}X}$, ${\color{blue}Y}$, or ${\color{green}Z}$) is represented as a coloured half-edge decoration, following the Pauli web notation of \cite{Bombin_2024}. This edge-level model has a higher error density than standard circuit-level noise, since errors can appear on every wire segment including between consecutive gates.

In the numerical methods sections (Sections~10--11), we instead apply uniform circuit-level depolarising noise via $\mathtt{tsim}$ \cite{QuEraComputing_tsim_2026} ($\mathtt{bloqade\text{-}tsim}$ v0.1.0), where noise is placed after each gate in the circuit (consistent with standard practice). This is applied to the injection and cultivation (double-checking) stages only; a noiseless projection circuit is used to extract logical error rates at the end.

\begin{landscape}
\section{Cutting decomposition diagrams}
\label{appendix:cutting_diagrams}
\begin{figure}[H]
\centering
\makeatletter
\begin{tikzpicture}[baseline=(m.center)]
  \matrix (m) [matrix of nodes,
               row sep=3mm, column sep=2mm,
               nodes={anchor=center, inner sep=0pt},
               ampersand replacement=\&]
  {
    \scalebox{0.42}{\includegraphics{MSC_d_3_full.pdf}} \& = \&
    \scalebox{0.9}{\input{cut2_term0_s.tikz}} \& + \&
    \scalebox{0.9}{\input{cut2_term1_s.tikz}} \\
  };
\end{tikzpicture}
\makeatother
\label{eq:large_d3}

\vspace{0.3cm}
\noindent\rule{\linewidth}{0.4pt}
\vspace{0.3cm}

\makeatletter
\begin{tikzpicture}[baseline=(m.center)]
  \matrix (m) [matrix of nodes,
               row sep=3mm, column sep=2mm,
               nodes={anchor=center, inner sep=0pt}]
  {
    \scalebox{0.12}{\includegraphics{d5_dashed.pdf}} & = &
    \scalebox{0.12}{\includegraphics{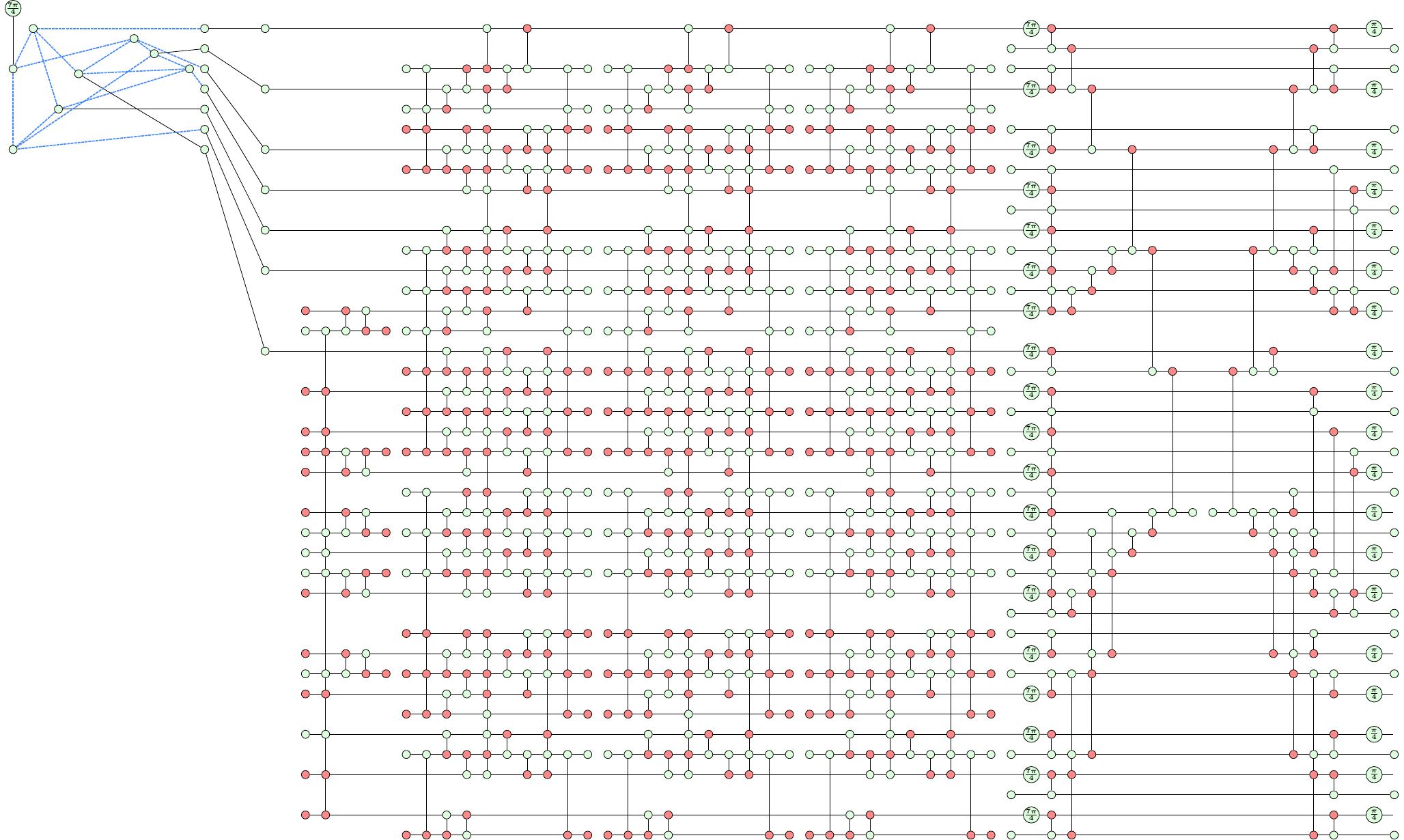}} & + &
    \scalebox{0.12}{\includegraphics{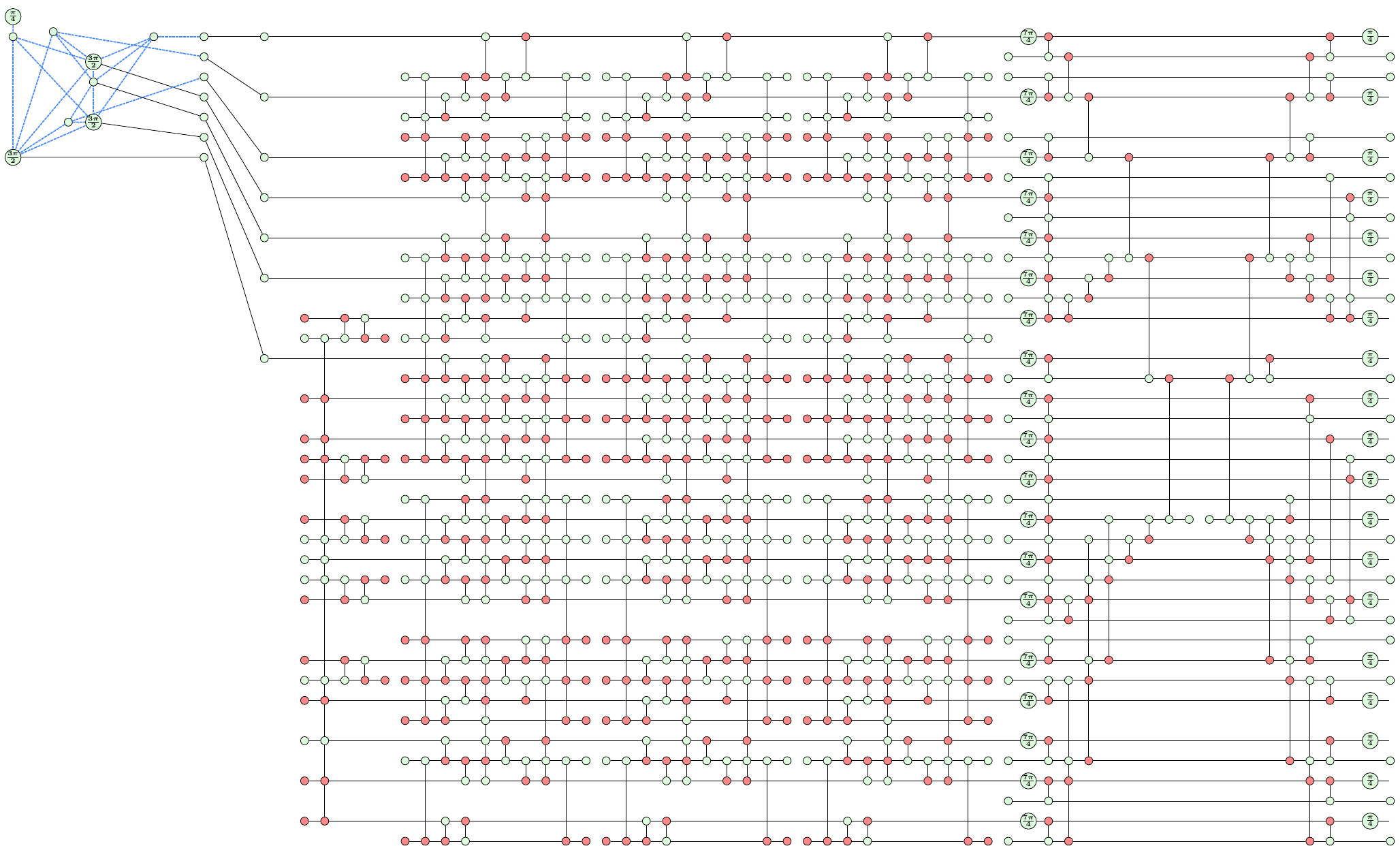}} & & & &\\
      & = &
    \scalebox{0.225}{\includegraphics{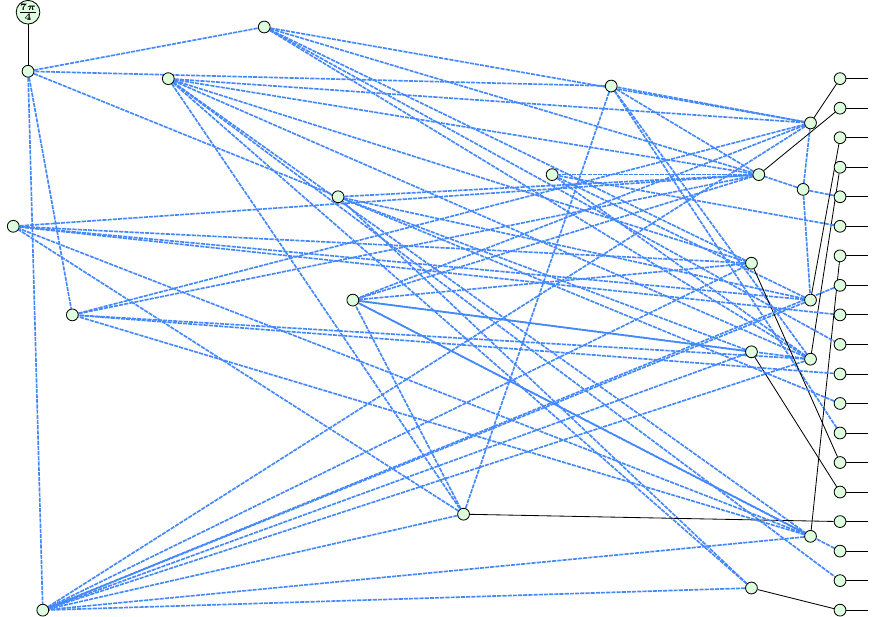}} & + &
    \scalebox{0.225}{\includegraphics{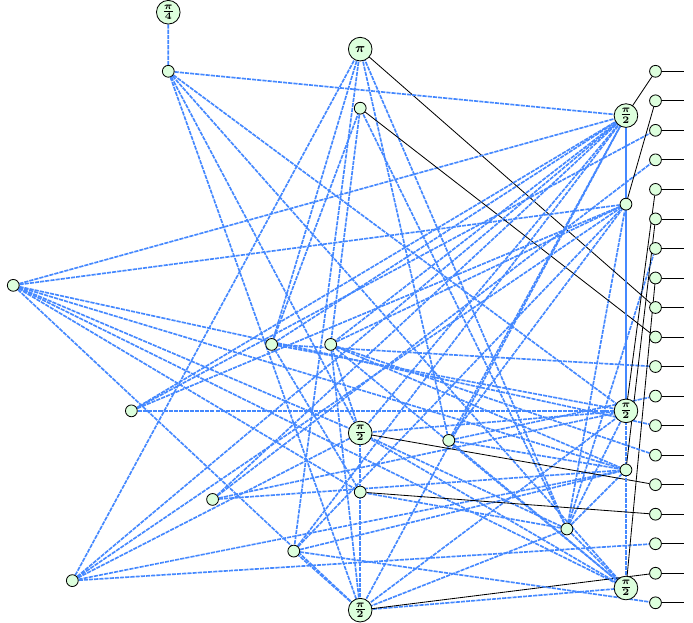}} & + &
    \scalebox{0.225}{\includegraphics{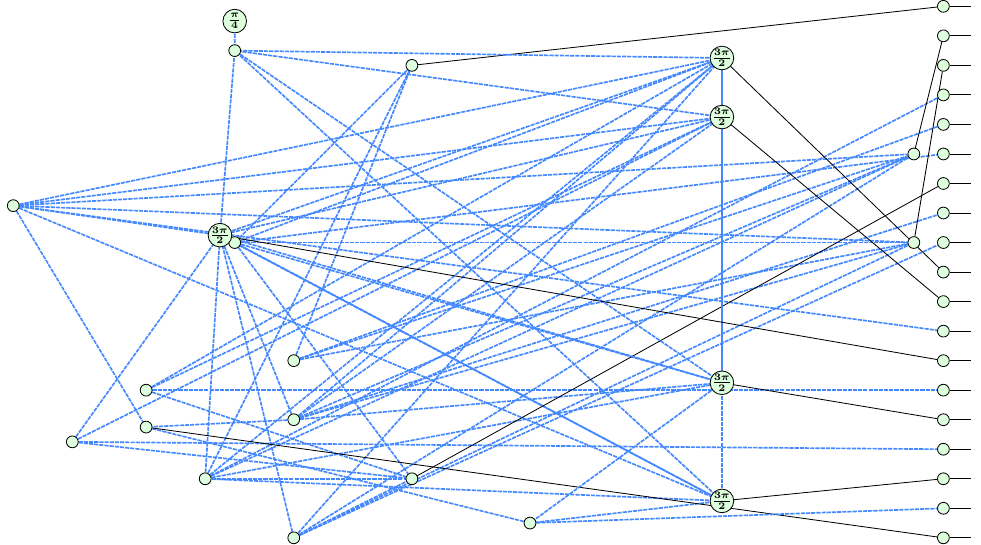}} & + &
    \scalebox{0.225}{\includegraphics{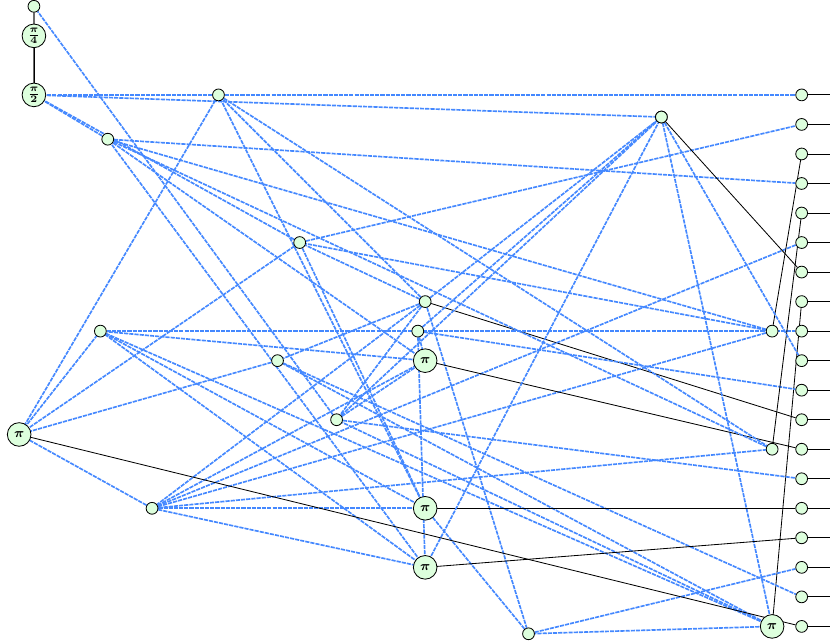}} \\
  };
\end{tikzpicture}
\makeatother
\label{eq:large_d5}
\caption{\label{fig:large_d3_d5}Cutting stabiliser decompositions of the magic state cultivation circuits. Top: the $d=3$ circuit decomposes into two Clifford ZX-diagrams. Bottom: the $d=5$ circuit decomposes into four Clifford ZX-diagrams via two intermediate terms.}
\end{figure}
\end{landscape}

\end{document}